\documentclass[aps,pra,twocolumn,groupedaddress]{revtex4-2}
\usepackage{blindtext}
\usepackage{enumitem}
\usepackage{braket}
\usepackage{float}
\usepackage{amssymb}
\usepackage{stackengine}
\usepackage{mathrsfs}
\usepackage{ stmaryrd }
\usepackage{ bbold }
\usepackage{graphicx}
\usepackage{epstopdf}
\usepackage{amsmath}
\usepackage{diagbox}
\usepackage[dvipsnames]{xcolor}
\usepackage{tikz}
\usepackage[bottom=0.95in, right=0.62in]{geometry}

\newcommand{\ssymbol}[1]{^{\@fnsymbol{#1}}}
\input{alphabet.tex}
\usepackage{ dsfont }

\begin{document}


\title{Unitary quantum process tomography with unreliable pure input states}


\author{François Verdeil}
\affiliation{Université de Toulouse, UPS, CNRS, CNES, OMP, IRAP, Toulouse, France}
\author{Yannick Deville}
\affiliation{Université de Toulouse, UPS, CNRS, CNES, OMP, IRAP, Toulouse, France}

\date{\today}

\begin{abstract}
Quantum process tomography (QPT) methods aim at identifying a given quantum process. QPT is a major quantum information processing tool, since it allows one to characterize the actual behavior of quantum gates, which are the building blocks of quantum computers. The present paper focuses on the estimation of a unitary process. This class is of particular interest because quantum mechanics postulates that the evolution of any closed quantum system is described by a unitary transformation. The standard approach of QTP is to measure copies of a particular set of predetermined (generally pure) states after they have been modified by the process to be identified. The main problem with this setup is that preparing an input state and setting it precisely to a predetermined value is challenging and thus yields errors. These errors can be decomposed into a sum of centred errors (i.e. whose average on all the copies is zero) and systematic errors that are the same on all the copies, the latter is often the main source of error in QPT. The algorithm we introduce in the current paper works for any input states that make QPT theoretically possible (i.e. unless there are several solutions due to e.g. a lack of diversity on the input states). The fact that we do not require the input states to be precisely set to predetermined values means that we can use a trick to remove the issue of systematic errors by considering that some states are unknown but measured before they go through the process to be identified. We achieve this by splitting the copies of each input state into several groups and measuring the copies of the $k$-th group after they have successively been transferred through $k$ instances of the process to be identified (each copy of each input state is only measured once). Using this trick we can compute estimates of the measured states before and after they go through the process without using the knowledge we might have on the initial states. We test our algorithm both on simulated data and experimentally to identify a CNOT gate on a trapped-ions qubit quantum computer.

\end{abstract}
\maketitle

\section{Introduction}
Quantum process tomography aims at identifying the quantum process associated with a physical quantum gate. It was first introduced in \cite{NC} and \cite{SQPT0}, the former came up with the name quantum process tomography (QPT). They use copies of a set of known input states that are transformed by the process. Those transformed states are then measured and estimated using quantum state tomography (QST aims at estimating a quantum state using measurements). This method scales really badly when the number of qubits increases. This is to be expected because, in general, a quantum process has $d^4-d^2$ independent real parameters, with $d$ the dimension of the Hilbert space (for an $n_{qb}$-qubit system $d=2^{n_{qb}}$ \cite{bookNC} p. 391). 

This method and others inspired by it would later be called standard QPT (SQPT), in contrast to non-standard QPT that uses ancilla qubits and weak measurements (see \cite{survey} for a survey). In \cite{Kosut}, Shabani et al. proposed a method that follows the SQPT approach and scales better with the number of qubits by assuming that the process matrix is sparse. This approach is very popular \cite{natureCS}, \cite{CS2}, \cite{CS3}, but we chose not to use it for the following reason. Almost all useful processes that we might want to identify have unitary target values (they are not unitary in practice because the implementation is not perfect but they are close to their target), and we think that, in general, assuming it is actually unitary (or equivalently the rank of its process matrix is $1$) is a better regularization hypothesis than assuming that its process matrix is sparse. A rank 2 process matrix with two similar eigenvalues can be considered sparse, but, the associated process is not a good approximation for a process that represents a unitary gate. 

Thus, like \cite{nearunitary}, \cite{reich}, we choose to restrict ourselves to unitary processes. This class is of particular interest because the evolution of any closed quantum system is described by a unitary transformation. In \cite{nearunitary}, Baldwin et al. study the tomography of unitary processes (and near-unitary ones). They use the work of Reich et al. in \cite{reich} to choose their input states. Reich et al. studied the tomography of unitary processes, they established a necessary and sufficient condition to be able to distinguish a unitary process from any other process (unitary or not), but did not explicitly propose a QPT algorithm.

 Unitary processes are easier to study in part because they involve ``only" $d^2$ independent real parameters ($d^2-1$ without the global phase). It still scales exponentially with the number of qubits, but is more reasonable than the $d^4-d^2$ real parameters that \cite{NC} identifies. The method of \cite{NC} is only experimentally realistic for 1 or 2 qubits, and even with two qubits there are $4^4-4^2=240$ real parameters. Another advantage of unitary processes is that they are relatively simple to parameterize. A unitary process is uniquely parameterized by a unitary matrix (the matrix $\Mv$ defined in Section \ref{section:State}, to be disntinguished from the above-mentioned process matrix), up to a global phase. In contrast the Kraus-operator representation of non-unitary processes is highly non-unique \cite{CPTPLM}, and assessing the differences between two Kraus-operator representations (target gate and estimated gate for example) is not as straightforward as comparing two unitary matrices up to a global phase.
 
Beyond the number of parameters of quantum processes, a major issue with algorithms that use the SQPT setup is the fact that the input states have to be known, and, for \cite{nearunitary} their value is imposed by the method. This is problematic because, in order to prepare a quantum state, one generally has to use a quantum gate, and in order to know with a high degree of accuracy which state is prepared, one must known the gate with a high degree of accuracy. This would require QPT. We are not the first to point out and to try to get around this problem. \cite{preGST} first identified this problem and tried to solve it by designing an algorithm that simultaneously identifies all the gates involved in the setup (including the gates used for state preparation) with no unitarity assumption. Later, gate set tomography (GST) was introduced \cite{GST0},\cite{gst_last}, it goes a step further, by identifying the initial state, the processes applied to it and the types of measurements performed at the same time. The identification is only possible up to significant indeterminacies (called gauge) and involves many parameters that make GST very difficult when $d$ increases. 

We also proposed our own solution to this problem: In 2015, we introduced the blind version of QPT (BQPT)  in \cite{TomoQuant1}, then detailed it in \cite{TomoQuant2} and more recently in \cite{TomoQuantmono}. In those papers, we focused on the tomography of the two-qubit cylindrical-symmetry Heisenberg coupling process. For those algorithms, the operator has to prepare one or several  copies of an unknown (hence the ``blind") set of initial states \footnote{This requires the preparation procedure to be known and reproducible, so that several copies of each used state may be prepared. It is not a violation of the no cloning theorem, the latter does not apply if we prepared the state that we want to reproduce.}. This removes the issue of systematic errors (with respect to a desired state) during the preparation. The system is identified by processing output measurements associated with $n_i$ different unknown input states going through the system. Generally, we need to perform QST or at least to estimate some measurement outcome probabilities for each of the $n_i$ output states. For the approaches of \cite{TomoQuant1}, \cite{TomoQuant2}, this kind of QST requires $n_c$ copies of each considered output state. Therefore for each one of the $n_i$ states the same experiment has to be repeated $n_c$ times with the same input state value, for $n_i \times n_c$ input state preparations in total. A more recent paper \cite{TomoQuantmono} also proposes “single-preparation BQPT methods”, i.e. methods which can operate with only one instance of each considered input state, i.e. $n_c = 1$.

In the current paper, we extend the semi-blind methods \cite{SSP} and \cite{MaxEnt}. The idea is to split the initial states into $n_s$ ($n_s = d$ for \cite{SSP} and $n_s = 2$ for \cite{MaxEnt}) groups, and measure the copies of the $k$-th group, after they have gone through the process $k$ (or $k-1$ for \cite{MaxEnt}) times. Thanks to this trick, we can use a QST algorithm in order to identify the states before and after they go through the process. Therefore, we have estimates of the input and output states, and we do not have to trust the value of the input states. We use the term ``semi-blind", because the constraints on the input states are 
 very loose
, but some copies of some states are measured before they are transformed by the process to be identified (each copy is only measured once though). We propose a setup that generalizes those of \cite{SSP} and \cite{MaxEnt} as follows. The QPT algorithm uses the quantum states estimated by the QST to estimate the process without using the prior knowledge we might have on the process (beyond unitarity) or the input states. We also give a necessary and sufficient condition (like in \cite{reich}) on the measured states that guarantee that the QPT is possible (i.e. $\Mv$ can be identified up to a global phase). This condition happens to be very loose, and it can be met with weakly constrained input states. 

Our QPT algorithm could work on an SQPT setup, i.e. with predetermined input state values. But since it works with (almost) any set of values, we chose to use the trick of measuring some of the copies of the states before they go through the process to remove the issue of systematic errors.

We intended to use a pure state QST algorithm from the literature for our QST-based QPT method, but there are surprisingly few articles about pure state QST. The work of Goyeneche et al. in \cite{Goyeneche} is closest to fit our needs, unfortunately, it uses entangled measurements that cannot be performed by measuring each qubit individually. We want to use measurements that are as simple as possible, because, as stressed in \cite{GST0}, we can never be sure that the model we have for the measurements is accurate. But the simpler the measurement the more we can rely on the model. We introduced two original QST setups in \cite{PRA} that only use unentangled measurements. In the present paper, we use one of them in order to perform QST for each measured state.  We chose unentangled measurements because, as Keith et al. remarked in \cite{QMTbaldwin} ``single-qubit gates have been demonstrated with high
fidelity, but entangling gates have lower fidelities", and unentangled measurements can be realized with single-qubit quantum gates and a measurement in the computational basis. 

Our contributions in the present paper are as follows. Section \ref{section:QPT_ls} explains how we can achieve QPT using only the results of the QST of the measured states. This QPT could work with any QST algorithm, but we recommend using the measurements and QST algorithm described in Appendix \ref{section:DataModel}. In Section \ref{section:conditions} we give a necessary and sufficient condition on the measured states for QPT to be possible. In Section \ref{section:setup} we give our recommendations on how to prepare the initial states to be sure that the condition is satisfied . In Section \ref{section:simu} we apply our algorithm to simulated data. Finally, in Section \ref{section:exp} we test our algorithm with experimental data on two trapped-ion qubits.

\section{Proposed QST-based quantum process tomography method}
\label{section:QPT_ls}
\subsection{Notations and data model}
\label{section:State}
An $n_{qb}$-qubit pure state $\ket{\varphi}$ can be decomposed in the $d$-element computational basis $\ket{0...0}$, $\ket{0...01}$, $\ket{0...10}$, ..., $\ket{1...1}$ ($d = 2^{n_{qb}}$). The components of $\ket{\varphi}$ in the basis can be stored in a $d$-element vector 
$
\vb = \begin{bmatrix}
v_1 \\ \vdots \\ v_d
\end{bmatrix}.
$
The components $v_j$ are complex and $\sum_{j=1}^{d}{|v_j|^2}=1$. The global phase of $\ket{\varphi}$ has no physical meaning when a state is considered individually.

In the rest of the paper we will use the vector notation instead of the kets (except for $\ket{0}$, we will still use it for one qubit). Contrary to most papers on QPT, we will not use mixed states or density matrices as we are dealing with pure states in a closed system. We need the system to be closed, because all processes of a closed system are unitary.

The unitary matrix that characterizes the process to be identified is denoted as $\Mv$, it is unique up to a global phase. Most often, the gate is physically realized by applying a constant Hamiltonian $\Hv$ to the input state in a closed system. After a time delay $\Delta_t$ the state vector is multiplied by $\Mv = \exp{\frac{-i}{\hbar} \Hv \Delta_t}$ according to the Schrödinger equation  ($\hbar$ is the reduced Plank constant, $i$ is the imaginary unit, $\exp$ is the matrix exponential).

The goal of QPT is to find $\Mv$ up to a global phase. When we write that ``QPT is possible" or ``the process is identifiable" with a given setup (with the states we measure and the types of measurements we perform on them), we mean that it is possible to find $\Mv$ up to a global phase with the sample frequencies of the outcomes of our measurements. If ``QPT is impossible" it means that there exist several unitary matrices $\Mv$ that differ by more than a global phase and that are indistinguishable with the states and measurements we chose. Other authors \cite{nearunitary}, \cite{reich} would write that the used states or measurements are ``informationally complete" (or not). This is the same idea.

The QPT algorithm that we use relies on estimates of the measured states. Those estimates are computed with a QST algorithm. In Appendix \ref{section:DataModel}, we summarize the QST algorithm that we recently introduced in \cite{PRA} that only relies on $1$-qubit Pauli measurements performed in parallel on each qubit. A total of $2 n_{qb}+1$ types of those measurements need to be performed on each measured state. 


\subsection{QPT setup}
\label{section:QPTsetup}
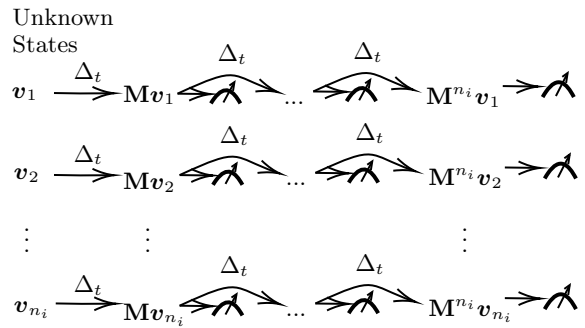
\begin{figure}[h]
\centering

\tikzset{every picture/.style={line width=0.75pt}} 

\begin{tikzpicture}[x=0.75pt,y=0.75pt,yscale=-1,xscale=1]

\draw    (342.23,124.69) -- (360.08,124.93) ;
\draw [shift={(360.08,124.93)}, rotate = 180] [color={rgb, 255:red, 0; green, 0; blue, 0 }  ][line width=0.75]    (10.93,-3.29) .. controls (6.95,-1.4) and (3.31,-0.3) .. (0,0) .. controls (3.31,0.3) and (6.95,1.4) .. (10.93,3.29)   ;
\draw [line width=1.5]    (362.33,130.25) .. controls (362.91,129.24) and (363.48,128.34) .. (364.03,127.55) .. controls (369.92,119.14) and (374.29,122.92) .. (378.11,130.25) ;
\draw    (369.04,129.72) -- (374.85,118.29) ;
\draw   (372.24,120.03) -- (374.94,118.1) -- (374.18,123.11) ;

\draw    (178.2,128.53) -- (195.01,128.74) ;
\draw [shift={(195.01,128.74)}, rotate = 180] [color={rgb, 255:red, 0; green, 0; blue, 0 }  ][line width=0.75]    (10.93,-3.29) .. controls (6.95,-1.4) and (3.31,-0.3) .. (0,0) .. controls (3.31,0.3) and (6.95,1.4) .. (10.93,3.29)   ;
\draw [line width=1.5]    (195.03,133.44) .. controls (195.57,132.55) and (196.11,131.75) .. (196.63,131.06) .. controls (202.17,123.62) and (206.28,126.96) .. (209.89,133.44) ;
\draw    (201.34,132.97) -- (206.82,122.87) ;
\draw   (204.35,124.41) -- (206.89,122.7) -- (206.19,127.13) ;

\draw    (178.2,128.53) .. controls (210.25,113.81) and (203.99,119.06) .. (227.07,127.16) ;
\draw [shift={(228.9,127.79)}, rotate = 198.5] [color={rgb, 255:red, 0; green, 0; blue, 0 }  ][line width=0.75]    (10.93,-3.29) .. controls (6.95,-1.4) and (3.31,-0.3) .. (0,0) .. controls (3.31,0.3) and (6.95,1.4) .. (10.93,3.29)   ;

\draw    (246.67,127.41) -- (263.48,127.62) ;
\draw [shift={(263.48,127.62)}, rotate = 180] [color={rgb, 255:red, 0; green, 0; blue, 0 }  ][line width=0.75]    (10.93,-3.29) .. controls (6.95,-1.4) and (3.31,-0.3) .. (0,0) .. controls (3.31,0.3) and (6.95,1.4) .. (10.93,3.29)   ;
\draw [line width=1.5]    (263.5,132.32) .. controls (264.04,131.43) and (264.57,130.64) .. (265.1,129.94) .. controls (270.64,122.5) and (274.75,125.84) .. (278.35,132.32) ;
\draw    (269.81,131.86) -- (275.29,121.75) ;
\draw   (272.82,123.29) -- (275.36,121.58) -- (274.65,126.01) ;

\draw    (246.67,127.41) .. controls (278.72,112.69) and (272.46,117.94) .. (295.54,126.05) ;
\draw [shift={(297.37,126.68)}, rotate = 198.5] [color={rgb, 255:red, 0; green, 0; blue, 0 }  ][line width=0.75]    (10.93,-3.29) .. controls (6.95,-1.4) and (3.31,-0.3) .. (0,0) .. controls (3.31,0.3) and (6.95,1.4) .. (10.93,3.29)   ;

\draw    (343.28,165.78) -- (361.13,166.02) ;
\draw [shift={(361.13,166.02)}, rotate = 180] [color={rgb, 255:red, 0; green, 0; blue, 0 }  ][line width=0.75]    (10.93,-3.29) .. controls (6.95,-1.4) and (3.31,-0.3) .. (0,0) .. controls (3.31,0.3) and (6.95,1.4) .. (10.93,3.29)   ;
\draw [line width=1.5]    (363.38,171.34) .. controls (363.96,170.33) and (364.53,169.43) .. (365.08,168.64) .. controls (370.97,160.23) and (375.34,164.01) .. (379.17,171.34) ;
\draw    (370.09,170.81) -- (375.91,159.38) ;
\draw   (373.29,161.13) -- (375.99,159.19) -- (375.24,164.2) ;

\draw    (179.25,169.62) -- (196.06,169.83) ;
\draw [shift={(196.06,169.83)}, rotate = 180] [color={rgb, 255:red, 0; green, 0; blue, 0 }  ][line width=0.75]    (10.93,-3.29) .. controls (6.95,-1.4) and (3.31,-0.3) .. (0,0) .. controls (3.31,0.3) and (6.95,1.4) .. (10.93,3.29)   ;
\draw [line width=1.5]    (196.08,174.53) .. controls (196.63,173.64) and (197.16,172.84) .. (197.68,172.15) .. controls (203.22,164.71) and (207.33,168.05) .. (210.94,174.53) ;
\draw    (202.39,174.06) -- (207.87,163.96) ;
\draw   (205.4,165.5) -- (207.95,163.79) -- (207.24,168.22) ;

\draw    (179.25,169.62) .. controls (211.3,154.9) and (205.04,160.15) .. (228.12,168.26) ;
\draw [shift={(229.96,168.88)}, rotate = 198.5] [color={rgb, 255:red, 0; green, 0; blue, 0 }  ][line width=0.75]    (10.93,-3.29) .. controls (6.95,-1.4) and (3.31,-0.3) .. (0,0) .. controls (3.31,0.3) and (6.95,1.4) .. (10.93,3.29)   ;

\draw    (247.72,168.5) -- (264.53,168.71) ;
\draw [shift={(264.53,168.71)}, rotate = 180] [color={rgb, 255:red, 0; green, 0; blue, 0 }  ][line width=0.75]    (10.93,-3.29) .. controls (6.95,-1.4) and (3.31,-0.3) .. (0,0) .. controls (3.31,0.3) and (6.95,1.4) .. (10.93,3.29)   ;
\draw [line width=1.5]    (264.55,173.41) .. controls (265.09,172.52) and (265.63,171.73) .. (266.15,171.03) .. controls (271.69,163.59) and (275.8,166.93) .. (279.41,173.41) ;
\draw    (270.86,172.95) -- (276.34,162.84) ;
\draw   (273.87,164.38) -- (276.42,162.67) -- (275.71,167.1) ;

\draw    (247.72,168.5) .. controls (279.77,153.78) and (273.51,159.03) .. (296.59,167.14) ;
\draw [shift={(298.42,167.77)}, rotate = 198.5] [color={rgb, 255:red, 0; green, 0; blue, 0 }  ][line width=0.75]    (10.93,-3.29) .. controls (6.95,-1.4) and (3.31,-0.3) .. (0,0) .. controls (3.31,0.3) and (6.95,1.4) .. (10.93,3.29)   ;

\draw    (343.28,231.72) -- (361.13,231.96) ;
\draw [shift={(361.13,231.96)}, rotate = 180] [color={rgb, 255:red, 0; green, 0; blue, 0 }  ][line width=0.75]    (10.93,-3.29) .. controls (6.95,-1.4) and (3.31,-0.3) .. (0,0) .. controls (3.31,0.3) and (6.95,1.4) .. (10.93,3.29)   ;
\draw [line width=1.5]    (363.38,237.28) .. controls (363.96,236.27) and (364.53,235.37) .. (365.08,234.58) .. controls (370.97,226.17) and (375.34,229.95) .. (379.17,237.28) ;
\draw    (370.09,236.75) -- (375.91,225.32) ;
\draw   (373.29,227.06) -- (375.99,225.13) -- (375.24,230.14) ;

\draw    (179.25,235.55) -- (196.06,235.77) ;
\draw [shift={(196.06,235.77)}, rotate = 180] [color={rgb, 255:red, 0; green, 0; blue, 0 }  ][line width=0.75]    (10.93,-3.29) .. controls (6.95,-1.4) and (3.31,-0.3) .. (0,0) .. controls (3.31,0.3) and (6.95,1.4) .. (10.93,3.29)   ;
\draw [line width=1.5]    (196.08,240.47) .. controls (196.63,239.57) and (197.16,238.78) .. (197.68,238.08) .. controls (203.22,230.65) and (207.33,233.99) .. (210.94,240.47) ;
\draw    (202.39,240) -- (207.87,229.9) ;
\draw   (205.4,231.44) -- (207.95,229.73) -- (207.24,234.15) ;

\draw    (179.25,235.55) .. controls (211.3,220.83) and (205.04,226.09) .. (228.12,234.19) ;
\draw [shift={(229.96,234.82)}, rotate = 198.5] [color={rgb, 255:red, 0; green, 0; blue, 0 }  ][line width=0.75]    (10.93,-3.29) .. controls (6.95,-1.4) and (3.31,-0.3) .. (0,0) .. controls (3.31,0.3) and (6.95,1.4) .. (10.93,3.29)   ;

\draw    (247.72,234.44) -- (264.53,234.65) ;
\draw [shift={(264.53,234.65)}, rotate = 180] [color={rgb, 255:red, 0; green, 0; blue, 0 }  ][line width=0.75]    (10.93,-3.29) .. controls (6.95,-1.4) and (3.31,-0.3) .. (0,0) .. controls (3.31,0.3) and (6.95,1.4) .. (10.93,3.29)   ;
\draw [line width=1.5]    (264.55,239.35) .. controls (265.09,238.46) and (265.63,237.66) .. (266.15,236.97) .. controls (271.69,229.53) and (275.8,232.87) .. (279.41,239.35) ;
\draw    (270.86,238.88) -- (276.34,228.78) ;
\draw   (273.87,230.32) -- (276.42,228.61) -- (275.71,233.04) ;

\draw    (247.72,234.44) .. controls (279.77,219.72) and (273.51,224.97) .. (296.59,233.07) ;
\draw [shift={(298.42,233.7)}, rotate = 198.5] [color={rgb, 255:red, 0; green, 0; blue, 0 }  ][line width=0.75]    (10.93,-3.29) .. controls (6.95,-1.4) and (3.31,-0.3) .. (0,0) .. controls (3.31,0.3) and (6.95,1.4) .. (10.93,3.29)   ;

\draw    (116.78,234.4) -- (129.92,234.61) -- (145.14,234.53) ;
\draw [shift={(147.14,234.52)}, rotate = 179.7] [color={rgb, 255:red, 0; green, 0; blue, 0 }  ][line width=0.75]    (10.93,-3.29) .. controls (6.95,-1.4) and (3.31,-0.3) .. (0,0) .. controls (3.31,0.3) and (6.95,1.4) .. (10.93,3.29)   ;
\draw    (115.9,169.56) -- (129.04,169.77) -- (144.27,169.69) ;
\draw [shift={(146.27,169.68)}, rotate = 179.7] [color={rgb, 255:red, 0; green, 0; blue, 0 }  ][line width=0.75]    (10.93,-3.29) .. controls (6.95,-1.4) and (3.31,-0.3) .. (0,0) .. controls (3.31,0.3) and (6.95,1.4) .. (10.93,3.29)   ;
\draw    (115.9,128.2) -- (129.04,128.41) -- (144.27,128.33) ;
\draw [shift={(146.27,128.32)}, rotate = 179.7] [color={rgb, 255:red, 0; green, 0; blue, 0 }  ][line width=0.75]    (10.93,-3.29) .. controls (6.95,-1.4) and (3.31,-0.3) .. (0,0) .. controls (3.31,0.3) and (6.95,1.4) .. (10.93,3.29)   ;

\draw (93.17,123.69) node [anchor=north west][inner sep=0.75pt]   [align=left] {$\vb_1$};
\draw (160,187) node [anchor=north west][inner sep=0.75pt]   [align=left] {\vdots};
\draw (93.55,78.22) node [anchor=north west][inner sep=0.75pt]   [align=left] {\begin{minipage}[lt]{13.5pt}\setlength\topsep{0pt}
\begin{center}
Unknown\\ States
\end{center}

\end{minipage}};
\draw (146.21+3,122.89) node [anchor=north west][inner sep=0.75pt]   [align=left] {$\Mv \vb_{1}$};
\draw (320,187) node [anchor=north west][inner sep=0.75pt]   [align=left] {\vdots};
\draw (302.17,123.69) node [anchor=north west][inner sep=0.75pt]   [align=left] {$\Mv^{n_i} \vb_{1}$};
\draw (99,187) node [anchor=north west][inner sep=0.75pt]   [align=left] {\vdots};

\draw (230.32,130.88) node [anchor=north west][inner sep=0.75pt]   [align=left] {...};
\draw (94.22,164.78) node [anchor=north west][inner sep=0.75pt]   [align=left] {$\vb_2$};
\draw (146.21+3,166.85) node [anchor=north west][inner sep=0.75pt]   [align=left] {$\Mv \vb_{2}$};
\draw (303.22,164.78) node [anchor=north west][inner sep=0.75pt]   [align=left] {$\Mv^{n_i} \vb_{2}$};
\draw (231.37,171.97) node [anchor=north west][inner sep=0.75pt]   [align=left] {...};
\draw (94.22,231.83) node [anchor=north west][inner sep=0.75pt]   [align=left] {$\vb_{n_i}$};
\draw (146.21+3,232.79) node [anchor=north west][inner sep=0.75pt]   [align=left] {$\Mv \vb_{n_i}$};
\draw (303.22,230.72) node [anchor=north west][inner sep=0.75pt]   [align=left] {$\Mv^{n_i} \vb_{n_i}$};
\draw (231.37,237.91) node [anchor=north west][inner sep=0.75pt]   [align=left] {...};
\draw (266.49,208.33) node [anchor=north west][inner sep=0.75pt]   [align=left] {$\Delta_t$};
\draw (198.02,209.45) node [anchor=north west][inner sep=0.75pt]   [align=left] {$\Delta_t$};
\draw (266.49,142.39) node [anchor=north west][inner sep=0.75pt]   [align=left] {$\Delta_t$};
\draw (198.02,143.51) node [anchor=north west][inner sep=0.75pt]   [align=left] {$\Delta_t$};
\draw (265.43,101.3) node [anchor=north west][inner sep=0.75pt]   [align=left] {$\Delta_t$};
\draw (196.96,102.42) node [anchor=north west][inner sep=0.75pt]   [align=left] {$\Delta_t$};
\draw (124.34,111.48) node [anchor=north west][inner sep=0.75pt]   [align=left] {$\Delta_t$};
\draw (124.34,153.17) node [anchor=north west][inner sep=0.75pt]   [align=left] {$\Delta_t$};
\draw (124.34,219.13) node [anchor=north west][inner sep=0.75pt]   [align=left] {$\Delta_t$};

\end{tikzpicture}

\caption{QPT setup. The ``double arrows" symbolize that some copies are measured (straight arrow) and the others are fed through the next gate.}
\label{fig:setup_hard}
\end{figure}

Our QPT algorithm is designed to be resistant to systematic errors on the initial states. In order to achieve this, we assume that the initial states are unknown but that measurements are performed at different time steps: see Fig. \ref{fig:setup_hard}.

The number of initial states is $n_i$. They are measured after waiting $\Delta_t$, $2 \Delta_t$, ... or $n_s \Delta_t$ ($n_s$ is the maximum number of time delays), and the state vector is multiplied by the matrix $\Mv$ associated with the process from one time delay to the next. The number of types of measurements that are performed on each qubit is called $n_t$. With the QST algorithm of Appendix \ref{section:DataModel}, we need $n_t = 2 n_{qb}+1$. One instance of a measurement alone does not bring much information on the state, we only know which one of the $d$ possible outcomes has occurred. This is why we choose to perform each type of measurement $n_c$ times in order to estimate the probabilities of all outcomes. Each copy of the input states can only be measured once, therefore, we need $n_s n_t n_c$ copies of each one of the $n_i$ input states, for a total of $n_i n_s n_t n_c$ prepared input states.

The initial states ($\vb_1, ..., \vb_{n_i}$) are never measured directly. We could imagine a similar setup where they are measured and the number of considered time delays is decreased by one (in fact we did this in \cite{MaxEnt} with $n_s = 2$ and $n_i = d$). We chose not to do this because some current quantum computers do not allow us to measure some states right after they are prepared. This is the case for the computer we used in Section \ref{section:exp}. This is explained in Section \ref{section:exp} (more precisely in the description of Fig. \ref{fig:setup_exp}) where we detail the experimental realization of our algorithm. 


The results of the experiment are the measurements counts. For each one of the $n_i\times n_s$ measured states with each one of the $n_t$ types of measurements, we count the number of times each one of the $d$ outcomes occurred. Those $d$ counts sum to $n_c$, and each one of those groups of $d$ measurement counts can be modelled as a random variable that follows a multinomial distribution with $n_c$ trials and the probabilities of the $d$ outcomes are determined by the value of the associated measured state and the type of measurement. Table 2 in Appendix \ref{section:tables} shows an example of measurement counts for a given QPT setup with $n_i = 4$, $n_s = 2$, $n_t = 5$.


\subsection{Main algorithm}
\label{section:idea}
We assume that QST is performed properly for the measured states of Fig. \ref{fig:setup_hard}. The measured states are denoted with the following convention:
\begin{equation} 
\vb_{j, k} = \Mv^k \vb_j, j\in \{1, ..., n_i\}, k\in \{1, ..., n_s\}
\end{equation} 
and $\{\widehat{\vb}_{j, k}\}_{j, k}$ are their estimates.
Those QST estimates each have a phase indeterminacy $\theta_{j, k}^{QST}$ and a QST error $\boldsymbol{\varepsilon}_{j, k}^{QST}$ such that  $E(||\boldsymbol{\varepsilon}_{j, k}^{QST}||_2) \underset{n_c \to +\infty}{\longrightarrow} 0$ ($E$ is the expected value):

\begin{equation} 
\label{eqn:QST}
\widehat{\vb}_{j, k} = \Mv^k \vb_j . e^{i\theta_{j, k}^{QST}} + \boldsymbol{\varepsilon}_{j, k}^{QST} \ \ \ \begin{matrix} j\in\{1, ..., n_i\} \\ k\in\{1, ..., n_s\}\end{matrix} .
\end{equation}

For the rest of this paper, when we refer to the ``QST error", we mean $\boldsymbol{\varepsilon}_{j, k}^{QST}$, not $\theta_{j, k}^{QST}$. For the rest of this section, we consider that there is no QST error: $\boldsymbol{\varepsilon}_{j, k}^{QST}=0$ unless stated otherwise. In Section \ref{section:State} we stated that the global phases of the states do not matter. This is true if the states are considered independently and this is the reason why the QST cannot recover the global phase, and why we do not consider $\theta_{j, k}^{QST}$ to be an ``error", it affects our estimate even in the ideal case with an infinite number of measurements and no source of error. But when the states are considered together (in order to find $\Mv$) the differences between the global phases of the different states matter.

\begin{figure}
\begin{centering}

\tikzset{every picture/.style={line width=0.75pt}} 

\begin{tikzpicture}[x=0.75pt,y=0.75pt,yscale=-1,xscale=1]

\draw    (204,88) -- (285,88) ;
\draw [shift={(287,88)}, rotate = 180] [color={rgb, 255:red, 0; green, 0; blue, 0 }  ][line width=0.75]    (10.93,-3.29) .. controls (6.95,-1.4) and (3.31,-0.3) .. (0,0) .. controls (3.31,0.3) and (6.95,1.4) .. (10.93,3.29)   ;
\draw    (204,156) -- (285,156) ;
\draw [shift={(287,156)}, rotate = 180] [color={rgb, 255:red, 0; green, 0; blue, 0 }  ][line width=0.75]    (10.93,-3.29) .. controls (6.95,-1.4) and (3.31,-0.3) .. (0,0) .. controls (3.31,0.3) and (6.95,1.4) .. (10.93,3.29)   ;

\draw (137,79) node [anchor=north west][inner sep=0.75pt]   [align=left] {$\xb_1$};
\draw (297,79) node [anchor=north west][inner sep=0.75pt]   [align=left] {$\Mv \xb_1$};
\draw (300,105) node [anchor=north west][inner sep=0.75pt]   [align=left] {$\vdots$};
\draw (240,105) node [anchor=north west][inner sep=0.75pt]   [align=left] {$\vdots$};
\draw (237,140) node [anchor=north west][inner sep=0.75pt]   [align=left] {$\Delta_t$};
\draw (237,72) node [anchor=north west][inner sep=0.75pt]   [align=left] {$\Delta_t$};

\draw (140,105) node [anchor=north west][inner sep=0.75pt]   [align=left] {$\vdots$};
\draw (137,147) node [anchor=north west][inner sep=0.75pt]   [align=left] {$\xb_{n_i(n_s-1)}$};
\draw (297,147) node [anchor=north west][inner sep=0.75pt]   [align=left] {$\Mv \xb_{n_i(n_s-1)}$};
\draw (112,47) node [anchor=north west][inner sep=0.75pt]   [align=left] {estimated};
\draw (274,47) node [anchor=north west][inner sep=0.75pt]   [align=left] {estimated};

\end{tikzpicture}

\end{centering}
\caption{virtual QPT setup}
\label{fig:setup_virtual}
\end{figure}
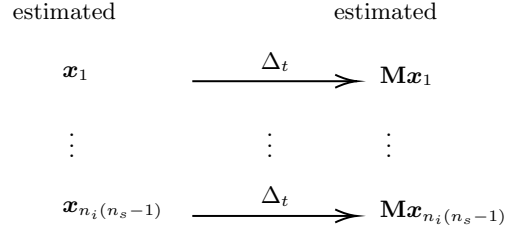

We know that 
\begin{equation} 
\label{eqn:vw}
\vb_{j, k+1} = \Mv \vb_{j, k} \ \ \  j\in\{1, ..., n_i\} ,\ k\in\{1, ..., n_s-1\}. 
\end{equation}
For the sake of simplicity, we define 
\begin{equation}
\label{eqn:XY0}
\begin{split}
\Xv & = [\vb_{1, 1}, ..., \vb_{1, n_s-1}, \vb_{2, 1}, ...., \vb_{n_i, n_s-1}] \\
\Yv & = \Mv \Xv =  [\vb_{1, 2}, ..., \vb_{1, n_s}, \vb_{2, 1}, ...., \vb_{n_i, n_s}] \\
\widehat{\Xv} & = [\widehat{\vb}_{1, 1}, ..., \widehat{\vb}_{1, n_s-1}, \widehat{\vb}_{2, 1}, ...., \widehat{\vb}_{n_i, n_s-1}] \\
\widehat{\Yv} & =  [\widehat{\vb}_{1, 2}, ..., \widehat{\vb}_{1, n_s}, \widehat{\vb}_{2, 1}, ...., \widehat{\vb}_{n_i, n_s}].
\end{split}
\end{equation}
With those notations, (\ref{eqn:vw}) becomes $\Yv = \Mv \Xv$, and we can forget about the setup of Fig. \ref{fig:setup_hard} and imagine that we are dealing with the simpler setup of Fig. \ref{fig:setup_virtual} (with the convention that $\xb_{\ell}$ is the $\ell$-th column of $\Xv$). 

This representation is closer to the SQPT setup (known initial states, measured output). In fact the only difference is that, here, the virtual input states (states on the left-hand side of Fig. \ref{fig:setup_virtual}) are not set to predetermined values, but are prepared with unknown quantum gates (including the gate that we are trying to identify). Those input states are estimated from the measurements. The algorithm that we will now describe could work on an SQPT setup with known predetermined input states $\Xv$ and measured output states $\Yv$. But not all SQPT algorithms would work on the ``real" original setup (Fig. \ref{fig:setup_hard}) because they often require the input states to be set to predetermined values.

One could think that we are now simply dealing with a simple least square problem with a unitarity constraint. But this is not the case as the input and output states are only ever measured individually; and when we consider them together in order to find $\Mv$, their relative phases (that are never measured) matter. 

Let us define the relative phases between the $\xb_{\ell}$ and the $\yb_{\ell}$

$\xi_{\ell}=\theta_{j, k}^{QST}-\theta_{j, k+1}^{QST},\ \ \ \begin{matrix}  j\in\{1, ..., n_i\}, k\in\{1, ..., n_s-1\} \\ \ell=k+(n_s-1)(j-1) \\ \ell \in \{1, ..., n_i (n_s-1)\}. \end{matrix}\ $,%

\noindent We can rewrite (\ref{eqn:vw}) with $\widehat{\Xv}$ and $\widehat{\Yv}$ (which we actually know thanks to the QST):
\begin{equation} 
\label{eqn:base_phase}
e^{i\xi_{\ell}} \widehat{\yb}_{\ell} = \Mv \widehat{\xb}_{\ell} \ \ \ \ell \in\{1, ..., n_i (n_s-1)\}, 
\end{equation}

\noindent where $\widehat{\xb}_{\ell}$ and $\widehat{\yb}_{\ell}$ are the $\ell$-th columns of $\widehat{\Xv}$ and $\widehat{\Yv}$ respectively.

For any index $\ell_0$, changing $\Mv$ to $\Mv.e^{-i\xi_{\ell_0}}$  and $\xi_{\ell}$ to $\xi_{\ell}-\xi_{\ell_0} \ \ \forall \ell $ does not change the equality (\ref{eqn:base_phase}). Therefore, we can also assume that a given $\ell_0$ (we explain how to choose it in Section \ref{section:phase}) is such that $\xi_{\ell_0} = 0$ taking into account the fact that $\Mv$ can only be recovered up to a global phase.

In the next section, we explain how to obtain the estimated phase factors $e^{i\widehat{\xi}_{\ell}}$. From that, we can define 
\begin{equation}
\widetilde{\yb}_{\ell}=\widehat{\yb}_{\ell} . e^{i\widehat{\xi}_{\ell}} \ \ \ell \in\{1, ..., n_i (n_s-1)\}, 
\end{equation}
with which an estimate of $\Mv$ can easily be found (if $n_i(n_s-1)\geq d$) as the problem becomes:
\begin{equation}
\label{eqn:problemU}
\widetilde{\Yv} = \Mv \widehat{\Xv}
\end{equation}
with
\begin{equation}
\label{eqn:Y}
\begin{matrix}
\widetilde{\Yv} = [\widetilde{\yb}_{1}, ..., \widetilde{\yb}_{n_i ( n_s - 1)}].
\end{matrix}
\end{equation}

$\widehat{\Mv} =\widetilde{\Yv}\widehat{\Xv}^{\dag}$ works as a solution ($.^{\dag}$ is the pseudo-inverse). But it is generally not a unitary solution because of the QST errors.
Finding $\widehat{\Mv}$ in the ensemble of $d$-dimensional unitray matrices (denoted as $\mathds{U}_d (\mathds{C})$) that is the total least square solution of (\ref{eqn:problemU}) has been solved By K. S. Arun in \cite{arun} (with the caveat that \cite{arun} solves (with different notations) $\widetilde{\Yv}^* =  \widehat{\Xv}^* \Mv$, where $.^*$ is the transconjugate, instead of $\widetilde{\Yv} = \Mv \widehat{\Xv}$). Arun's solution can be rewritten as:
\begin{equation}
\label{eqn:Mgreedy}
\begin{matrix}
 \Bv = \widetilde{\Yv} \widehat{\Xv}^* \\
 \Uv\ \Sv\ \Vv^* = \Bv \\
 \widehat{\Mv}_{LS} = \Uv\ \Vv^*
\end{matrix}
\end{equation}
$\Uv\ \Sv\ \Vv^*$ is the singular value decomposition of $\Bv$. This is optimal in the total least square sense, meaning that $\widehat{\Mv}_{LS} (\widehat{\Xv} + \Delta_X^0) = (\widetilde{\Yv} + \Delta_Y^0) $ where $\Delta_X, \Delta_Y$ are the solutions to the following optimization problem:
\begin{equation}
\label{eqn:TLS}
\{ \Delta_X^0, \Delta_Y^0\}= \underset{ \{\widehat{\Xv}+\Delta_X, \widetilde{\Yv}+\Delta_Y\} \in \mathcal{L} }{\arg\min}  || \Delta_X ||^2+||\Delta_Y ||^2
\end{equation}
where $\mathcal{L}$ is the ensemble of the pairs of $d\times n_i (n_s-1)$ matrices linked by a unitary transformation:

\noindent $\mathcal{L}= \begin{Bmatrix} \Xv, \Yv \in \mathds{C}^{d\times n_i (n_s-1)}, \exists \Pv \in \mathds{U}_d(\mathds{C}), \Yv = \Pv \Xv \end{Bmatrix}$.

We use the total least square approach because both $\widehat{\Xv}$ and $\widetilde{\Yv}$ are subject to errors, $\Delta_X^0$ and  $\Delta_Y^0$ can be understood as our estimate of those errors. This approach would be optimal in the maximum likelihood sense if the errors on $\widehat{\Xv}$ and $\widetilde{\Yv}$ were Gaussian iid on every component. This is not the case in practice (especially if $n_s >2$, in this case, some columns of $\widehat{\Xv}$ are also in $\widetilde{\Yv}$ up to a phase, thus their errors are highly correlated), but minimizing the norm of the error is never a bad idea as a first approximation.  

The solution $\widehat{\Mv}_{LS}$ is unique if and only if both $\widehat{\Xv}$ and $\widetilde{\Yv}$ are of full rank. This is not explicitly stated in \cite{arun} but it is proven for the orthogonal Procrustes problem, in \cite{Golum}, which \cite{arun} showed to be equivalent to the total least squared problem that we consider.
\subsection{Phase recovery algorithm}
\label{section:phase}
The aim of the current section is to find $\{e^{i\widehat{\xi}_{\ell}}\}_{\ell}$ given the vectors, $\widehat{\xb}_{\ell}, \widehat{\yb}_{\ell}\ \  \ell \in\{1, ..., n_i (n_s-1)\}$ such that there exists a unitary matrix $\Mv$ that realizes (\ref{eqn:base_phase}). 

We will use the fact that unitary matrices preserve the dot product:

 \begin{equation}
\label{eqn:propriete1}
\begin{split}
\widehat{\xb}_{\ell_1}^* \widehat{\xb}_{\ell_2} & = (\Mv \widehat{\xb}_{\ell_1})^*\ (\Mv \widehat{\xb}_{\ell_2}) \\
  & =  \widehat{\yb}_{\ell_1}^* \widehat{\yb}_{\ell_2} e^{i(\xi_{\ell_2}-\xi_{\ell_1})}.
\end{split}
\end{equation}
Therefore, for any $\{\ell_1, \ell_2\}$ pair in $\{1, ..., n_i (n_s-1)\}^2$, such that $\widehat{\yb}_{\ell_1}^* \widehat{\yb}_{\ell_2}\neq 0$, we have the following estimate of $\xi_{\ell_2}-\xi_{\ell_1}$:
\begin{equation}
\label{eqn:argument}
\widehat{\xi}_{\ell_1, \ell_2} = arg \left(\frac{\widehat{\xb}_{\ell_1}^*\widehat{\xb}_{\ell_2}}{ \widehat{\yb}_{\ell_1}^* \widehat{\yb}_{\ell_2} }\right)
\end{equation}
where $arg$ is the phase of a complex number. Using (\ref{eqn:argument}), we can compute estimates $\widehat{\xi}_{\ell_2}$ of all the phases $\xi_{\ell_2} \forall \ell_2 \in\{1, ..., n_i (n_s-1)\}$ relative to a single phase $\xi_{\ell_0}$ (by setting $l_1$ to $l_0$). Using the fact that $\xi_{l_0}=0$ the estimate of $\xi_{\ell_2}$ is 
\begin{equation}
\label{eqn:duh}
\widehat{\xi}_{\ell_2}= \widehat{\xi}_{\ell_0, \ell_2}.
\end{equation}
The problem of the choice of $l_0$ remains. In order to solve it, let us look at (\ref{eqn:argument}) with $\ell_1$ replaced by $\ell_0$ (this is how we compute $\widehat{\xi}_{\ell_0, \ell_2}$):
\begin{equation}
\label{eqn:argument0}
\widehat{\xi}_{\ell_0, \ell_2} = arg \left(\frac{\widehat{\xb}_{\ell_0}^*\widehat{\xb}_{\ell_2}}{ \widehat{\yb}_{\ell_0}^* \widehat{\yb}_{\ell_2} }\right).
\end{equation}
It assumes that the dot products $\widehat{\xb}_{\ell_0}^*\widehat{\xb}_{\ell_2}$ and $\widehat{\yb}_{\ell_0}^*\widehat{\yb}_{\ell_2}$ are not zero. In practice, in order for $\widehat{\xi}_{\ell_0, \ell_2}$ to be a good estimate, we need both dot products to be as far from zero as possible. Interestingly, the two dot products are supposed to have the same modulus (see (\ref{eqn:propriete1})), therefore, we choose the index $\ell_0$ solution of:
\begin{equation}
\label{eqn:maxl0}
\ell_0 = \underset{\ell}{\arg\max} \left( \underset{\ell_2}{\min} \left|\widehat{\yb}_{\ell}^*\widehat{\yb}_{\ell_2} \right| \right) .
\end{equation}
With this choice, $\ell_0$ is such that the smallest dot product is as high as possible. The optimization is performed by exhaustive search. 

In practice, if the corresponding maximum is $0$ i.e. if, for all the $\widehat{\yb}_{\ell}$, we can find an orthogonal $\widehat{\yb}_{\ell_2}$, then this does not work as (\ref{eqn:argument0}) cannot be written for all the phases. In practice, two vectors are never going to be truly orthogonal, so in our implementation of our method, we say that two unit-norm vectors are orthogonal when the modulus of their dot product is smaller than $b_{orth}$ initially set to $0.05$. And if the quantity maximized in (\ref{eqn:maxl0}) is greater than this $b_{orth}$, we simply compute all the phases with (\ref{eqn:argument0}) and then (\ref{eqn:duh}). Otherwise, we apply the following algorithm:
\begin{enumerate}[leftmargin=*]
\item We start by computing the phase differences $\widehat{\xi}_{\ell_0, \ell_2}$ (\ref{eqn:argument0}) such that $\widehat{\yb}_{\ell_2}$ are not orthogonal (modulus of dot product $> b_{orth}$) to $\widehat{\yb}_{\ell_0}$. We then have the absolute phase of $\widehat{\yb}_{\ell_2}$ using (\ref{eqn:duh}).
\item We call $\mathcal{F}$ the set of $\{\widehat{\yb}_{\ell_2}\}_{\ell_2}$ for which we do not have the phase yet. We define $\mathcal{S}$ as the complement of $\mathcal{F}$ in $\{\widehat{\yb}_{\ell_2}\}_{\ell_2}$.
\item For each element $\widehat{\yb}_{\ell_f}$ of $\mathcal{F}$
\begin{enumerate}[leftmargin=*]
\item If all elements of $\mathcal{S}$ are orthogonal (modulus of dot product $<b_{orth}$) to $\widehat{\yb}_{\ell_f}$ we change nothing and go to the next $\widehat{\yb}_{\ell_f}$.
\item Else, we define $\widehat{\yb}_{\ell_s}$ as the element of $\mathcal{S}$ that is the least orthogonal (greatest modulus of dot product) to $\widehat{\yb}_{\ell_f}$. 
\item We compute the relative phase  $\widehat{\xi}_{\ell_s, \ell_f}$ with (\ref{eqn:argument}) and deduce the phase $\widehat{\xi}_{\ell_f}$:

$\widehat{\xi}_{\ell_f} = \widehat{\xi}_{\ell_0, \ell_f} = \widehat{\xi}_{\ell_0, \ell_s}+\widehat{\xi}_{\ell_s, \ell_f}$.
\item We remove $\widehat{\yb}_{\ell_f}$ from $\mathcal{F}$ and add it to $\mathcal{S}$.
\end{enumerate}
\item If $\mathcal{F}$ is empty then we are finished.
\item If $\mathcal{F}$ is not empty but the number of elements in it has decreased since Step 3, then we go to Step 3.
\item If $\mathcal{F}$ is not empty and the number of elements did not change, but $\mathds{C}^d$ is spanned by the elements of $\mathcal{S}$, then we remove the elements of $\mathcal{F}$ from $\widehat{\Yv}$ and the corresponding elements of $\widehat{\Xv}$, exit the phase recovery algorithm and go on to solve (\ref{eqn:Mgreedy}) without the elements we removed.
\item If the conditions in Step 5 and 6 are false, and $b_{orth}$ is still $0.05$, then we change $b_{orth}$ to $0$ and go to Step 3. 
\item If $b_{orth}$ was already $0$ when reaching Step 7, then this is a failure case and we exit the algorithm.
\end{enumerate}
This algorithm cannot be stuck in an infinite loop as:
\begin{itemize} [leftmargin=*]
\item The maximum number of times we can go from Step 5 to Step 3 without going to Step 6 is the cardinal of $\mathcal{F}$ at Step 2 (because the cardinal of $\mathcal{F}$ decreases by at least one every time) which is strictly smaller than $n_i (n_s-1)$.
\item The maximum number of times we can go from Step 7 to Step 3 is $1$.
\end{itemize}
The algorithm can fail at Step 8, but we will show in Section \ref{section:CNS} that it always works if the QPT is possible in the first place.
\subsection{Application to the standard QPT setup}
\label{section:SQPT}
The reader might be more comfortable with the SQPT setup where the input states are known, and only one time delay is considered. Our algorithm can easily be adapted to this setup, because, as we stated in Section  \ref{section:idea}, the virtual setup of Fig. \ref{fig:setup_virtual} is almost the SQPT setup, with $n_x = n_i (n_s-1)$ input states: $x_1, ..., x_{n_x}$. The only difference  between the two is that, for the SQPT setup, the input states are considered to be known (and not estimated by QST). This is not a problem for our QPT algorithm: in the SQPT setup, we can still define the matrices $\widehat{X}$ (because the input states are known) and $\widehat{Y}$ (because the output states are estimated by the QST algorithm), compute $\widetilde{Y}$ with the algorithm of Section \ref{section:phase} and, finally, compute $\widehat{M}_{LS}$ with (\ref{eqn:Mgreedy}).
\section{Idetifiability condtions}
\label{section:conditions}
\subsection{A necessary and sufficient condition for process identifiability}
\label{section:CNS}
The QST error $\{\boldsymbol{\varepsilon}_{j, k}^{QST}\}_{j, k}$ is neglected in the current section, therefore $\widetilde{\Yv} = \Mv \widehat{\Xv}$, with $\widetilde{\Yv} = \widehat{\Yv} \Dv(\xib)$, and $\Dv(\xib)$ is the diagonal matrix defined by the phases contained in $\xib$: $\Dv(\xib) = \begin{pmatrix} e^{i \xi_1} & & \\ & \ddots & \\ & & e^{i \xi_{n_x}} \end{pmatrix} $ ($n_x$ is the number of columns in $\widehat{\Xv}$, so $n_i (n_s-1)$ in the base problem).

Therefore, the QPT problem after QST consists of finding the unitary matrix $\Mv$ subject to the following condition:
\begin{equation}
\label{eqn:avec_phase}
 \widehat{\Yv} \Dv(\xib) = \Mv \widehat{\Xv}
\end{equation}
where $\widehat{\Xv}$ and $\widehat{\Yv}$ are known from the QST and the elements $\xi_1, ..., \xi_{n_x}$ of $\xib$ that define $\Dv(\xib)$ are unknown before the QPT. $\widehat{\Xv}$ contains our estimates of the virtual input states. Since there is no QST error, the columns of $\widehat{\Xv}$ are the same as those of $\Xv$ up to global phases.

The following condition on $\Xv$ is necessary and sufficient for QPT to be possible with the setup of Fig. \ref{fig:setup_hard}:
\begin{equation}
\label{eqn:CNS}
 \forall \ell \in \{1, ..., n_x\}, \  rank\left(\Fv_S^{n_x}(\xb_{\ell})\right) = d
\end{equation}

\noindent where $\Fv_S$ is the function that takes as its input a matrix $\Xv_{in}$ whose columns are columns of $\Xv$ and that returns the columns of $\Xv$ (grouped in a matrix in the order in which they appear in $\Xv$) that are not orthogonal to at least one column of $\Xv_{in}$. $\Fv_S^{n_x}$ is $\Fv_S$ applied $n_x$ times, and $\xb_{\ell}$ is the $\ell$-th column of $\Xv$. This definition is closely linked to our phase recovery algorithm. If we change the starting $b_{orth}$ to $0$, then, at Step 2, the columns of $\mathcal{S}$ match the columns of $\Fv_S^{1}(\xb_{\ell_0})$. They match in the sense that there are as many columns and their positions in $\widehat{\Yv}$ and $\Xv$ respectively are the same, this is because $\widehat{\yb}_{\ell_1} \perp \widehat{\yb}_{\ell_2} \Leftrightarrow \xb_{\ell_1} \perp \xb_{\ell_2}$ ($\perp$ means that two vectors are orthogonal). The $k$-th time we go to Step 5, the columns of $\mathcal{S}$ match the columns of $\Fv_S^{k+1}(\xb_{\ell_0})$, and, importantly, the subspaces they span have the same dimension.

Equation (\ref{eqn:CNS}) is a condition on the columns of the matrix $\Xv$. However, it is very easy to check that we could have set this condition on the columns of $\widehat{\Xv}$, $\Yv$ or $\widehat{\Yv}$ and have an equivalent condition. This is because multiplying all the columns by the same unitary matrix on the left, or multiplying each of them by a different scalar phase factor does not change the rank nor the orthogonality between the columns. We use $\Xv$ because it is a matrix that contains the input states (of the virtual setup in Fig. \ref{fig:setup_virtual}), and we prefer to have a condition on the input states.

Appendix \ref{section:math} shows that, in the absence of QST error, (\ref{eqn:CNS}) is a necessary (Section \ref{section:math2}) and sufficient (Section \ref{section:math1}) condition for QPT to be possible (i.e. for $\Mv$ to be identifiable up to a global phase). Furthermore, the proof of the sufficiency of (\ref{eqn:CNS}) in Section \ref{section:math1} also shows that our QPT algorithm always work if (\ref{eqn:CNS}) is true. 

This is a strong validation of our algorithm, it always achieves QPT if the setup (i.e. the measured states) makes it possible and only fails if QPT was impossible in the first place. Of course, this is only true if there are no QST errors. When considering non-null QST errors, we can have issues with setups for which $\Xv$ is poorly conditioned and for which $\Xv$ contains groups of columns that are too close to being orthogonal for the phase recovery to succeed.
\subsection{A simpler sufficient condition}
We have established that (\ref{eqn:CNS}) is a necessary and sufficient condition for QPT to be possible. 
The condition of (\ref{eqn:CNS}) is quite cumbersome to check however, and we prefer to use the following sufficient condition (it may be shown that it is not necessary):
\begin{equation}
\label{eqn:CS}
\begin{matrix}
 rank(\Xv) = d \ \ \text{ and }\\
 \exists \ell_0 \in \{1, ..., n_x\},  \forall \ell \in \{1, ..., n_x\}  \ \xb_{\ell_0} \not\perp \xb_{\ell}.
 \end{matrix}
\end{equation}
In plain words, $\Xv$ is of full rank and there exists a column of $\Xv$ that is not orthogonal to any of the others. 

It is really easy to check that (\ref{eqn:CS})$\Rightarrow$(\ref{eqn:CNS}): if (\ref{eqn:CS}) is met, then all the columns of $\Xv$ are in $\Fv_S^{k}(\xb_{\ell})$ for any $k\geq 2$ and any $\ell$.

Therefore, when we design the QPT setup of Fig. \ref{fig:setup_hard} we have to hope (or to make sure) that the quantum states that are represented by the columns of $\Xv$ satisfy (\ref{eqn:CNS}) or (\ref{eqn:CS}). 

In practice, (\ref{eqn:CS}) is a very reasonable condition. The probability of two random states (with any non-degenerated density function) being orthogonal is $0$ and the probability of $d$ random vectors or more (in a $d$-dimensional Hilbert space) being in a subspace of dimension $\leq d-1$ (this relates to the condition on the rank of $\Xv$) is also $0$. Essentially the states that make our algorithm fail are in a set of zero measure. 

Therefore the conditions of (\ref{eqn:CS}) (and the condition of (\ref{eqn:CNS})) will always be satisfied in practice. Even if we try to prepare states that make our QPT method fail, the small random error in their preparation will ensure that the actual states satisfy (\ref{eqn:CS}). This does not mean that we can safely ignore the conditions though. Because if $\Xv$ is of full rank but is almost singular, or if there are too many columns close to being orthogonal, then our algorithm (that works on $\widehat{\Xv}$ instead of the unknown $\Xv$) is expected to work poorly if there are QST errors.

\section{QPT setups compatble with identifiability}
\label{section:setup}
\subsection{Our recommendations for suitable QPT setups}
\label{section:recommendations}

If possible, we recommend to consider what the gate that we are trying to identify is supposed to do (otherwise we hereafter propose a solution that works for any unitary gate without prior knowledge). For example, if the gate to be identified is supposed to be a 2-qubit CNOT gate, $\Mv_{tg} = \begin{pmatrix} 1&0&0&0\\0&1&0&0\\0&0&0&1\\0&0&1&0\end{pmatrix}$ (``tg" stands for target), then, choosing a single input state ($n_i = 1$) and $d+1 = 5$ time steps ($n_s = 5$) is a very bad idea. This is because the CNOT gate applied twice is supposed to return the initial state. Therefore, $\Xv$ has two identical pairs of columns. In practice, it will make $\widehat{\Xv}$ close to being singular and the quality of the estimate of $\Mv$ will be very poor.

In order to have a matrix $\Xv$ such that (\ref{eqn:CS}) is ``comfortably" satisfied with the CNOT gate, let us consider $n_s = 2$ time delays, and $n_i = d = 4$ input states that form a basis of the Hilbert space with one of them far from being orthogonal to all the others. For example, we can aim for those targets: \small
\begin{equation}
\label{eqn:v_init}
\vb_1^{tg} = \begin{pmatrix} 1 \\ 0 \\ 0 \\ 0 \end{pmatrix}, \vb_2^{tg} =  \begin{pmatrix} \frac{1}{\sqrt{2}} \\ \frac{1}{\sqrt{2}} \\ 0 \\ 0 \end{pmatrix}, \vb_3^{tg} =  \begin{pmatrix} \frac{1}{\sqrt{2}} \\ 0 \\ \frac{1}{\sqrt{2}} \\ 0 \end{pmatrix}, \vb_4^{tg} = \frac{1}{2} \begin{pmatrix} 1 \\ 1 \\ 1 \\ 1 \end{pmatrix}.
\end{equation}\normalsize 

The actual states $\vb_1, \vb_2, \vb_3, \vb_4$ should be reasonably close to those targets but our QST algorithm will behave as if the states were totally unknown (so that we are resilient to systematic errors).

Using the fact that multiplying two vectors by the same unitary matrix preserves their dot product, it is really easy to check that the $\Xv = \Mv \begin{bmatrix} \vb_1 & \vb_2 & \vb_3 & \vb_4 \end{bmatrix}$ satisfies (\ref{eqn:CS}) and thus (\ref{eqn:CNS}) with a comfortable margin for any unitary matrix $\Mv$ if the $\{\vb_k\}_k$ are equal (or fairly close) to their target.

The target states of (\ref{eqn:v_init}) can be generalized to any number $n_{qb}$ of qubits:
\begin{equation}
\label{eqn:V_INIT}
[\vb_1^{tg}, ..., \vb_d^{tg}] = \underbrace{\begin{pmatrix}1 & \frac{1}{\sqrt{2}} \\ 0 & \frac{1}{\sqrt{2}}\end{pmatrix} \otimes ... \otimes \begin{pmatrix}1 & \frac{1}{\sqrt{2}} \\ 0 & \frac{1}{\sqrt{2}}\end{pmatrix}}_\textrm{$n_{qb}$ times}
\end{equation}

\noindent in the sense that the target vectors on the left-hand-side are defined as the columns of the matrix on the right-hand-side. For all $n_{qb}$-qubit unitary gates, using the states of (\ref{eqn:V_INIT}) with $n_i = d, n_s = 2$ generates a matrix $\Xv$ that satisfies (\ref{eqn:CS}) because its columns form a basis and none of them are orthogonal. These states have the advantages of being unentangled and really easy to prepare: If all qubits are initialized at $\ket{0}$, then the $k$-th state can be prepared by applying a 1-qubit Hadamard gate to the qubits with indices for which there is a $1$ in the binary decomposition of $k-1$. For example, for $\vb_1^{tg}$, all qubits stay on $\ket{0}$ and no Hadamard gate is applied, for $\vb_2^{tg}$, all qubits are initialized at $\ket{0}$ and a Hadamard gate is applied to the last qubit, etc. The Hadamard gate is represented by the unitary matrix $\Hv_d = \frac{1}{\sqrt{2}} \begin{pmatrix} 1 & 1 \\ 1 & -1 \end{pmatrix}$, it transforms $\ket{0}$ into $\frac{1}{\sqrt{2}}\ket{0}+\frac{1}{\sqrt{2}}\ket{1}$.

\begin{figure}
\begin{centering}
\tikzset{every picture/.style={line width=0.75pt}} 

\begin{tikzpicture}[x=0.75pt,y=0.75pt,yscale=-1,xscale=1]

\draw   (215.25,72.02) -- (242.65,72.02) -- (242.65,174.38) -- (215.25,174.38) -- cycle ;
\draw    (242,168.33) -- (264.62,168.58) ;
\draw [shift={(264.62,168.58)}, rotate = 180] [color={rgb, 255:red, 0; green, 0; blue, 0 }  ][line width=0.75]    (10.93,-3.29) .. controls (6.95,-1.4) and (3.31,-0.3) .. (0,0) .. controls (3.31,0.3) and (6.95,1.4) .. (10.93,3.29)   ;
\draw    (242,83.33) -- (264.62,83.58) ;
\draw [shift={(264.62,83.58)}, rotate = 180] [color={rgb, 255:red, 0; green, 0; blue, 0 }  ][line width=0.75]    (10.93,-3.29) .. controls (6.95,-1.4) and (3.31,-0.3) .. (0,0) .. controls (3.31,0.3) and (6.95,1.4) .. (10.93,3.29)   ;
\draw    (341,165.33) -- (363.62,165.58) ;
\draw [shift={(363.62,165.58)}, rotate = 180] [color={rgb, 255:red, 0; green, 0; blue, 0 }  ][line width=0.75]    (10.93,-3.29) .. controls (6.95,-1.4) and (3.31,-0.3) .. (0,0) .. controls (3.31,0.3) and (6.95,1.4) .. (10.93,3.29)   ;
\draw    (341,83.33) -- (363.62,83.58) ;
\draw [shift={(363.62,83.58)}, rotate = 180] [color={rgb, 255:red, 0; green, 0; blue, 0 }  ][line width=0.75]    (10.93,-3.29) .. controls (6.95,-1.4) and (3.31,-0.3) .. (0,0) .. controls (3.31,0.3) and (6.95,1.4) .. (10.93,3.29)   ;
\draw [line width=1.5]    (364.65,171.15) .. controls (365.39,170.09) and (366.1,169.15) .. (366.8,168.33) .. controls (374.27,159.52) and (379.8,163.48) .. (384.65,171.15) ;
\draw    (373.15,170.6) -- (380.52,158.64) ;
\draw   (377.2,160.46) -- (380.62,158.44) -- (379.67,163.68) ;

\draw [line width=1.5]    (364.65,88.15) .. controls (365.39,87.09) and (366.1,86.15) .. (366.8,85.33) .. controls (374.27,76.52) and (379.8,80.48) .. (384.65,88.15) ;
\draw    (373.15,87.6) -- (380.52,75.64) ;
\draw   (377.2,77.46) -- (380.62,75.44) -- (379.67,80.68) ;

\draw [line width=1.5]    (265.65,174.15) .. controls (266.39,173.09) and (267.1,172.15) .. (267.8,171.33) .. controls (275.27,162.52) and (280.8,166.48) .. (285.65,174.15) ;
\draw    (274.15,173.6) -- (281.52,161.64) ;
\draw   (278.2,163.46) -- (281.62,161.44) -- (280.67,166.68) ;

\draw [line width=1.5]    (265.65,88.15) .. controls (266.39,87.09) and (267.1,86.15) .. (267.8,85.33) .. controls (275.27,76.52) and (280.8,80.48) .. (285.65,88.15) ;
\draw    (274.15,87.6) -- (281.52,75.64) ;
\draw   (278.2,77.46) -- (281.62,75.44) -- (280.67,80.68) ;

\draw    (242,83.33) .. controls (285.59,65.73) and (276.54,72.26) .. (308.75,82.02) ;
\draw [shift={(310.25,82.47)}, rotate = 196.39] [color={rgb, 255:red, 0; green, 0; blue, 0 }  ][line width=0.75]    (10.93,-3.29) .. controls (6.95,-1.4) and (3.31,-0.3) .. (0,0) .. controls (3.31,0.3) and (6.95,1.4) .. (10.93,3.29)   ;
\draw    (242,168.33) .. controls (285.59,150.73) and (276.54,157.26) .. (308.75,167.02) ;
\draw [shift={(310.25,167.47)}, rotate = 196.39] [color={rgb, 255:red, 0; green, 0; blue, 0 }  ][line width=0.75]    (10.93,-3.29) .. controls (6.95,-1.4) and (3.31,-0.3) .. (0,0) .. controls (3.31,0.3) and (6.95,1.4) .. (10.93,3.29)   ;
\draw   (127.62,72.02) -- (180,72.02) -- (180,98.12) -- (127.62,98.12) -- cycle ;
\draw    (106.65,83.15) -- (127.62,83.58) ;
\draw [shift={(127.62,83.58)}, rotate = 180] [color={rgb, 255:red, 0; green, 0; blue, 0 }  ][line width=0.75]    (10.93,-3.29) .. controls (6.95,-1.4) and (3.31,-0.3) .. (0,0) .. controls (3.31,0.3) and (6.95,1.4) .. (10.93,3.29)   ;
\draw   (127.62,148.02) -- (181,148.02) -- (181,174.12) -- (127.62,174.12) -- cycle ;
\draw    (181,160.33) -- (203.62,160.58) -- (210.25,161.01) ;
\draw [shift={(212.25,161.13)}, rotate = 183.64] [color={rgb, 255:red, 0; green, 0; blue, 0 }  ][line width=0.75]    (10.93,-3.29) .. controls (6.95,-1.4) and (3.31,-0.3) .. (0,0) .. controls (3.31,0.3) and (6.95,1.4) .. (10.93,3.29)   ;
\draw    (106.65,159.15) -- (127.62,159.58) ;
\draw [shift={(127.62,159.58)}, rotate = 180] [color={rgb, 255:red, 0; green, 0; blue, 0 }  ][line width=0.75]    (10.93,-3.29) .. controls (6.95,-1.4) and (3.31,-0.3) .. (0,0) .. controls (3.31,0.3) and (6.95,1.4) .. (10.93,3.29)   ;
\draw   (311.25,72.02) -- (338.65,72.02) -- (338.65,174.38) -- (311.25,174.38) -- cycle ;
\draw    (180,84.33) -- (202.62,84.58) -- (209.25,85.01) ;
\draw [shift={(211.25,85.13)}, rotate = 183.64] [color={rgb, 255:red, 0; green, 0; blue, 0 }  ][line width=0.75]    (10.93,-3.29) .. controls (6.95,-1.4) and (3.31,-0.3) .. (0,0) .. controls (3.31,0.3) and (6.95,1.4) .. (10.93,3.29)   ;

\draw (89,74.4) node [anchor=north west][inner sep=0.75pt]    {$\ket{0}$};
\draw (220,115) node [anchor=north west][inner sep=0.75pt]   [align=left] {\begin{minipage}[lt]{11.22pt}\setlength\topsep{0pt}
\begin{center}
$\Mv$
\end{center}

\end{minipage}};
\draw (165,32) node [anchor=north west][inner sep=0.75pt]   [align=left] {\begin{minipage}[lt]{50.6pt}\setlength\topsep{0pt}
\begin{center}
$\begin{matrix}
\vb_k \\
\overbrace{}
\end{matrix}$
\end{center}
\end{minipage}};

\draw (193,32) node [anchor=north west][inner sep=0.75pt]   [align=left] {\begin{minipage}[lt]{50.6pt}\setlength\topsep{0pt}
\begin{center}
$\simeq \vb_k^{tg} $
\end{center}
\end{minipage}};

\draw (135,75) node [anchor=north west][inner sep=0.75pt]   [align=left] {$\Hv^{b_1(k)}_d$};
\draw (90,150.4) node [anchor=north west][inner sep=0.75pt]    {$\ket{0}$};
\draw (129,148) node [anchor=north west][inner sep=0.75pt]   [align=left] {$\Hv^{b_{n_{qb}}(k)}_d$};
\draw (274,110) node [anchor=north west][inner sep=0.75pt]   [align=left] {$\vdots$};
\draw (355,110) node [anchor=north west][inner sep=0.75pt]   [align=left] {$\vdots$};
\draw (100,110) node [anchor=north west][inner sep=0.75pt]   [align=left] {$\vdots$};
\draw (316,115) node [anchor=north west][inner sep=0.75pt]   [align=left] {\begin{minipage}[lt]{11.22pt}\setlength\topsep{0pt}
\begin{center}
$\Mv$
\end{center}

\end{minipage}};
\draw (261,32) node [anchor=north west][inner sep=0.75pt]   [align=left] {\begin{minipage}[lt]{29.93pt}\setlength\topsep{0pt}
\begin{center}
$\begin{matrix}
\Mv \vb_k \\
\overbrace{}
\end{matrix}$
\end{center}

\end{minipage}};
\draw (327,32) node [anchor=north west][inner sep=0.75pt]   [align=left] {\begin{minipage}[lt]{40.39pt}\setlength\topsep{0pt}
\begin{center}
$\begin{matrix}
\Mv^2 \vb_k \\
\overbrace{}
\end{matrix}$
\end{center}

\end{minipage}};

\end{tikzpicture}
\end{centering}

\caption{Quantum circuit representing the preparation of a state $\vb_k^{tg}$ of (\ref{eqn:V_INIT}) and the semi-blind QPT of $\Mv$ using this state. The circuit has to be realized for all $k\in\{1, ..., d\}$. The $2\times 2$ matrix $\Hv_d$ is the unitary matrix associated with the one-qubit Hadamard gate, and $b_j(k) \in \{0, 1\}$ is the $j$-th element of the binary decomposition of $k-1$ on $n_{qb}$ bits. Thus $\Hv_d^{b_j(k)} = \Iv_2$ if $b_j(k) = 0$ and $\Hv_d^{b_j(k)} = \Hv_d$ if $b_j(k) = 1$. This setup works well for QST for any value of the unitary matrix $\Mv$ that represents the gate to be identified. Our algorithm is robust to poor implementation of the Hadamard gates. The ``double arrows" in the middle symbolize that half the copies are measured (straight arrow) and the other half fed through the next gate (in practice, this is achieved by waiting $2\Delta_t$ instead of $\Delta_t$).}
\label{fig:deux_td}
\end{figure}
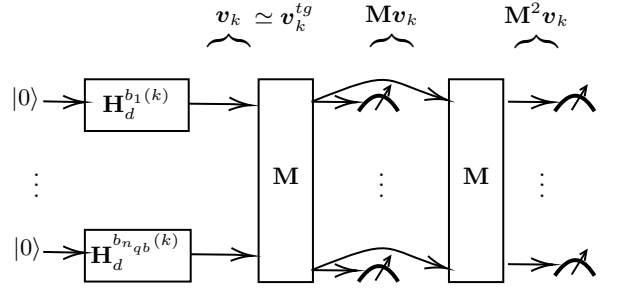

The setup of Fig. \ref{fig:deux_td} with $n_i = d$, $n_s=2$ generates the initial states of (\ref{eqn:V_INIT}), it is very interesting:
\begin{itemize} [leftmargin=*]
\item As explained above, it can identify  any type of unitary gate without prior knowledge (the sufficient condition (\ref{eqn:CS}) is always satisfied).
\item It is fairly easy to prepare, as we only require a single type of gate (Hadamard) other that the gate that we want to identify.
\item We can also tolerate errors in the Hadamard gates: they do not have to be perfect and they do not have to be all the same, we simply require the behaviour of each gate to remain the same while we prepare copies of each state.
\item Using such simple gates at most once for each qubit should limit the decoherence issues.
\end{itemize}

One drawback is that there can be issues with the conditioning of $\Xv$ when $n_{qb}$ increases. The matrix $\Xv$ is always invertible with the states of (\ref{eqn:V_INIT}). But we observed that, experimentally, its smallest singular value decreases exponentially (with $n_{qb}$) towards $0$. For more than 4 qubits we recommend to consider more than $d$ input states, or to use 1-qubit rotation gates (instead of Hadamard gates) with adapted angles that provide a decent conditioning and maintain the non-orthogonality condition in (\ref{eqn:CS}). The conditioning and the inner product of the columns of $\Xv$ do not depend on the value of the unitary matrix $\Mv$, they only depend on the values of the initial states.

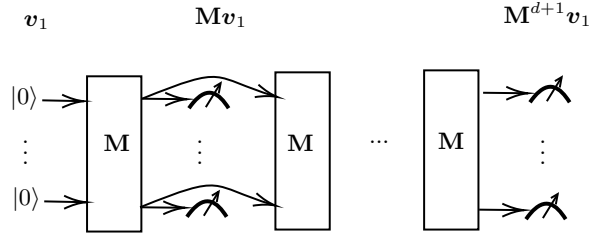
\begin{figure}
\begin{centering}

\tikzset{every picture/.style={line width=0.75pt}} 

\begin{tikzpicture}[x=0.75pt,y=0.75pt,yscale=-1,xscale=1]

\draw    (242,83.33) -- (264.62,83.58) ;
\draw [shift={(264.62,83.58)}, rotate = 180] [color={rgb, 255:red, 0; green, 0; blue, 0 }  ][line width=0.75]    (10.93,-3.29) .. controls (6.95,-1.4) and (3.31,-0.3) .. (0,0) .. controls (3.31,0.3) and (6.95,1.4) .. (10.93,3.29)   ;
\draw    (414,80.33) -- (436.62,80.58) ;
\draw [shift={(436.62,80.58)}, rotate = 180] [color={rgb, 255:red, 0; green, 0; blue, 0 }  ][line width=0.75]    (10.93,-3.29) .. controls (6.95,-1.4) and (3.31,-0.3) .. (0,0) .. controls (3.31,0.3) and (6.95,1.4) .. (10.93,3.29)   ;
\draw [line width=1.5]    (437.65,85.15) .. controls (438.39,84.09) and (439.1,83.15) .. (439.8,82.33) .. controls (447.27,73.52) and (452.8,77.48) .. (457.65,85.15) ;
\draw    (446.15,84.6) -- (453.52,72.64) ;
\draw   (450.2,74.46) -- (453.62,72.44) -- (452.67,77.68) ;

\draw [line width=1.5]    (265.65,88.15) .. controls (266.39,87.09) and (267.1,86.15) .. (267.8,85.33) .. controls (275.27,76.52) and (280.8,80.48) .. (285.65,88.15) ;
\draw    (274.15,87.6) -- (281.52,75.64) ;
\draw   (278.2,77.46) -- (281.62,75.44) -- (280.67,80.68) ;

\draw    (242,83.33) .. controls (285.59,65.73) and (276.54,72.26) .. (308.75,82.02) ;
\draw [shift={(310.25,82.47)}, rotate = 196.39] [color={rgb, 255:red, 0; green, 0; blue, 0 }  ][line width=0.75]    (10.93,-3.29) .. controls (6.95,-1.4) and (3.31,-0.3) .. (0,0) .. controls (3.31,0.3) and (6.95,1.4) .. (10.93,3.29)   ;
\draw    (191.65,84.15) -- (212.62,84.58) ;
\draw [shift={(212.62,84.58)}, rotate = 180] [color={rgb, 255:red, 0; green, 0; blue, 0 }  ][line width=0.75]    (10.93,-3.29) .. controls (6.95,-1.4) and (3.31,-0.3) .. (0,0) .. controls (3.31,0.3) and (6.95,1.4) .. (10.93,3.29)   ;
\draw   (384.25,69.02) -- (411.65,69.02) -- (411.65,149.38) -- (384.25,149.38) -- cycle ;
\draw   (214.25,72.02) -- (241.65,72.02) -- (241.65,150.38) -- (214.25,150.38) -- cycle ;
\draw   (309.25,70.02) -- (336.65,70.02) -- (336.65,149.38) -- (309.25,149.38) -- cycle ;
\draw    (192.65,135.15) -- (213.62,135.58) ;
\draw [shift={(213.62,135.58)}, rotate = 180] [color={rgb, 255:red, 0; green, 0; blue, 0 }  ][line width=0.75]    (10.93,-3.29) .. controls (6.95,-1.4) and (3.31,-0.3) .. (0,0) .. controls (3.31,0.3) and (6.95,1.4) .. (10.93,3.29)   ;
\draw    (241,138.13) -- (263.62,138.38) ;
\draw [shift={(263.62,138.38)}, rotate = 180] [color={rgb, 255:red, 0; green, 0; blue, 0 }  ][line width=0.75]    (10.93,-3.29) .. controls (6.95,-1.4) and (3.31,-0.3) .. (0,0) .. controls (3.31,0.3) and (6.95,1.4) .. (10.93,3.29)   ;
\draw [line width=1.5]    (264.65,142.95) .. controls (265.39,141.89) and (266.1,140.95) .. (266.8,140.13) .. controls (274.27,131.32) and (279.8,135.28) .. (284.65,142.95) ;
\draw    (273.15,142.4) -- (280.52,130.44) ;
\draw   (277.2,132.26) -- (280.62,130.24) -- (279.67,135.48) ;

\draw    (241,138.13) .. controls (284.59,120.53) and (275.54,127.06) .. (307.75,136.82) ;
\draw [shift={(309.25,137.27)}, rotate = 196.39] [color={rgb, 255:red, 0; green, 0; blue, 0 }  ][line width=0.75]    (10.93,-3.29) .. controls (6.95,-1.4) and (3.31,-0.3) .. (0,0) .. controls (3.31,0.3) and (6.95,1.4) .. (10.93,3.29)   ;
\draw    (412,139.33) -- (434.62,139.58) ;
\draw [shift={(434.62,139.58)}, rotate = 180] [color={rgb, 255:red, 0; green, 0; blue, 0 }  ][line width=0.75]    (10.93,-3.29) .. controls (6.95,-1.4) and (3.31,-0.3) .. (0,0) .. controls (3.31,0.3) and (6.95,1.4) .. (10.93,3.29)   ;
\draw [line width=1.5]    (435.65,144.15) .. controls (436.39,143.09) and (437.1,142.15) .. (437.8,141.33) .. controls (445.27,132.52) and (450.8,136.48) .. (455.65,144.15) ;
\draw    (444.15,143.6) -- (451.52,131.64) ;
\draw   (448.2,133.46) -- (451.62,131.44) -- (450.67,136.68) ;

\draw (174,75.4) node [anchor=north west][inner sep=0.75pt]    {$\ket{0}$};
\draw (389,99) node [anchor=north west][inner sep=0.75pt]   [align=left] {\begin{minipage}[lt]{11.22pt}\setlength\topsep{0pt}
\begin{center}
$\Mv$
\end{center}

\end{minipage}};
\draw (260,36) node [anchor=north west][inner sep=0.75pt]   [align=left] {\begin{minipage}[lt]{30.51pt}\setlength\topsep{0pt}
\begin{center}
$\Mv \vb_1$
\end{center}

\end{minipage}};
\draw (410,33) node [anchor=north west][inner sep=0.75pt]   [align=left] {\begin{minipage}[lt]{52.61pt}\setlength\topsep{0pt}
\begin{center}
$\Mv^{d+1} \vb_1$
\end{center}

\end{minipage}};
\draw (220.25,100.02) node [anchor=north west][inner sep=0.75pt]   [align=left] {\begin{minipage}[lt]{11.22pt}\setlength\topsep{0pt}
\begin{center}
$\Mv$
\end{center}

\end{minipage}};
\draw (313,100) node [anchor=north west][inner sep=0.75pt]   [align=left] {\begin{minipage}[lt]{11.22pt}\setlength\topsep{0pt}
\begin{center}
$\Mv$
\end{center}

\end{minipage}};
\draw (355,103) node [anchor=north west][inner sep=0.75pt]   [align=left] {...};
\draw (268,95) node [anchor=north west][inner sep=0.75pt]   [align=left] {$\vdots$};
\draw (441,95) node [anchor=north west][inner sep=0.75pt]   [align=left] {$\vdots$};
\draw (175,126.4) node [anchor=north west][inner sep=0.75pt]    {$\ket{0}$};
\draw (180,95) node [anchor=north west][inner sep=0.75pt]    {$\vdots$};
\draw (167,40) node [anchor=north west][inner sep=0.75pt]   [align=left] {\begin{minipage}[lt]{30.51pt}\setlength\topsep{0pt}
\begin{center}
$\vb_1$
\end{center}

\end{minipage}};

\end{tikzpicture}

\end{centering}

\caption{Quantum circuit that can be used to perform QPT of the gate represented by $\Mv$ with a single input state and $d+1$ time steps: $n_i=1, n_s = d+1$. This setup works if and only if the matrix $\Mv$ generates states $\Xv = [\Mv \vb_1, ..., \Mv^d \vb_1]$ that satisfy (\ref{eqn:CNS}). This setup is very easy to realize as there is only one value of the initial state. The ``double arrows" symbolize that the $n_c n_t$ copies are measured (straight arrow) and the others are fed through the next gate.}

\label{fig:bcp_td}
\end{figure}

This setup can be less efficient than the one described below (Fig. \ref{fig:bcp_td}) if we have a rough idea of what the gate that we are trying to identify is supposed to do. For example, with $n_{qb} = 2$, if the unitary process that we want to identify is supposed to be represented by 
\begin{equation}
\label{eqn:wierd}\Mv_{tg} = \frac{1}{2}\begin{pmatrix} 1&-\sqrt{2}&0&1\\1&\sqrt{2}&0&1\\1&0&-\sqrt{2}&-1\\1&0&\sqrt{2}&-1\end{pmatrix}, 
\end{equation}
we hereafter show how to exploit this knowledge to design a more efficient setup.
If $\Mv$ is close enough to $\Mv_{tg}$, then we can use a single initial state, $n_i = 1$ and $n_s = 5$ time delays, and $\Mv$ should (unless it is really far from the target) make the initial state evolve in a way that makes (\ref{eqn:CS}) true. If the initial state is $\vb_1  = \begin{pmatrix} 1 \\ 0 \\ 0 \\ 0 \end{pmatrix}$ for example, then, we can compute the $\Xv$ we would have if $\Mv$ was exactly the target:
\begin{equation}
\label{eqn:Xtg}
\Xv_{tg} = \begin{bmatrix} \Mv_{tg}\vb_1 & ... & \Mv_{tg}^d\vb_1\end{bmatrix}. 
\end{equation}
The matrix $\Xv_{tg}$ is well conditioned (greatest singular value only $~2.5$ times greater than the smallest) and its first column is reasonably far from being orthogonal to the others (smallest modulus of dot product of $0.14$). This setup (generalized for any number of qubits) is represented in Fig. \ref{fig:bcp_td}.
Considering fewer states and more time delays than in the setup of Fig. \ref{fig:deux_td} means that we partly rely on the gate we are trying to identify to create the states that we will use. The drawbacks of doing this are the following:
\begin{itemize} [leftmargin=*]
\item  We are never sure that (\ref{eqn:CNS}) is true with a comfortable margin, unless we have a decent idea of what the gate that we want to characterize does. In contrast, with $n_i = d, n_s = 2$ and the initial states of (\ref{eqn:V_INIT}), we are sure that (\ref{eqn:CS}) and (\ref{eqn:CNS}) are comfortably satisfied for any $\Mv$.
\item It does not work with most of the ``classic" (i.e. most often considered and used in the literature) quantum gates because those gates often involve rotations of $90^{\circ}$ (which can make too many columns of $\Xv$ close to being orthogonal) or do not change some directions of the Hilbert space (which can make $\Xv$ poorly conditioned).
\item  If all the gates to be identified ($\Mv$) represented in Fig. \ref{fig:bcp_td} are not the same, the estimate will suffer, whereas if the two gates to be identified in Fig. \ref{fig:deux_td} are different, it is not really a problem, because the second gate is identified, the first one is only used for the state preparation. We think that this is not a big issue because the way the gates are physically realized (see Section \ref{section:State}) makes it easy to apply the same gate several times. 
\item Having higher values of $n_s$ can create decoherence issues for some architectures, because the state is observed after waiting $n_s \Delta_t$.
\end{itemize}
But the setup of Fig. \ref{fig:bcp_td} has the following advantages:
\begin{itemize} [leftmargin=*]
\item Preparing copies of a single input state value is a lot simpler for the operator. Only one type of gate (the gate to be identified) is used. Using more types of gates is problematic because, even though the QPT algorithm of Section \ref{section:QPT_ls} makes no assumption on the values of the unitary matrices that represent each gate ($\Mv$ and the gates used for the initial states preparation), it still assumes that they are unitary gates that do exactly the same thing every time we repeat the experiment. If this is not the case, the quality of our estimate of $\Mv$ will suffer.
\item Having a higher $n_s$ means that more of the estimated output states are being ``re-used" as estimates of input states and vice versa. This means that fewer states are measured overall, and that we can afford to make more copies of each state that we measure. For example, with $n_{qb} = 2$, we can use the $n_i = 4$ different initial states values of (\ref{eqn:V_INIT}) and $n_s = 2$ time steps to estimate a 2-qubit gate, this require $8$ different states to be measured. If the same gate can be estimated with a single input state value ($n_i = 1$) and $n_s = 5$ time delays, this only require $5$ different states to be measured.
\item The example of the target gate (\ref{eqn:wierd}) that we used to illustrate a case where $n_i = 1, n_s = d+1$ might seem far fetched and make the reader think that $n_i = 1, n_s = d+1$ works very rarely, but it works fairly well for random unitary gates (see Section \ref{section:multiQB}).
\end{itemize}

We could consider an intermediate setup with $d>n_i>1$. It can be useful if $\Mv$ is the identity (up to a global phase) on a subspace of the Hilbert space but brings enough diversity to the supplement of this subspace. For example, let us assume that we want to perform QPT for a gate that is supposed to be represented by $\Mv_{tg} = \begin{pmatrix} 1 & 0 & 0 & 0 \\ 0 & 1 & 0 & 0 \\ 0 & 0 & \frac{1}{\sqrt{2}} & -\frac{1}{\sqrt{2}} \\ 0 & 0 & \frac{1}{\sqrt{2}} & \frac{1}{\sqrt{2}}\end{pmatrix}$. If we consider a single arbitrary input state value $\vb_1$ and if $\Mv = \Mv_{tg}$, then, the $\Xv_{tg}$ of (\ref{eqn:Xtg}) will never be of full rank because its first two rows will contain the same value in all columns, so that these rows with be colinear. But if we consider $n_i = 2$ input states and $n_s = 3$ time delays, we can make QPT possible with $\vb_1^{tg} = \frac{1}{\sqrt{2}} \begin{pmatrix} 1 \\ 0 \\ 1 \\ 0 \end{pmatrix}, \vb_2^{tg} = \frac{1}{\sqrt{2}} \begin{pmatrix} 0 \\ 1 \\ 0 \\ 1 \end{pmatrix}$. It is very easy to check that $\Xv = \begin{bmatrix}\Mv \vb_1 & \Mv^2 \vb_1 & \Mv \vb_2 & \Mv^2 \vb_2\end{bmatrix}$ satisfies (\ref{eqn:CNS}) comfortably if $\Mv$ and the input states are equal to their target. And therefore (\ref{eqn:CNS}) should still be satisfied if they are close to their target.
\subsection{Comparison with the literature on unitary QPT}
\label{section:comparaison_setup}
In Appendix 1 of \cite{reich}, Reich et al. give a necessary and sufficient condition on the input states for a unitary process to be uniquely (up to a global phase) determined among all processes (unitary or not). This condition is
\begin{equation}
\label{eqn:CNSreichMixed}
 Com \left( \left\{ \rhob_{\ell}\right\}_\ell \right) = \{ e^{i\theta}\Iv_d\}_{\theta \in \mathds{R}}
\end{equation}
where $\rhob_{\ell}$ is the density matrix of the $\ell$-th mixed input state, $Com$ is the comutant, $Com \left( \left\{ \rhob_{\ell}\right\}_\ell \right)$ refers to the ensemble of the $\emph{unitary}$ matrices that commute with all $\left\{\rhob_{\ell} \right\}_\ell$, and $\Iv_d$ is the $d\times d$ identity matrix. 

If we write (\ref{eqn:CNSreichMixed}) for pure input states: $\rhob_{\ell} = \xb_{\ell} \xb_{\ell}^*$, the condition becomes:
\begin{equation}
\label{eqn:CNSreich}
 Com \left( \left\{ \xb_{\ell} \xb_{\ell}^*\right\}_\ell \right) = \{ e^{i\theta}\Iv_d\}_{\theta \in \mathds{R}}.
\end{equation}

The conditions (\ref{eqn:CNS}) and (\ref{eqn:CNSreich}) are equivalent, we show it in Appendix \ref{section:math3}. This does not mean that (\ref{eqn:CNS}) is equivalent to the original condition (\ref{eqn:CNSreichMixed}) of Reich et al.  because the latter was defined with mixed input states, and (\ref{eqn:CNSreich}) is its reformulation with pure input states. Therefore we only have the equivalence between (\ref{eqn:CNS}) and (\ref{eqn:CNSreichMixed}) when the input states are pure.

In \cite{reich}, the authors give two examples of sets of input states that satisfy (\ref{eqn:CNSreichMixed}). The first one is a set of $d+1$ pure input states: $d$ states that form an orthonormal basis and a last state that is the average of the first $d$ states (it works with any orthonormal basis). We see the same idea as (\ref{eqn:CS}): one of the states is not orthogonal to the others. 

Reich et al. also show that if we allow for the use of mixed states, then, only 2 input states are required for QPT to be possible. This is because, if a mixed state with $d$ distinct eigenvalues goes through the unitary process, then the output has the same eigenvalues but the eigenvectors are multiplied by the unitary matrix associated with the process. Therefore it is possible to evaluate the unitary matrix on the orthonormal basis of the eigenvectors by using a single mixed input state. This is not enough though because all the states are orthogonal to one another, so they need a second input state. We chose not to consider mixed states because they are more complex to produce and set to a predetermined value. Baldwin et al. remark in \cite{nearunitary} when they use the results of \cite{reich}, ``in practice we do not have reliable procedures to produce a desired, reproducible, mixed state". We do not really agree with that statement, since any mixed state can be seen as a statistical mixture of at most $d$ pure states (represented by the eigenvectors of the density matrix). If we can generate all the pure states of the mixture and then randomly select one of them for each copy of the mixed state that we generate (with the probability of choosing each eigenvector given by the associated eigenvalue of the density matrix of the target mixed state), then, we can generate the desired mixed state. Generating a mixed state this way would not be optimal however, as it involves preparing copies of $d$ different pure states and then ``mixing" them (not recording which pure state is used for which copy of the mixed state). Using the $d$ pure states directly is more efficient than using this ``mixture". 

Beyond reformulating (\ref{eqn:CNSreich}) as (\ref{eqn:CNS}) and providing a set of simple input states, our contribution is that, contrary to \cite{reich}, we do provide a QPT algorithm that works with every set of input states that verifies our necessary and sufficient condition. We also took advantage of the fact that (\ref{eqn:CNS}) is loose to introduce our semi-blind setup that removes the issue of systematic errors.

In \cite{nearunitary}, Baldwin et al. propose a unitary QPT algorithm (after their Equation (20)) adapted to the following specific set of $d$ input states: 
\begin{equation}
\label{eqn:v_init_baldwin}
\left\{ \deltab_1, \{\frac{1}{\sqrt{2}} (\deltab_1 + \deltab_k)\}_{k\in\{2, ..., d\}}\right\}
\end{equation}
(where $\deltab_k$ is the $d$-dimensional vector that contains $d-1$ zeros and a single $1$ on the $k$-th element). The last $d-1$ states of the set are entangled, they can be more challenging to prepare precisely than unentangled states. This algorithm is a lot simpler than ours but it only works for this particular set of $d$ states, and the quality of the QPT's estimate  will be limited by the systematic errors on the initial states. Our algorithm works for any set of states for which QPT is possible, and we take advantage of this to propose a semi-blind setup that eliminates the issue of systematic errors and works with the unentangled initial states of (\ref{eqn:V_INIT}). Baldwin et al. also go further and propose an algorithm adapted to processes close to being unitary. This algorithm could actually be used with arbitrary input states, as long as they satisfy (\ref{eqn:CNS}). But it is a $l_1$ minimization algorithm with a regularization parameter, the solution (and its rank) will depend on the value of the regularization parameter. 

\section{Performance on simulated data}
\label{section:simu}

We chose to first focus on the setup with the four initial states (\ref{eqn:v_init}) to estimate random two-qubit gates in order to test our resiliency to different kinds of errors. We consider this setup because two-qubit QPT is a classic problem and the four initial states (\ref{eqn:v_init}) are versatile. We aim to test the resiliency to the multinomial error (error generated by the finite number of measurements) in Section \ref{section:multinom}. After that, in Section \ref{section:errC0} and \ref{section:simuBaldwin}, we test our resiliency to the systematic errors on the input states, and we compare our algorithm to that of Baldwin et al. \cite{nearunitary}. Finally, in Section \ref{section:multiQB}, we expand the tested setup by increasing the number of qubits. We also compare the performance of the setup of Fig. \ref{fig:deux_td} to that of the setup of Fig. \ref{fig:bcp_td} when the number of qubits is varied.

\subsection{Impact of the number of measurements}
\label{section:multinom}

We designed simulations to determine the number $n_c$ (number of copies of each state that need to be prepared and measured for each measurement type) that is required to obtain an accurate estimate of $\Mv$ with the setup of Fig. \ref{fig:deux_td}. We start by considering the case when the finite number of measurements is the only source of errors. There is no decoherence, the measurements follow the model, and the initial states that we consider are prepared with perfect Hadamard gates (as we will see in Section \ref{section:errC0}, the latter point does not really matter). We make $n_c$ vary from $20$ to $25000$, and for each value of $n_c$, we generate $500$ random quantum gates defined by unitary matrices created by applying the Gram-Schmidt algorithm to a random complex matrix with each coefficient i.i.d. following the circularly-symmetric centered complex normal distribution. For each $n_c$ we also generate $500$ (one per gate to be identified) sets of $n_s n_i n_t$ random measurement counts (the measurement counts are defined at the end of Section \ref{section:QPTsetup}) associated with all $n_c$ copies. They are simulated with a $d$-outcome multinomial distribution with the probabilities of all outcomes set to their theoretical values (Born rule). For example, if one of the measured states is $\vb = \begin{pmatrix} 0.5 & 0.5 & 0.5 i & 0.5 i\end{pmatrix}^T$ (to be identified with the measurements) and we perform measurements in the computational basis (measurement type $ZZ$ in Appendix \ref{section:DataModel}) with $n_c = 50$ copies of this state measured in this basis, we simulate measurement counts by sampling a multinomial distribution with probability parameters given by the Born rule $\pv = |\Iv_4 \vb|^2 =\begin{pmatrix} 0.25 & 0.25 & 0.25 & 0.25 \end{pmatrix}^T$ for the 4 outcomes and $n_c = 50$ trials. The resulting empirical measurement counts could be $\begin{pmatrix} 13 & 10 & 15 & 12 \end{pmatrix}^T$ for example.

For each $n_c$, we therefore have $500$ unitary matrices $\{\Mv\}$ to be estimated and $500$ sets of measurement counts to perform the QPT. We run our algorithm and get $500$ estimates $\{\widehat{\Mv}_{LS}\}$. We then have to quantify the error between the $\{\Mv\}$ and the $\{\widehat{\Mv}_{LS}\}$. We choose to use the following metric denoted $\epsilon (\widehat{\Mv}_{LS}, \Mv)$ and called the error:

\begin{equation}
\label{eqn:ourerr}
\epsilon (\widehat{\Mv}_{LS}, \Mv) = \frac{1}{\sqrt{2d}} ||\Mv-\widehat{\Mv}_{LS}e^{i\phi}||
\end{equation}

\noindent where $\phi$ is the angle that minimizes the error (it accounts for the fact that $\Mv$ can only be recovered up to a global phase) and $||.||$ is the Frobenius norm. This metric is between $0$ (if $\widehat{\Mv}$ and $\Mv_{true}$ are the same up to a global phase) and $1$ (if they are orthogonal with respect to the Hilbert–Schmidt inner product). It can be shown that 
\begin{equation}
\label{eqn:phaseM}
\phi = arg\big(tr(\widehat{\Mv}_{LS}^* \Mv)\big)
\end{equation}
($arg$ is the phase of a complex number and $tr$ is the trace).

\begin{figure}[h]
    \centering
    \includegraphics[width=7.5cm,height=6cm]{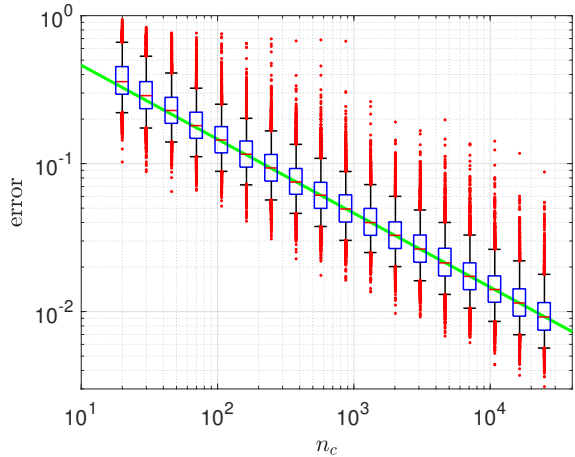}
    \caption{Box-plots of the QPT error when the number $n_c$ of measured state copies per state and per type of measurement increases. The light (green) line from top left to bottom right represents the function $n \longrightarrow \frac{C}{\sqrt{n}}$ where $C$ is computed so that the line fits (in the least square sense) the median in the box-plots associated with $n_c \geq 1000$. The central (blue) rectangle represents the space between the first and last quartiles where half of the observations are. The narrow (red) line in the middle of each rectangle is the median, and the (black) whiskers range from the $5$ percentile to the $95$ percentile. By definition, $90\%$ of the samples are within the range of whiskers, the ones that are not are called ``outliers" and represented as small (red) dots above and below each box-plot.}
	\label{fig:eps1}
\vspace{-1mm}
\end{figure}

In the literature, the fidelity $f(\widehat{\Mv}_{LS}, \Mv)$ is more often used (see e.g. Eq. (24) in \cite{nearunitary}). It is defined for all quantum processes (not only unitary processes). It can be shown that the two metrics are linked when we are dealing with unitary processes: $f(\widehat{\Mv}_{LS}, \Mv) = 1-\epsilon (\widehat{\Mv}_{LS}, \Mv)^2$. We consider that $\epsilon$ is a lot more attractive than $f$ because $\epsilon$ is a distance and has a real meaning (a value $\epsilon$ of $0.1$ can be seen as a $10\%$ error). It is also a lot more informative when the estimate $\widehat{\Mv}_{LS}$ starts to become close to $\Mv$: clearly an error $\epsilon$ of $0.1$ is a lot worse than an error of $0.01$, but comparing the associated values of $f$, equal to $0.99$ and $0.9999$, is harder.

Fig. \ref{fig:eps1} shows the box-plots (first and last quartiles, median, $5$ and $95$ percentile and outliers) of the 500 samples of the error for each value of $n_c$. We also display the line proportional to $\frac{1}{\sqrt{n_c}}$ that fits some of the measured medians (see caption).

With the log scale,  we find the classic inverse relation between the estimation error and the square root of the number of samples (the line accurately fits the last medians) if $n_c$ is high enough ($n_c \geq 100$).
\subsection{Impact of the systematic errors}
\label{section:errC0}
\begin{figure}[h]
    \centering
    \includegraphics[width=7.5cm,height=6cm]{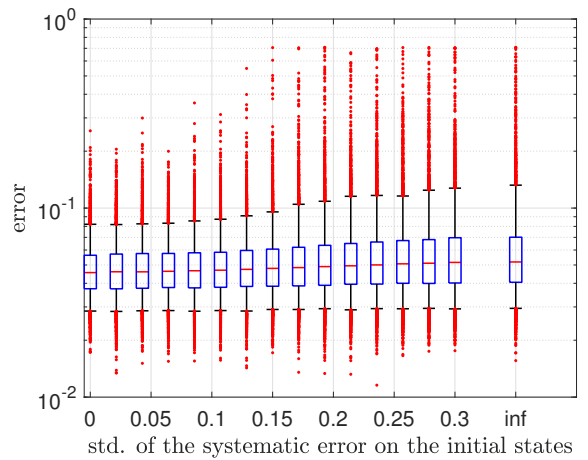}
    \caption{Box-plots of the QPT error versus the standard deviation of the systematic error on the initial states with $n_c=1000$. Same box-plot representation as in Fig. \ref{fig:eps1}.}
	\label{fig:eps2}
\vspace{-1mm}
\end{figure}

We design a second simulation in order to see what happens if we add systematic errors to the initial states. We simulate the initial states of (\ref{eqn:v_init}), we fix $n_c = 1000$, and we simulate systematic errors, each of the $n_t n_c n_s$ copies of a given state has the same error. Physically this means the initialization of the states at $\ket{0}$ or the Hadamard gate used to transform them have errors, but, for a given gate or initialization, this error is the same on each copy. We have to consider the multinomial error ($n_c = 1000$) in addition to the systematic error because the QPT error would always be zero otherwise. The systematic errors are modelled as random complex vectors with each coefficient i.i.d. following the circularly-symmetric centered complex normal distribution added to each vector of (\ref{eqn:v_init}). After adding this error we re-normalize the initial vectors. The standard deviation of this systematic error is the same for all components, and we vary it from $0$ to $0.3$. We also test our algorithm with totally random states uniformly sampled on the ensemble of pure states. Fig. \ref{fig:eps2} displays the box-plots of the QPT error versus the standard deviation (std.) of the error. The infinite std. (inf) refers to the totally random initial states.

There is a significant trend of greater $95$  percentile and greater outliers when the systematic error increases. The $95$  percentile goes from $0.083$ to $0.13$. For the lower part of the box-plots, the difference is less noticeable. Even when the initial states are totally random, the median of the error ($0.052$) is close to the median we have with no systematic error ($0.046$). The QPT algorithm is resilient to the systematic errors on the initial states because it makes no assumption on the values of the initial states. The only drawback of the systematic errors is that they can make the QPT problem harder by making too many initial states close to being orthogonal or making the matrix $\Xv$ poorly conditioned. The chance of those problems occurring is fairly low, even with totally random input states, this is the reason why the impact of the systematic errors is mostly visible in the upper part of the box-plots.
\subsection{Comparison with the literature on the SQPT setup}
\label{section:simuBaldwin}
\begin{figure}[H]

\tikzset{every picture/.style={line width=0.75pt}} 

\begin{tikzpicture}[x=0.75pt,y=0.75pt,yscale=-1,xscale=1]

\draw    (234,77.33) -- (249.17,77.58) ;
\draw [shift={(249.17,77.58)}, rotate = 180] [color={rgb, 255:red, 0; green, 0; blue, 0 }  ][line width=0.75]    (10.93,-3.29) .. controls (6.95,-1.4) and (3.31,-0.3) .. (0,0) .. controls (3.31,0.3) and (6.95,1.4) .. (10.93,3.29)   ;
\draw   (206.25,66.02) -- (233.65,66.02) -- (233.65,90) -- (206.25,90) -- cycle ;
\draw    (234,104.33) -- (249.17,104.58) ;
\draw [shift={(249.17,104.58)}, rotate = 180] [color={rgb, 255:red, 0; green, 0; blue, 0 }  ][line width=0.75]    (10.93,-3.29) .. controls (6.95,-1.4) and (3.31,-0.3) .. (0,0) .. controls (3.31,0.3) and (6.95,1.4) .. (10.93,3.29)   ;
\draw   (206.25,93.02) -- (233.65,93.02) -- (233.65,117) -- (206.25,117) -- cycle ;
\draw    (234,131.33) -- (249.17,131.58) ;
\draw [shift={(249.17,131.58)}, rotate = 180] [color={rgb, 255:red, 0; green, 0; blue, 0 }  ][line width=0.75]    (10.93,-3.29) .. controls (6.95,-1.4) and (3.31,-0.3) .. (0,0) .. controls (3.31,0.3) and (6.95,1.4) .. (10.93,3.29)   ;
\draw   (206.25,120.02) -- (233.65,120.02) -- (233.65,144) -- (206.25,144) -- cycle ;
\draw    (234,158.33) -- (249.17,158.58) ;
\draw [shift={(249.17,158.58)}, rotate = 180] [color={rgb, 255:red, 0; green, 0; blue, 0 }  ][line width=0.75]    (10.93,-3.29) .. controls (6.95,-1.4) and (3.31,-0.3) .. (0,0) .. controls (3.31,0.3) and (6.95,1.4) .. (10.93,3.29)   ;
\draw   (206.25,147.02) -- (233.65,147.02) -- (233.65,171) -- (206.25,171) -- cycle ;
\draw    (99,77.33) -- (114.17,77.58) ;
\draw [shift={(114.17,77.58)}, rotate = 180] [color={rgb, 255:red, 0; green, 0; blue, 0 }  ][line width=0.75]    (10.93,-3.29) .. controls (6.95,-1.4) and (3.31,-0.3) .. (0,0) .. controls (3.31,0.3) and (6.95,1.4) .. (10.93,3.29)   ;
\draw    (54.17,78.15) -- (69.62,78.58) ;
\draw [shift={(69.62,78.58)}, rotate = 180] [color={rgb, 255:red, 0; green, 0; blue, 0 }  ][line width=0.75]    (10.93,-3.29) .. controls (6.95,-1.4) and (3.31,-0.3) .. (0,0) .. controls (3.31,0.3) and (6.95,1.4) .. (10.93,3.29)   ;
\draw   (71.25,66.02) -- (98.65,66.02) -- (98.65,90) -- (71.25,90) -- cycle ;
\draw    (99,104.33) -- (114.17,104.58) ;
\draw [shift={(114.17,104.58)}, rotate = 180] [color={rgb, 255:red, 0; green, 0; blue, 0 }  ][line width=0.75]    (10.93,-3.29) .. controls (6.95,-1.4) and (3.31,-0.3) .. (0,0) .. controls (3.31,0.3) and (6.95,1.4) .. (10.93,3.29)   ;
\draw    (54.17,105.15) -- (69.62,105.58) ;
\draw [shift={(69.62,105.58)}, rotate = 180] [color={rgb, 255:red, 0; green, 0; blue, 0 }  ][line width=0.75]    (10.93,-3.29) .. controls (6.95,-1.4) and (3.31,-0.3) .. (0,0) .. controls (3.31,0.3) and (6.95,1.4) .. (10.93,3.29)   ;
\draw   (71.25,93.02) -- (98.65,93.02) -- (98.65,117) -- (71.25,117) -- cycle ;
\draw    (99,131.33) -- (114.17,131.58) ;
\draw [shift={(114.17,131.58)}, rotate = 180] [color={rgb, 255:red, 0; green, 0; blue, 0 }  ][line width=0.75]    (10.93,-3.29) .. controls (6.95,-1.4) and (3.31,-0.3) .. (0,0) .. controls (3.31,0.3) and (6.95,1.4) .. (10.93,3.29)   ;
\draw    (54.17,132.15) -- (69.62,132.58) ;
\draw [shift={(69.62,132.58)}, rotate = 180] [color={rgb, 255:red, 0; green, 0; blue, 0 }  ][line width=0.75]    (10.93,-3.29) .. controls (6.95,-1.4) and (3.31,-0.3) .. (0,0) .. controls (3.31,0.3) and (6.95,1.4) .. (10.93,3.29)   ;
\draw   (71.25,120.02) -- (98.65,120.02) -- (98.65,144) -- (71.25,144) -- cycle ;
\draw    (99,158.33) -- (114.17,158.58) ;
\draw [shift={(114.17,158.58)}, rotate = 180] [color={rgb, 255:red, 0; green, 0; blue, 0 }  ][line width=0.75]    (10.93,-3.29) .. controls (6.95,-1.4) and (3.31,-0.3) .. (0,0) .. controls (3.31,0.3) and (6.95,1.4) .. (10.93,3.29)   ;
\draw    (54.17,159.15) -- (69.62,159.58) ;
\draw [shift={(69.62,159.58)}, rotate = 180] [color={rgb, 255:red, 0; green, 0; blue, 0 }  ][line width=0.75]    (10.93,-3.29) .. controls (6.95,-1.4) and (3.31,-0.3) .. (0,0) .. controls (3.31,0.3) and (6.95,1.4) .. (10.93,3.29)   ;
\draw   (71.25,147.02) -- (98.65,147.02) -- (98.65,171) -- (71.25,171) -- cycle ;
\draw    (149,78.33) -- (171.62,78.58) ;
\draw [shift={(171.62,78.58)}, rotate = 180] [color={rgb, 255:red, 0; green, 0; blue, 0 }  ][line width=0.75]    (10.93,-3.29) .. controls (6.95,-1.4) and (3.31,-0.3) .. (0,0) .. controls (3.31,0.3) and (6.95,1.4) .. (10.93,3.29)   ;
\draw [line width=1.5]    (169.18,84.15) .. controls (169.78,83.16) and (170.36,82.29) .. (170.93,81.52) .. controls (176.98,73.31) and (181.47,77) .. (185.4,84.15) ;
\draw    (176.08,83.63) -- (182.05,72.49) ;
\draw   (179.36,74.19) -- (182.14,72.3) -- (181.36,77.18) ;

\draw    (149,78.33) .. controls (184.17,60.82) and (177.09,67.2) .. (202.74,76.87) ;
\draw [shift={(204.36,77.47)}, rotate = 199.93] [color={rgb, 255:red, 0; green, 0; blue, 0 }  ][line width=0.75]    (10.93,-3.29) .. controls (6.95,-1.4) and (3.31,-0.3) .. (0,0) .. controls (3.31,0.3) and (6.95,1.4) .. (10.93,3.29)   ;
\draw [line width=1.5]    (169.18,110.15) .. controls (169.78,109.25) and (170.36,108.45) .. (170.93,107.74) .. controls (176.98,100.23) and (181.47,103.6) .. (185.4,110.15) ;
\draw    (176.08,109.68) -- (182.05,99.47) ;
\draw   (179.36,101.03) -- (182.14,99.3) -- (181.36,103.77) ;

\draw    (149,104.33) .. controls (184.17,86.82) and (177.09,93.2) .. (202.74,102.87) ;
\draw [shift={(204.36,103.47)}, rotate = 199.93] [color={rgb, 255:red, 0; green, 0; blue, 0 }  ][line width=0.75]    (10.93,-3.29) .. controls (6.95,-1.4) and (3.31,-0.3) .. (0,0) .. controls (3.31,0.3) and (6.95,1.4) .. (10.93,3.29)   ;
\draw [line width=1.5]    (169.18,137.15) .. controls (169.78,136.25) and (170.36,135.45) .. (170.93,134.74) .. controls (176.98,127.23) and (181.47,130.6) .. (185.4,137.15) ;
\draw    (176.08,136.68) -- (182.05,126.47) ;
\draw   (179.36,128.03) -- (182.14,126.3) -- (181.36,130.77) ;

\draw    (149,131.33) .. controls (184.17,113.82) and (177.09,120.2) .. (202.74,129.87) ;
\draw [shift={(204.36,130.47)}, rotate = 199.93] [color={rgb, 255:red, 0; green, 0; blue, 0 }  ][line width=0.75]    (10.93,-3.29) .. controls (6.95,-1.4) and (3.31,-0.3) .. (0,0) .. controls (3.31,0.3) and (6.95,1.4) .. (10.93,3.29)   ;
\draw [line width=1.5]    (169.99,164.15) .. controls (170.59,163.25) and (171.17,162.45) .. (171.74,161.74) .. controls (177.79,154.23) and (182.28,157.6) .. (186.21,164.15) ;
\draw    (176.89,163.68) -- (182.86,153.47) ;
\draw   (180.17,155.03) -- (182.95,153.3) -- (182.18,157.77) ;

\draw    (149.81,158.33) .. controls (184.98,140.82) and (177.9,147.2) .. (203.55,156.87) ;
\draw [shift={(205.17,157.47)}, rotate = 199.93] [color={rgb, 255:red, 0; green, 0; blue, 0 }  ][line width=0.75]    (10.93,-3.29) .. controls (6.95,-1.4) and (3.31,-0.3) .. (0,0) .. controls (3.31,0.3) and (6.95,1.4) .. (10.93,3.29)   ;
\draw    (285,77.33) -- (303.36,77.58) ;
\draw [shift={(303.36,77.58)}, rotate = 180] [color={rgb, 255:red, 0; green, 0; blue, 0 }  ][line width=0.75]    (10.93,-3.29) .. controls (6.95,-1.4) and (3.31,-0.3) .. (0,0) .. controls (3.31,0.3) and (6.95,1.4) .. (10.93,3.29)   ;
\draw [line width=1.5]    (303.65,83.15) .. controls (304.39,82.09) and (305.1,81.15) .. (305.8,80.33) .. controls (313.27,71.52) and (318.8,75.48) .. (323.65,83.15) ;
\draw    (312.15,82.6) -- (319.52,70.64) ;
\draw   (316.2,72.46) -- (319.62,70.44) -- (318.67,75.68) ;

\draw    (149,104.33) -- (171.62,104.58) ;
\draw [shift={(171.62,104.58)}, rotate = 180] [color={rgb, 255:red, 0; green, 0; blue, 0 }  ][line width=0.75]    (10.93,-3.29) .. controls (6.95,-1.4) and (3.31,-0.3) .. (0,0) .. controls (3.31,0.3) and (6.95,1.4) .. (10.93,3.29)   ;
\draw    (285,103.33) -- (303.36,103.58) ;
\draw [shift={(303.36,103.58)}, rotate = 180] [color={rgb, 255:red, 0; green, 0; blue, 0 }  ][line width=0.75]    (10.93,-3.29) .. controls (6.95,-1.4) and (3.31,-0.3) .. (0,0) .. controls (3.31,0.3) and (6.95,1.4) .. (10.93,3.29)   ;
\draw [line width=1.5]    (303.65,109.15) .. controls (304.39,108.09) and (305.1,107.15) .. (305.8,106.33) .. controls (313.27,97.52) and (318.8,101.48) .. (323.65,109.15) ;
\draw    (312.15,108.6) -- (319.52,96.64) ;
\draw   (316.2,98.46) -- (319.62,96.44) -- (318.67,101.68) ;

\draw    (149,131.33) -- (171.62,131.58) ;
\draw [shift={(171.62,131.58)}, rotate = 180] [color={rgb, 255:red, 0; green, 0; blue, 0 }  ][line width=0.75]    (10.93,-3.29) .. controls (6.95,-1.4) and (3.31,-0.3) .. (0,0) .. controls (3.31,0.3) and (6.95,1.4) .. (10.93,3.29)   ;
\draw    (285,130.33) -- (303.36,130.58) ;
\draw [shift={(303.36,130.58)}, rotate = 180] [color={rgb, 255:red, 0; green, 0; blue, 0 }  ][line width=0.75]    (10.93,-3.29) .. controls (6.95,-1.4) and (3.31,-0.3) .. (0,0) .. controls (3.31,0.3) and (6.95,1.4) .. (10.93,3.29)   ;
\draw [line width=1.5]    (303.65,136.15) .. controls (304.39,135.09) and (305.1,134.15) .. (305.8,133.33) .. controls (313.27,124.52) and (318.8,128.48) .. (323.65,136.15) ;
\draw    (312.15,135.6) -- (319.52,123.64) ;
\draw   (316.2,125.46) -- (319.62,123.44) -- (318.67,128.68) ;

\draw    (150,158.33) -- (172.62,158.58) ;
\draw [shift={(172.62,158.58)}, rotate = 180] [color={rgb, 255:red, 0; green, 0; blue, 0 }  ][line width=0.75]    (10.93,-3.29) .. controls (6.95,-1.4) and (3.31,-0.3) .. (0,0) .. controls (3.31,0.3) and (6.95,1.4) .. (10.93,3.29)   ;
\draw    (285.81,157.33) -- (304.17,157.58) ;
\draw [shift={(304.17,157.58)}, rotate = 180] [color={rgb, 255:red, 0; green, 0; blue, 0 }  ][line width=0.75]    (10.93,-3.29) .. controls (6.95,-1.4) and (3.31,-0.3) .. (0,0) .. controls (3.31,0.3) and (6.95,1.4) .. (10.93,3.29)   ;
\draw [line width=1.5]    (304.65,163.15) .. controls (305.39,162.09) and (306.1,161.15) .. (306.8,160.33) .. controls (314.27,151.52) and (319.8,155.48) .. (324.65,163.15) ;
\draw    (313.15,162.6) -- (320.52,150.64) ;
\draw   (317.2,152.46) -- (320.62,150.44) -- (319.67,155.68) ;

\draw (23,46) node [anchor=north west][inner sep=0.75pt]   [align=left] {unknown};
\draw (211.25,72.02) node [anchor=north west][inner sep=0.75pt]   [align=left] {\begin{minipage}[lt]{11.22pt}\setlength\topsep{0pt}
\begin{center}
$\Mv$
\end{center}

\end{minipage}};
\draw (112,72) node [anchor=north west][inner sep=0.75pt]   [align=left] {\begin{minipage}[lt]{27.68pt}\setlength\topsep{0pt}
\begin{center}
$\Mv \vb_1$
\end{center}

\end{minipage}};
\draw (211.25,99.02) node [anchor=north west][inner sep=0.75pt]   [align=left] {\begin{minipage}[lt]{11.22pt}\setlength\topsep{0pt}
\begin{center}
$\Mv$
\end{center}

\end{minipage}};
\draw (112,99) node [anchor=north west][inner sep=0.75pt]   [align=left] {\begin{minipage}[lt]{27.68pt}\setlength\topsep{0pt}
\begin{center}
$\Mv \vb_2$
\end{center}

\end{minipage}};
\draw (211.25,126.02) node [anchor=north west][inner sep=0.75pt]   [align=left] {\begin{minipage}[lt]{11.22pt}\setlength\topsep{0pt}
\begin{center}
$\Mv$
\end{center}

\end{minipage}};
\draw (112,126) node [anchor=north west][inner sep=0.75pt]   [align=left] {\begin{minipage}[lt]{27.68pt}\setlength\topsep{0pt}
\begin{center}
$\Mv \vb_3$
\end{center}

\end{minipage}};
\draw (211.25,153.02) node [anchor=north west][inner sep=0.75pt]   [align=left] {\begin{minipage}[lt]{11.22pt}\setlength\topsep{0pt}
\begin{center}
$\Mv$
\end{center}

\end{minipage}};
\draw (112,153) node [anchor=north west][inner sep=0.75pt]   [align=left] {\begin{minipage}[lt]{27.68pt}\setlength\topsep{0pt}
\begin{center}
$\Mv \vb_4$
\end{center}

\end{minipage}};
\draw (76.25,72.02) node [anchor=north west][inner sep=0.75pt]   [align=left] {\begin{minipage}[lt]{11.22pt}\setlength\topsep{0pt}
\begin{center}
$\Mv$
\end{center}

\end{minipage}};
\draw (27,72) node [anchor=north west][inner sep=0.75pt]   [align=left] {\begin{minipage}[lt]{19.18pt}\setlength\topsep{0pt}
\begin{center}
$\vb_1$
\end{center}

\end{minipage}};
\draw (76.25,99.02) node [anchor=north west][inner sep=0.75pt]   [align=left] {\begin{minipage}[lt]{11.22pt}\setlength\topsep{0pt}
\begin{center}
$\Mv$
\end{center}

\end{minipage}};
\draw (27,99) node [anchor=north west][inner sep=0.75pt]   [align=left] {\begin{minipage}[lt]{19.18pt}\setlength\topsep{0pt}
\begin{center}
$\vb_2$
\end{center}

\end{minipage}};
\draw (76.25,126.02) node [anchor=north west][inner sep=0.75pt]   [align=left] {\begin{minipage}[lt]{11.22pt}\setlength\topsep{0pt}
\begin{center}
$\Mv$
\end{center}

\end{minipage}};
\draw (27,126) node [anchor=north west][inner sep=0.75pt]   [align=left] {\begin{minipage}[lt]{19.18pt}\setlength\topsep{0pt}
\begin{center}
$\vb_3$
\end{center}

\end{minipage}};
\draw (76.25,153.02) node [anchor=north west][inner sep=0.75pt]   [align=left] {\begin{minipage}[lt]{11.22pt}\setlength\topsep{0pt}
\begin{center}
$\Mv$
\end{center}

\end{minipage}};
\draw (27,153) node [anchor=north west][inner sep=0.75pt]   [align=left] {\begin{minipage}[lt]{19.18pt}\setlength\topsep{0pt}
\begin{center}
$\vb_4$
\end{center}

\end{minipage}};
\draw (242,72) node [anchor=north west][inner sep=0.75pt]   [align=left] {\begin{minipage}[lt]{36.19pt}\setlength\topsep{0pt}
\begin{center}
$\Mv^2 \vb_1$
\end{center}

\end{minipage}};
\draw (242,99) node [anchor=north west][inner sep=0.75pt]   [align=left] {\begin{minipage}[lt]{36.19pt}\setlength\topsep{0pt}
\begin{center}
$\Mv^2 \vb_2$
\end{center}

\end{minipage}};
\draw (242,126) node [anchor=north west][inner sep=0.75pt]   [align=left] {\begin{minipage}[lt]{36.19pt}\setlength\topsep{0pt}
\begin{center}
$\Mv^2 \vb_3$
\end{center}

\end{minipage}};
\draw (242,153) node [anchor=north west][inner sep=0.75pt]   [align=left] {\begin{minipage}[lt]{36.19pt}\setlength\topsep{0pt}
\begin{center}
$\Mv^2 \vb_4$
\end{center}

\end{minipage}};

\end{tikzpicture}

\caption{Semi-blind setup with $n_{qb} = 2, n_i = 4, n_s = 2$}
\label{fig:2qbSB}
\end{figure}
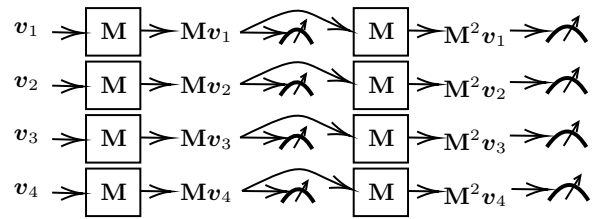

In the current section, we aim to compare our algorithm with that of \cite{nearunitary} in the case when there are two qubits and four initial states ($n_{qb} = 2, n_i = 4$). It is not a trivial task as the algorithm of \cite{nearunitary} has been designed for the SQPT setup of Fig. \ref{fig:2qbB} (in the case $n_{qb} = 2, n_i = 4$). Our algorithm was defined for the semi-blind setup of Fig. \ref{fig:2qbSB}. We can adapt it to the SQPT setup of Fig. \ref{fig:2qbB} (see Section \ref{section:SQPT}) but we would lose our resiliency to systematic errors, which was one of our goals. We chose to compare the following three algorithms:
\begin{enumerate}[leftmargin=*]
    \item Our algorithm running on the setup of Fig. \ref{fig:2qbSB} with $n_c = 1000$.
    \item The algorithm of \cite{nearunitary} running on the setup of Fig. \ref{fig:2qbB}, we chose to set $n_c = 2000$ in order to account for the fact that Fig. \ref{fig:2qbB} requires four states to be measured instead of eight.
    \item Our algorithm running on the setup of Fig. \ref{fig:2qbB} with the adaptations of Section \ref{section:SQPT} and with $n_c = 2000$.
\end{enumerate}
For those 3 options, the initial states are the states defined in Equation (20) of \cite{nearunitary} or in (\ref{eqn:v_init_baldwin}) in the current paper. We introduce a systematic error modelled in the same way as the systematic error in the previous section. For the sake of simplicity we use the QST algorithm of Appendix \ref{section:DataModel} for all QPT algorithms.

Let us first compare the left-hand side plot with the other two in Fig. \ref{fig:eps3}. When the systematic error increases, the error is smaller in the left-hand side plot. This is unsurprising, since our algorithm for the semi-blind setup (whose performance is represented on the left-hand side plot) has been designed to be resilient to the systematic errors, it does not use the values of the initial state. In contrast the other two algorithms assume that the values of the initial states are exactly known. It also makes sense that for lower systematic errors, the non-blind algorithms running on the SQPT setup (they are represented on the middle and right-hand side of Fig. \ref{fig:eps3}) perform better, because they use the values of the initial states rather than wasting half the measurements to get a noisy estimate of two sets of states before and after the gate has been applied.

\begin{figure}[H]

\tikzset{every picture/.style={line width=0.75pt}} 

\tikzset{every picture/.style={line width=0.75pt}} 

\begin{tikzpicture}[x=0.75pt,y=0.75pt,yscale=-1,xscale=1]

\draw    (418,80.33) -- (440.62,80.58) ;
\draw [shift={(440.62,80.58)}, rotate = 180] [color={rgb, 255:red, 0; green, 0; blue, 0 }  ][line width=0.75]    (10.93,-3.29) .. controls (6.95,-1.4) and (3.31,-0.3) .. (0,0) .. controls (3.31,0.3) and (6.95,1.4) .. (10.93,3.29)   ;
\draw    (367.65,81.15) -- (388.62,81.58) ;
\draw [shift={(388.62,81.58)}, rotate = 180] [color={rgb, 255:red, 0; green, 0; blue, 0 }  ][line width=0.75]    (10.93,-3.29) .. controls (6.95,-1.4) and (3.31,-0.3) .. (0,0) .. controls (3.31,0.3) and (6.95,1.4) .. (10.93,3.29)   ;
\draw   (390.25,69.02) -- (417.65,69.02) -- (417.65,93) -- (390.25,93) -- cycle ;
\draw    (418,107.33) -- (440.62,107.58) ;
\draw [shift={(440.62,107.58)}, rotate = 180] [color={rgb, 255:red, 0; green, 0; blue, 0 }  ][line width=0.75]    (10.93,-3.29) .. controls (6.95,-1.4) and (3.31,-0.3) .. (0,0) .. controls (3.31,0.3) and (6.95,1.4) .. (10.93,3.29)   ;
\draw    (367.65,108.15) -- (388.62,108.58) ;
\draw [shift={(388.62,108.58)}, rotate = 180] [color={rgb, 255:red, 0; green, 0; blue, 0 }  ][line width=0.75]    (10.93,-3.29) .. controls (6.95,-1.4) and (3.31,-0.3) .. (0,0) .. controls (3.31,0.3) and (6.95,1.4) .. (10.93,3.29)   ;
\draw   (390.25,96.02) -- (417.65,96.02) -- (417.65,120) -- (390.25,120) -- cycle ;
\draw    (418,134.33) -- (440.62,134.58) ;
\draw [shift={(440.62,134.58)}, rotate = 180] [color={rgb, 255:red, 0; green, 0; blue, 0 }  ][line width=0.75]    (10.93,-3.29) .. controls (6.95,-1.4) and (3.31,-0.3) .. (0,0) .. controls (3.31,0.3) and (6.95,1.4) .. (10.93,3.29)   ;
\draw    (367.65,135.15) -- (388.62,135.58) ;
\draw [shift={(388.62,135.58)}, rotate = 180] [color={rgb, 255:red, 0; green, 0; blue, 0 }  ][line width=0.75]    (10.93,-3.29) .. controls (6.95,-1.4) and (3.31,-0.3) .. (0,0) .. controls (3.31,0.3) and (6.95,1.4) .. (10.93,3.29)   ;
\draw   (390.25,123.02) -- (417.65,123.02) -- (417.65,147) -- (390.25,147) -- cycle ;
\draw    (418,161.33) -- (440.62,161.58) ;
\draw [shift={(440.62,161.58)}, rotate = 180] [color={rgb, 255:red, 0; green, 0; blue, 0 }  ][line width=0.75]    (10.93,-3.29) .. controls (6.95,-1.4) and (3.31,-0.3) .. (0,0) .. controls (3.31,0.3) and (6.95,1.4) .. (10.93,3.29)   ;
\draw    (367.65,162.15) -- (388.62,162.58) ;
\draw [shift={(388.62,162.58)}, rotate = 180] [color={rgb, 255:red, 0; green, 0; blue, 0 }  ][line width=0.75]    (10.93,-3.29) .. controls (6.95,-1.4) and (3.31,-0.3) .. (0,0) .. controls (3.31,0.3) and (6.95,1.4) .. (10.93,3.29)   ;
\draw   (390.25,150.02) -- (417.65,150.02) -- (417.65,174) -- (390.25,174) -- cycle ;
\draw    (482,80.33) -- (504.62,80.58) ;
\draw [shift={(504.62,80.58)}, rotate = 180] [color={rgb, 255:red, 0; green, 0; blue, 0 }  ][line width=0.75]    (10.93,-3.29) .. controls (6.95,-1.4) and (3.31,-0.3) .. (0,0) .. controls (3.31,0.3) and (6.95,1.4) .. (10.93,3.29)   ;
\draw    (482,107.33) -- (504.62,107.58) ;
\draw [shift={(504.62,107.58)}, rotate = 180] [color={rgb, 255:red, 0; green, 0; blue, 0 }  ][line width=0.75]    (10.93,-3.29) .. controls (6.95,-1.4) and (3.31,-0.3) .. (0,0) .. controls (3.31,0.3) and (6.95,1.4) .. (10.93,3.29)   ;
\draw    (482,134.33) -- (504.62,134.58) ;
\draw [shift={(504.62,134.58)}, rotate = 180] [color={rgb, 255:red, 0; green, 0; blue, 0 }  ][line width=0.75]    (10.93,-3.29) .. controls (6.95,-1.4) and (3.31,-0.3) .. (0,0) .. controls (3.31,0.3) and (6.95,1.4) .. (10.93,3.29)   ;
\draw    (482,161.33) -- (504.62,161.58) ;
\draw [shift={(504.62,161.58)}, rotate = 180] [color={rgb, 255:red, 0; green, 0; blue, 0 }  ][line width=0.75]    (10.93,-3.29) .. controls (6.95,-1.4) and (3.31,-0.3) .. (0,0) .. controls (3.31,0.3) and (6.95,1.4) .. (10.93,3.29)   ;
\draw [line width=1.5]    (505.65,86.15) .. controls (506.39,85.09) and (507.1,84.15) .. (507.8,83.33) .. controls (515.27,74.52) and (520.8,78.48) .. (525.65,86.15) ;
\draw    (514.15,85.6) -- (521.52,73.64) ;
\draw   (518.2,75.46) -- (521.62,73.44) -- (520.67,78.68) ;

\draw [line width=1.5]    (505.65,114.15) .. controls (506.39,113.09) and (507.1,112.15) .. (507.8,111.33) .. controls (515.27,102.52) and (520.8,106.48) .. (525.65,114.15) ;
\draw    (514.15,113.6) -- (521.52,101.64) ;
\draw   (518.2,103.46) -- (521.62,101.44) -- (520.67,106.68) ;

\draw [line width=1.5]    (505.65,139.15) .. controls (506.39,138.09) and (507.1,137.15) .. (507.8,136.33) .. controls (515.27,127.52) and (520.8,131.48) .. (525.65,139.15) ;
\draw    (514.15,138.6) -- (521.52,126.64) ;
\draw   (518.2,128.46) -- (521.62,126.44) -- (520.67,131.68) ;

\draw [line width=1.5]    (505.65,166.15) .. controls (506.39,165.09) and (507.1,164.15) .. (507.8,163.33) .. controls (515.27,154.52) and (520.8,158.48) .. (525.65,166.15) ;
\draw    (514.15,165.6) -- (521.52,153.64) ;
\draw   (518.2,155.46) -- (521.62,153.44) -- (520.67,158.68) ;

\draw (441,75) node [anchor=north west][inner sep=0.75pt]   [align=left] {\begin{minipage}[lt]{30.51pt}\setlength\topsep{0pt}
\begin{center}
$\Mv \vb_1$
\end{center}

\end{minipage}};
\draw (395.25,75.02) node [anchor=north west][inner sep=0.75pt]   [align=left] {\begin{minipage}[lt]{11.22pt}\setlength\topsep{0pt}
\begin{center}
$\Mv$
\end{center}

\end{minipage}};
\draw (340,75) node [anchor=north west][inner sep=0.75pt]   [align=left] {\begin{minipage}[lt]{19.18pt}\setlength\topsep{0pt}
\begin{center}
$\vb_1$
\end{center}

\end{minipage}};
\draw (441,102) node [anchor=north west][inner sep=0.75pt]   [align=left] {\begin{minipage}[lt]{30.51pt}\setlength\topsep{0pt}
\begin{center}
$\Mv \vb_2$
\end{center}

\end{minipage}};
\draw (395.25,102.02) node [anchor=north west][inner sep=0.75pt]   [align=left] {\begin{minipage}[lt]{11.22pt}\setlength\topsep{0pt}
\begin{center}
$\Mv$
\end{center}

\end{minipage}};
\draw (340,102) node [anchor=north west][inner sep=0.75pt]   [align=left] {\begin{minipage}[lt]{19.18pt}\setlength\topsep{0pt}
\begin{center}
$\vb_2$
\end{center}

\end{minipage}};
\draw (441,129) node [anchor=north west][inner sep=0.75pt]   [align=left] {\begin{minipage}[lt]{30.51pt}\setlength\topsep{0pt}
\begin{center}
$\Mv \vb_3$
\end{center}

\end{minipage}};
\draw (395.25,129.02) node [anchor=north west][inner sep=0.75pt]   [align=left] {\begin{minipage}[lt]{11.22pt}\setlength\topsep{0pt}
\begin{center}
$\Mv$
\end{center}

\end{minipage}};
\draw (340,129) node [anchor=north west][inner sep=0.75pt]   [align=left] {\begin{minipage}[lt]{19.18pt}\setlength\topsep{0pt}
\begin{center}
$\vb_3$
\end{center}

\end{minipage}};
\draw (441,156) node [anchor=north west][inner sep=0.75pt]   [align=left] {\begin{minipage}[lt]{30.51pt}\setlength\topsep{0pt}
\begin{center}
$\Mv \vb_4$
\end{center}

\end{minipage}};
\draw (395.25,156.02) node [anchor=north west][inner sep=0.75pt]   [align=left] {\begin{minipage}[lt]{11.22pt}\setlength\topsep{0pt}
\begin{center}
$\Mv$
\end{center}

\end{minipage}};
\draw (340,156) node [anchor=north west][inner sep=0.75pt]   [align=left] {\begin{minipage}[lt]{19.18pt}\setlength\topsep{0pt}
\begin{center}
$\vb_4$
\end{center}

\end{minipage}};
\draw (332,45) node [anchor=north west][inner sep=0.75pt]   [align=left] {trusted};

\end{tikzpicture}

\caption{SQPT setup with $n_{qb} = 2$}
\label{fig:2qbB}
\end{figure}
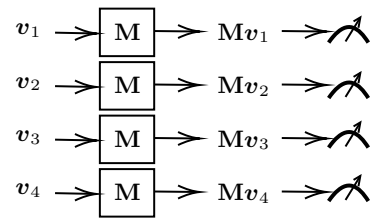

\begin{figure*}
    \centering
    \includegraphics[width=15cm,height=5cm]{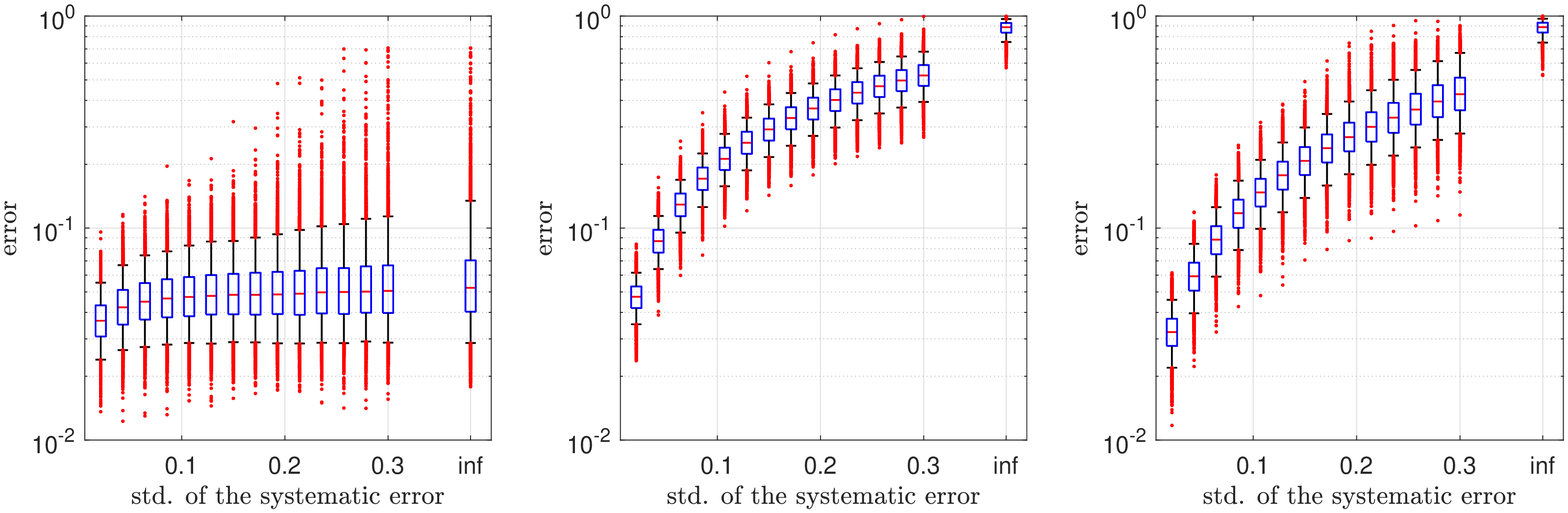}
    \caption{Box-plots of the QPT error in the presence of systematic errors with the initial states of \cite{nearunitary}. The first plot (left-hand side) represents the error with our semi-blind algorithm running on the semi-blind setup of Fig. \ref{fig:2qbSB}. The second plot (middle) represents the error with the algorithm proposed in \cite{nearunitary} on the standard setup considered in \cite{nearunitary} (represented in Fig. \ref{fig:2qbB} for two qubits). The third plot (right-hand side) represents the performance of our algorithm adapted to run on the SQPT setup of Fig. \ref{fig:2qbB}}
	\label{fig:eps3}
\vspace{-1mm}
\end{figure*}

The middle plot of Fig. \ref{fig:eps3} (algorithm of \cite{nearunitary} applied to the SQPT setup of Fig. \ref{fig:2qbB}) has higher errors than the right-hand plot for all values (except for the ``infinite std", where they are the same). This means that our algorithm adapted for the SQPT setup yields a better estimate than that of \cite{nearunitary}. This is because the latter is very simple and elegant, but it was not designed to mitigate the effect of the errors that we model here (systematic and multinomial error). It directly estimates the coefficients of the unitary matrix from the QST estimates of the states. The resulting matrix has no reason to be unitary if there are errors. In contrast, our algorithm finds the unitary matrix that fits our QST estimate best (in the least square sense). The algorithm of \cite{nearunitary} is simpler and faster but as we will see in Section \ref{section:multiQB}, our QPT algorithm is really quick, at least compared to the QST algorithm we use (it is so quick in fact, that, its execution time is negligible compared to the QST).
 
Overall, the performances of our algorithm on the semi-blind setup are very satisfying. It yields lower errors than the algorithm of \cite{nearunitary} (resp. than our algorithm with the adaptations of Section \ref{section:SQPT}) when the standard deviation of the Gaussian systematic error is roughly $0.007$ (resp. $0.025$) on each component of the initial states. Those values ($0.007$ and $0.025$) are obtained by linearly interpolating the median of the box-plots in the 3 graphs and computing the values of systematic errors for which the interpolated lines of the median error cross.

\subsection{Other setups}
\label{section:multiQB}
Now that we studied the impacts of the different types of errors, we want to see how well our QPT algorithm works with more than two qubits and with the setups of Fig. \ref{fig:deux_td} and Fig. \ref{fig:bcp_td}.

For $1$ to $6$ qubits, we simulate $500$ random $d \times d$  unitary matrices by applying the Gram Schmidt process to random Gaussian matrices (like in the previous subsections). And we try to identify the processes associated with each matrix with the following two setups: 
\begin{enumerate}
    \item With the $d$ initial states of (\ref{eqn:V_INIT}) and two time delays ($n_i = d$, $n_s = 2$, displayed on Fig. \ref{fig:deux_td}). We simulate $n_c = 2500$ measurements per measurement type and per measured state. We consider that the states are prepared with the setup of Fig. \ref{fig:deux_td}. The Hadamard gates used for the initial state preparation are considered to be imperfect and are represented by $\begin{pmatrix} \cos(\theta_r) & -\sin(\theta_r) e^{i\phi_r}\\ \sin(\theta_r) & \cos(\theta_r) e^{i\phi_r} \end{pmatrix}\Hv_d$ instead of $\Hv_d$, where $\theta_r$ and $\phi_r$ are two random i.i.d. Gaussian centered angles with a standard deviation of $0.05$ radians (two different values are sampled for each gate of Fig. \ref{fig:deux_td} and for each one of the $500$ gates to be identified).
    \item With a single random initial state and $d+1$ time delays ($n_i = 1$, $n_s = d+1$, like the setup of Fig. \ref{fig:bcp_td} but with a random initial state). We use $n_c = [2500 \frac{2d}{d+1}]$ ($[.]$ refers to the closest integer), so that the total number of measurements $n_c n_i n_s n_t$ is the same (or almost the same if $2500 \frac{2d}{d+1}$ is not an integer) in both setups for a given number of qubits. 
\end{enumerate}
As explained in Section \ref{section:recommendations}, the first setup is more robust and the second can be easier to prepare and allows for higher $n_c$ (since $\frac{2d}{d+1}>1$).

The chosen $n_c$ is really high for small values of $n_{qb}$. If $n_{qb}=2$, for example, there are only $d=4$ outcomes whose probabilities we want to estimate. Then, having $n_c = 2500$ or $n_c = \left[2500 \frac{2\times 4}{4+1}\right] = 4000$ is more than enough to have very good estimates. But if $n_{qb} = 6$, there are $d = 64$ outcomes, and the associated probabilities are much smaller (since they sum to one), then the number of measurements does not seem that excessive.  

\begin{figure}[h]
    \centering
    \includegraphics[width=7.5cm,height=6cm]{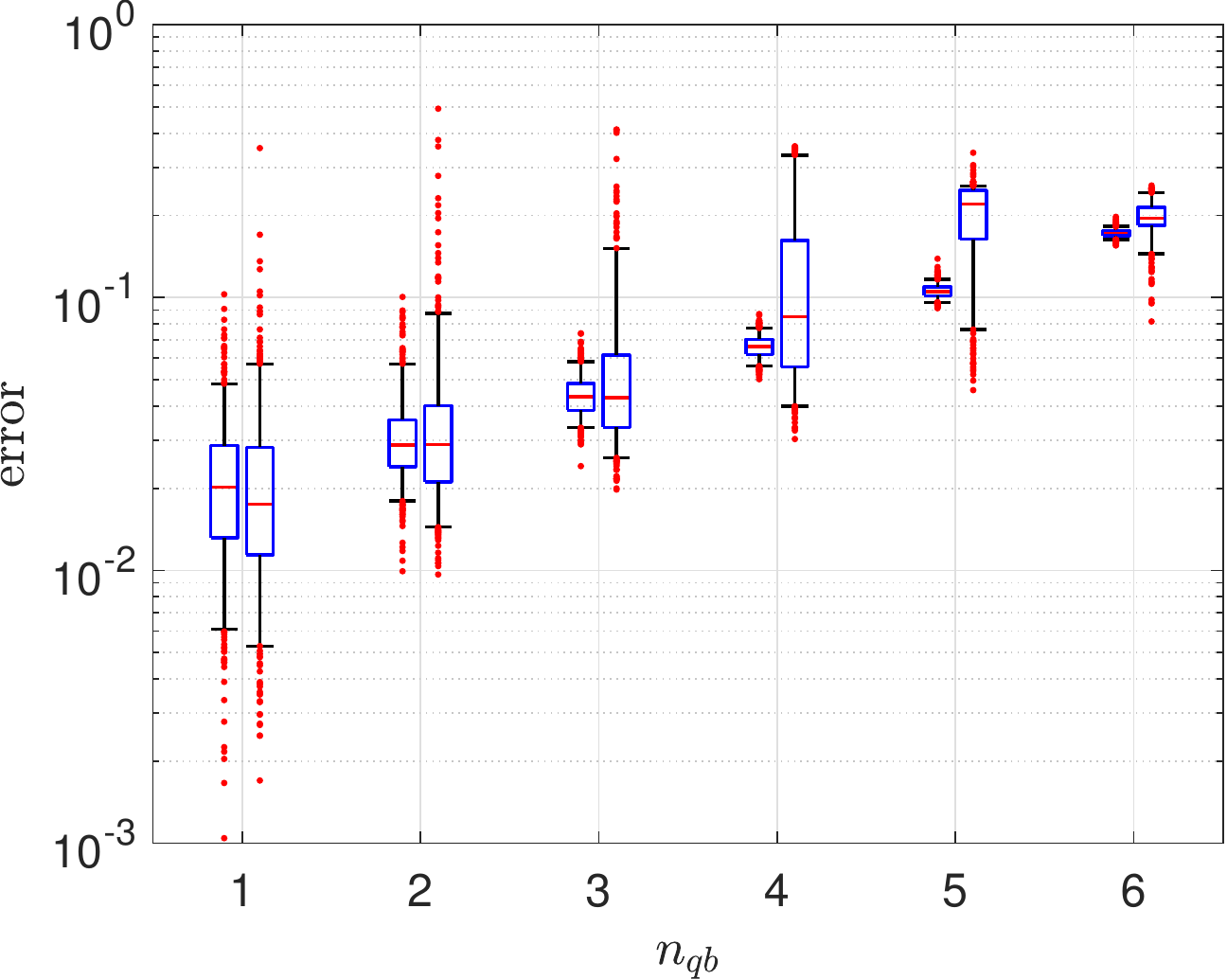}
    \caption{Box-plots of the QPT error on quantum gates acting on $1$ to $6$ qubits. There are two box-plots for each number of qubits. The one on the left represents the errors with the first setup ($n_i=d$, $n_s=2$) and the one on the right represents the errors with the second setup ($n_i=1$, $n_s=d+1$). Both setups have the same total number of measurements}
	\label{fig:eps4}
\vspace{-1mm}
\end{figure}

Fig. \ref{fig:eps4} shows the box-plots of the error for both setups and for all numbers of qubits. Unsurprisingly the error increases with $n_{qb}$. This is the case despite the fact that the total number of measurements $n_c n_i n_s n_t$ also increases with $n_{qb}$ ($n_c$ is constant but $n_i$ increases for the first setup (for which $n_i = d$), $n_s$ increases for the second setup (for which $n_s = d+1$) and $n_t = 2 n_{qb}+1$ increases for both setups). This is not surprising because the number of parameters of the unitary matrix we estimate is $d^2 = 2^{2n_{qb}}$, it increases exponentially with $n_{qb}$. 

For the first setup, the range of the error decreases (in log-scale) when the number of qubits increases. We think this is because, when the number of parameters to be estimated (in $\Mv$) increases, the error on the matrix becomes more predictable as the errors on the parameters get ``averaged" by the norm in (\ref{eqn:ourerr}). We do not see this phenomenon with the second setup because, for this setup, we rely on $\Mv$ to create a $\Xv$ that satisfies (\ref{eqn:CNS}) with a good margin, and the fact that we choose $500$ different random values of $\Mv$ for each number of qubits makes the error vary much more. In contrast, the $\Xv$ of the second setup always satisfies (\ref{eqn:CNS}) and does not depend on $\Mv$. 

The relative performances of the two setups are interesting. For one to four qubits, with the second setup, the largest errors are larger and the smallest errors are smaller than with the first setup. This is in line with what we expected, the first setup is supposed to be more reliable, it makes sense that its greatest outliers are smaller. The second setup uses the measurements more efficiently (see end of Section \ref{section:recommendations}), it makes sense that, when we sampled a random $\Mv$ that makes the second setup work, it works better than the first setup. And if we have an idea of the value of $\Mv$ before performing the QPT, we will know in which part of the box-plot the error is likely to be. Hopefully, for our $\Mv$, (\ref{eqn:CNS}) will be satisfied with a comfortable margin, and we will know that the error is in the lower part of the box-plot. The more qubits there are the worse the second setup gets compared to the first. We think this is because, for a random matrix (and $\Xv$ is basically a random matrix for the second setup but is deterministic for the first if the systematic errors are neglected) (\ref{eqn:CNS}) becomes closer to being false  when dimension increases. 

The average execution times of the QST and QPT algorithms are reported in Table 1. The simulations were coded on Matlab \footnote{https://www.mathworks.com/products/matlab.html} on a 210 Intel Xeon silver 4214 2.4-GHz. We allowed the script to run on $10$ threads, but each simulation ran sequentially, Matlab only parallelized the linear algebra computation on large matrices.

\begin{table}[H]
\centering
\caption{Average execution time of the QST of all measured states and of the QPT of the process for the setup of Fig. \ref{fig:deux_td} (setup 1) and the setup of Fig. \ref{fig:bcp_td} (setup 2).}
\setlength{\tabcolsep}{1pt}
 \begin{tabular}{| >{\centering}p{2cm} | >{\centering}p{0.9cm} | >{\centering}p{0.9cm} | >{\centering}p{0.9cm} | >{\centering}p{0.9cm} | >{\centering}p{0.9cm} | p{0.9cm} |}
\hline
\backslashbox{algo.}{$n_{qb}$} &$1$&$2$&$3$& $4$&$5$& $\ \ \ \ 6$\\ \hline 
QST setup 1 & $0.06s$ &$0.14s $&$ 0.37s $& $1.67s$ & $12s$ & $188s$  \\ \hline
QPT setup 1 & $7$e-5$s$ &$9$e-5$s$&$1$e-4$s$& $3$e-4$s$ & $6$e-4$s$ & $2$e-3$s$   \\ \hline
QST setup 2&  $0.05s$ &$0.10s $&$ 0.21s $& $0.85s$ & $6.9s$ & $123s$ \\ \hline
QPT setup 2&  $1$e-4$s$ &$1$e-4$s$&$2$e-4$s$& $3$e-4$s$ & $8$e-4$s$ & $3$e-3$s$  \\ \hline

\end{tabular}
\end{table}

Clearly, the execution time of the QPT is not significant, it is the QST that takes the longest. There are faster QST algorithms in the literature, and we could shorten the QST time greatly by not implementing the fine tuning ML based approach (see \cite{PRA}). But we choose to sacrifice execution time for precision. In contrast, for the QPT, we only need to compute a few dot products for the phase recovery, and then a few products of $d\times d$ matrices, and then perform a singular value decomposition to solve the total least square problem under a unitarity constraint.
\section{Experimental results for a 2-qubit CNOT gate}
\label{section:exp}
In the current section, we test our QPT algorithm  experimentally using a trapped-ion quantum computer \footnote{arn:aws:braket:::device/qpu/ionq/ionQdevice} on Amazon Web Services (AWS). 
We want to perform QPT on a 2-qubit CNOT gate. As stated in Section \ref{section:recommendations}, the use of two time steps is adapted to this gate. Therefore we chose $n_i=d=4$, $n_s=2$ with the four initial states $\vb_1^{tg}$ to $\vb_4^{tg}$ of (\ref{eqn:v_init}) which are the same as (\ref{eqn:V_INIT}) and Fig. \ref{fig:deux_td}, with $n_{qb} = 2$. As explained in Section \ref{section:recommendations}, those initial states can be realized with Hadamard gates (see Fig. \ref{fig:deux_td}). In order not to rely on the implementation of the Hadamard gates, our algorithm will behave as if the input states were totally unknown. The target value of the process matrix of the CNOT gate to be identified is:

\begin{equation}
\label{eqn:targetM}
\Mv_{tg}=\begin{pmatrix} 1&0&0&0\\0&1&0&0\\0&0&0&1\\0&0&1&0\end{pmatrix}
\end{equation}

This value is not used by our QPT method either. The process to be identified is considered to be totally unknown. 
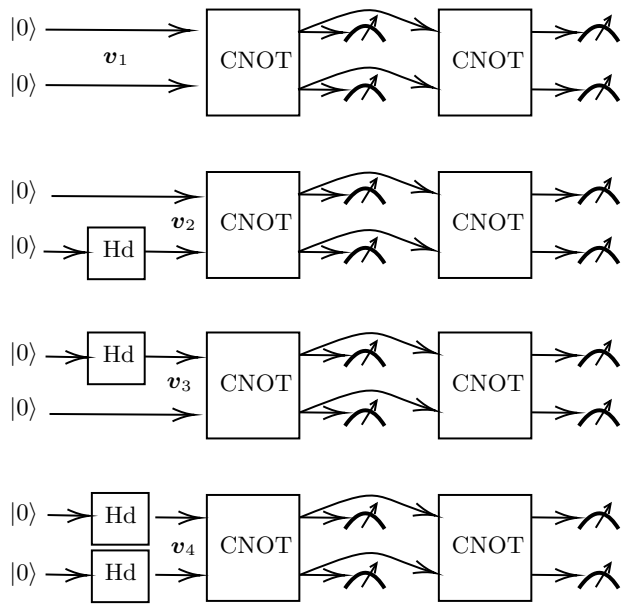
\begin{figure}[h]
\begin{centering}

\tikzset{every picture/.style={line width=0.75pt}} 

\begin{tikzpicture}[x=0.75pt,y=0.75pt,yscale=-1,xscale=1]

\draw   (93,286.02) -- (121.65,286.02) -- (121.65,312.12) -- (93,312.12) -- cycle ;
\draw   (151,287.02) -- (197.65,287.02) -- (197.65,340.5) -- (151,340.5) -- cycle ;
\draw   (93,315.02) -- (121.65,315.02) -- (121.65,341.12) -- (93,341.12) -- cycle ;
\draw    (125,327.33) -- (147.62,327.58) ;
\draw [shift={(147.62,327.58)}, rotate = 180] [color={rgb, 255:red, 0; green, 0; blue, 0 }  ][line width=0.75]    (10.93,-3.29) .. controls (6.95,-1.4) and (3.31,-0.3) .. (0,0) .. controls (3.31,0.3) and (6.95,1.4) .. (10.93,3.29)   ;
\draw    (125,298.33) -- (147.62,298.58) ;
\draw [shift={(147.62,298.58)}, rotate = 180] [color={rgb, 255:red, 0; green, 0; blue, 0 }  ][line width=0.75]    (10.93,-3.29) .. controls (6.95,-1.4) and (3.31,-0.3) .. (0,0) .. controls (3.31,0.3) and (6.95,1.4) .. (10.93,3.29)   ;
\draw    (198,327.33) -- (220.62,327.58) ;
\draw [shift={(220.62,327.58)}, rotate = 180] [color={rgb, 255:red, 0; green, 0; blue, 0 }  ][line width=0.75]    (10.93,-3.29) .. controls (6.95,-1.4) and (3.31,-0.3) .. (0,0) .. controls (3.31,0.3) and (6.95,1.4) .. (10.93,3.29)   ;
\draw    (198,298.33) -- (220.62,298.58) ;
\draw [shift={(220.62,298.58)}, rotate = 180] [color={rgb, 255:red, 0; green, 0; blue, 0 }  ][line width=0.75]    (10.93,-3.29) .. controls (6.95,-1.4) and (3.31,-0.3) .. (0,0) .. controls (3.31,0.3) and (6.95,1.4) .. (10.93,3.29)   ;
\draw   (268,287.02) -- (314.65,287.02) -- (314.65,340.5) -- (268,340.5) -- cycle ;
\draw    (315,327.33) -- (337.62,327.58) ;
\draw [shift={(337.62,327.58)}, rotate = 180] [color={rgb, 255:red, 0; green, 0; blue, 0 }  ][line width=0.75]    (10.93,-3.29) .. controls (6.95,-1.4) and (3.31,-0.3) .. (0,0) .. controls (3.31,0.3) and (6.95,1.4) .. (10.93,3.29)   ;
\draw    (315,298.33) -- (337.62,298.58) ;
\draw [shift={(337.62,298.58)}, rotate = 180] [color={rgb, 255:red, 0; green, 0; blue, 0 }  ][line width=0.75]    (10.93,-3.29) .. controls (6.95,-1.4) and (3.31,-0.3) .. (0,0) .. controls (3.31,0.3) and (6.95,1.4) .. (10.93,3.29)   ;
\draw   (151,205.02) -- (197.65,205.02) -- (197.65,258.5) -- (151,258.5) -- cycle ;
\draw   (91,205.02) -- (119.65,205.02) -- (119.65,231.12) -- (91,231.12) -- cycle ;
\draw    (198,245.33) -- (220.62,245.58) ;
\draw [shift={(220.62,245.58)}, rotate = 180] [color={rgb, 255:red, 0; green, 0; blue, 0 }  ][line width=0.75]    (10.93,-3.29) .. controls (6.95,-1.4) and (3.31,-0.3) .. (0,0) .. controls (3.31,0.3) and (6.95,1.4) .. (10.93,3.29)   ;
\draw    (198,216.33) -- (220.62,216.58) ;
\draw [shift={(220.62,216.58)}, rotate = 180] [color={rgb, 255:red, 0; green, 0; blue, 0 }  ][line width=0.75]    (10.93,-3.29) .. controls (6.95,-1.4) and (3.31,-0.3) .. (0,0) .. controls (3.31,0.3) and (6.95,1.4) .. (10.93,3.29)   ;
\draw   (268,205.02) -- (314.65,205.02) -- (314.65,258.5) -- (268,258.5) -- cycle ;
\draw    (315,245.33) -- (337.62,245.58) ;
\draw [shift={(337.62,245.58)}, rotate = 180] [color={rgb, 255:red, 0; green, 0; blue, 0 }  ][line width=0.75]    (10.93,-3.29) .. controls (6.95,-1.4) and (3.31,-0.3) .. (0,0) .. controls (3.31,0.3) and (6.95,1.4) .. (10.93,3.29)   ;
\draw    (315,216.33) -- (337.62,216.58) ;
\draw [shift={(337.62,216.58)}, rotate = 180] [color={rgb, 255:red, 0; green, 0; blue, 0 }  ][line width=0.75]    (10.93,-3.29) .. controls (6.95,-1.4) and (3.31,-0.3) .. (0,0) .. controls (3.31,0.3) and (6.95,1.4) .. (10.93,3.29)   ;
\draw   (91,152.02) -- (119.65,152.02) -- (119.65,178.12) -- (91,178.12) -- cycle ;
\draw   (151,124.02) -- (197.65,124.02) -- (197.65,177.5) -- (151,177.5) -- cycle ;
\draw    (68.65,164.15) -- (89.62,164.58) ;
\draw [shift={(89.62,164.58)}, rotate = 180] [color={rgb, 255:red, 0; green, 0; blue, 0 }  ][line width=0.75]    (10.93,-3.29) .. controls (6.95,-1.4) and (3.31,-0.3) .. (0,0) .. controls (3.31,0.3) and (6.95,1.4) .. (10.93,3.29)   ;
\draw    (119.65,164.15) -- (146.62,164.58) ;
\draw [shift={(146.62,164.58)}, rotate = 180] [color={rgb, 255:red, 0; green, 0; blue, 0 }  ][line width=0.75]    (10.93,-3.29) .. controls (6.95,-1.4) and (3.31,-0.3) .. (0,0) .. controls (3.31,0.3) and (6.95,1.4) .. (10.93,3.29)   ;
\draw    (197,164.33) -- (219.62,164.58) ;
\draw [shift={(219.62,164.58)}, rotate = 180] [color={rgb, 255:red, 0; green, 0; blue, 0 }  ][line width=0.75]    (10.93,-3.29) .. controls (6.95,-1.4) and (3.31,-0.3) .. (0,0) .. controls (3.31,0.3) and (6.95,1.4) .. (10.93,3.29)   ;
\draw    (197,135.33) -- (219.62,135.58) ;
\draw [shift={(219.62,135.58)}, rotate = 180] [color={rgb, 255:red, 0; green, 0; blue, 0 }  ][line width=0.75]    (10.93,-3.29) .. controls (6.95,-1.4) and (3.31,-0.3) .. (0,0) .. controls (3.31,0.3) and (6.95,1.4) .. (10.93,3.29)   ;
\draw   (268,124.02) -- (314.65,124.02) -- (314.65,177.5) -- (268,177.5) -- cycle ;
\draw [line width=1.5]    (338.65,170.15) .. controls (339.39,169.09) and (340.1,168.15) .. (340.8,167.33) .. controls (348.27,158.52) and (353.8,162.48) .. (358.65,170.15) ;
\draw    (347.15,169.6) -- (354.52,157.64) ;
\draw   (351.2,159.46) -- (354.62,157.44) -- (353.67,162.68) ;

\draw    (315,164.33) -- (337.62,164.58) ;
\draw [shift={(337.62,164.58)}, rotate = 180] [color={rgb, 255:red, 0; green, 0; blue, 0 }  ][line width=0.75]    (10.93,-3.29) .. controls (6.95,-1.4) and (3.31,-0.3) .. (0,0) .. controls (3.31,0.3) and (6.95,1.4) .. (10.93,3.29)   ;
\draw    (315,135.33) -- (337.62,135.58) ;
\draw [shift={(337.62,135.58)}, rotate = 180] [color={rgb, 255:red, 0; green, 0; blue, 0 }  ][line width=0.75]    (10.93,-3.29) .. controls (6.95,-1.4) and (3.31,-0.3) .. (0,0) .. controls (3.31,0.3) and (6.95,1.4) .. (10.93,3.29)   ;
\draw   (151,42.02) -- (197.65,42.02) -- (197.65,95.5) -- (151,95.5) -- cycle ;
\draw    (69.65,80.15) -- (142.62,80.58) ;
\draw [shift={(142.62,80.58)}, rotate = 180] [color={rgb, 255:red, 0; green, 0; blue, 0 }  ][line width=0.75]    (10.93,-3.29) .. controls (6.95,-1.4) and (3.31,-0.3) .. (0,0) .. controls (3.31,0.3) and (6.95,1.4) .. (10.93,3.29)   ;
\draw    (197,82.33) -- (219.62,82.58) ;
\draw [shift={(219.62,82.58)}, rotate = 180] [color={rgb, 255:red, 0; green, 0; blue, 0 }  ][line width=0.75]    (10.93,-3.29) .. controls (6.95,-1.4) and (3.31,-0.3) .. (0,0) .. controls (3.31,0.3) and (6.95,1.4) .. (10.93,3.29)   ;
\draw    (197,53.33) -- (219.62,53.58) ;
\draw [shift={(219.62,53.58)}, rotate = 180] [color={rgb, 255:red, 0; green, 0; blue, 0 }  ][line width=0.75]    (10.93,-3.29) .. controls (6.95,-1.4) and (3.31,-0.3) .. (0,0) .. controls (3.31,0.3) and (6.95,1.4) .. (10.93,3.29)   ;
\draw   (268,42.02) -- (314.65,42.02) -- (314.65,95.5) -- (268,95.5) -- cycle ;
\draw    (315,82.33) -- (337.62,82.58) ;
\draw [shift={(337.62,82.58)}, rotate = 180] [color={rgb, 255:red, 0; green, 0; blue, 0 }  ][line width=0.75]    (10.93,-3.29) .. controls (6.95,-1.4) and (3.31,-0.3) .. (0,0) .. controls (3.31,0.3) and (6.95,1.4) .. (10.93,3.29)   ;
\draw    (315,53.33) -- (337.62,53.58) ;
\draw [shift={(337.62,53.58)}, rotate = 180] [color={rgb, 255:red, 0; green, 0; blue, 0 }  ][line width=0.75]    (10.93,-3.29) .. controls (6.95,-1.4) and (3.31,-0.3) .. (0,0) .. controls (3.31,0.3) and (6.95,1.4) .. (10.93,3.29)   ;
\draw    (69.65,52.15) -- (142.62,52.58) ;
\draw [shift={(142.62,52.58)}, rotate = 180] [color={rgb, 255:red, 0; green, 0; blue, 0 }  ][line width=0.75]    (10.93,-3.29) .. controls (6.95,-1.4) and (3.31,-0.3) .. (0,0) .. controls (3.31,0.3) and (6.95,1.4) .. (10.93,3.29)   ;
\draw    (72.65,136.15) -- (145.62,136.58) ;
\draw [shift={(145.62,136.58)}, rotate = 180] [color={rgb, 255:red, 0; green, 0; blue, 0 }  ][line width=0.75]    (10.93,-3.29) .. controls (6.95,-1.4) and (3.31,-0.3) .. (0,0) .. controls (3.31,0.3) and (6.95,1.4) .. (10.93,3.29)   ;
\draw    (71.65,246.15) -- (144.62,246.58) ;
\draw [shift={(144.62,246.58)}, rotate = 180] [color={rgb, 255:red, 0; green, 0; blue, 0 }  ][line width=0.75]    (10.93,-3.29) .. controls (6.95,-1.4) and (3.31,-0.3) .. (0,0) .. controls (3.31,0.3) and (6.95,1.4) .. (10.93,3.29)   ;
\draw    (119.65,217.15) -- (146.62,217.58) ;
\draw [shift={(146.62,217.58)}, rotate = 180] [color={rgb, 255:red, 0; green, 0; blue, 0 }  ][line width=0.75]    (10.93,-3.29) .. controls (6.95,-1.4) and (3.31,-0.3) .. (0,0) .. controls (3.31,0.3) and (6.95,1.4) .. (10.93,3.29)   ;
\draw    (69.65,217.15) -- (90.62,217.58) ;
\draw [shift={(90.62,217.58)}, rotate = 180] [color={rgb, 255:red, 0; green, 0; blue, 0 }  ][line width=0.75]    (10.93,-3.29) .. controls (6.95,-1.4) and (3.31,-0.3) .. (0,0) .. controls (3.31,0.3) and (6.95,1.4) .. (10.93,3.29)   ;
\draw    (70.65,297.15) -- (91.62,297.58) ;
\draw [shift={(91.62,297.58)}, rotate = 180] [color={rgb, 255:red, 0; green, 0; blue, 0 }  ][line width=0.75]    (10.93,-3.29) .. controls (6.95,-1.4) and (3.31,-0.3) .. (0,0) .. controls (3.31,0.3) and (6.95,1.4) .. (10.93,3.29)   ;
\draw    (69.65,327.15) -- (90.62,327.58) ;
\draw [shift={(90.62,327.58)}, rotate = 180] [color={rgb, 255:red, 0; green, 0; blue, 0 }  ][line width=0.75]    (10.93,-3.29) .. controls (6.95,-1.4) and (3.31,-0.3) .. (0,0) .. controls (3.31,0.3) and (6.95,1.4) .. (10.93,3.29)   ;
\draw [line width=1.5]    (338.65,140.15) .. controls (339.39,139.09) and (340.1,138.15) .. (340.8,137.33) .. controls (348.27,128.52) and (353.8,132.48) .. (358.65,140.15) ;
\draw    (347.15,139.6) -- (354.52,127.64) ;
\draw   (351.2,129.46) -- (354.62,127.44) -- (353.67,132.68) ;

\draw [line width=1.5]    (338.65,88.15) .. controls (339.39,87.09) and (340.1,86.15) .. (340.8,85.33) .. controls (348.27,76.52) and (353.8,80.48) .. (358.65,88.15) ;
\draw    (347.15,87.6) -- (354.52,75.64) ;
\draw   (351.2,77.46) -- (354.62,75.44) -- (353.67,80.68) ;

\draw [line width=1.5]    (338.65,58.15) .. controls (339.39,57.09) and (340.1,56.15) .. (340.8,55.33) .. controls (348.27,46.52) and (353.8,50.48) .. (358.65,58.15) ;
\draw    (347.15,57.6) -- (354.52,45.64) ;
\draw   (351.2,47.46) -- (354.62,45.44) -- (353.67,50.68) ;

\draw [line width=1.5]    (338.65,334.15) .. controls (339.39,333.09) and (340.1,332.15) .. (340.8,331.33) .. controls (348.27,322.52) and (353.8,326.48) .. (358.65,334.15) ;
\draw    (347.15,333.6) -- (354.52,321.64) ;
\draw   (351.2,323.46) -- (354.62,321.44) -- (353.67,326.68) ;

\draw [line width=1.5]    (338.65,304.15) .. controls (339.39,303.09) and (340.1,302.15) .. (340.8,301.33) .. controls (348.27,292.52) and (353.8,296.48) .. (358.65,304.15) ;
\draw    (347.15,303.6) -- (354.52,291.64) ;
\draw   (351.2,293.46) -- (354.62,291.44) -- (353.67,296.68) ;

\draw [line width=1.5]    (338.65,252.15) .. controls (339.39,251.09) and (340.1,250.15) .. (340.8,249.33) .. controls (348.27,240.52) and (353.8,244.48) .. (358.65,252.15) ;
\draw    (347.15,251.6) -- (354.52,239.64) ;
\draw   (351.2,241.46) -- (354.62,239.44) -- (353.67,244.68) ;

\draw [line width=1.5]    (338.65,222.15) .. controls (339.39,221.09) and (340.1,220.15) .. (340.8,219.33) .. controls (348.27,210.52) and (353.8,214.48) .. (358.65,222.15) ;
\draw    (347.15,221.6) -- (354.52,209.64) ;
\draw   (351.2,211.46) -- (354.62,209.44) -- (353.67,214.68) ;

\draw [line width=1.5]    (220.65,170.15) .. controls (221.39,169.09) and (222.1,168.15) .. (222.8,167.33) .. controls (230.27,158.52) and (235.8,162.48) .. (240.65,170.15) ;
\draw    (229.15,169.6) -- (236.52,157.64) ;
\draw   (233.2,159.46) -- (236.62,157.44) -- (235.67,162.68) ;

\draw [line width=1.5]    (220.65,140.15) .. controls (221.39,139.09) and (222.1,138.15) .. (222.8,137.33) .. controls (230.27,128.52) and (235.8,132.48) .. (240.65,140.15) ;
\draw    (229.15,139.6) -- (236.52,127.64) ;
\draw   (233.2,129.46) -- (236.62,127.44) -- (235.67,132.68) ;

\draw [line width=1.5]    (220.65,88.15) .. controls (221.39,87.09) and (222.1,86.15) .. (222.8,85.33) .. controls (230.27,76.52) and (235.8,80.48) .. (240.65,88.15) ;
\draw    (229.15,87.6) -- (236.52,75.64) ;
\draw   (233.2,77.46) -- (236.62,75.44) -- (235.67,80.68) ;

\draw [line width=1.5]    (220.65,58.15) .. controls (221.39,57.09) and (222.1,56.15) .. (222.8,55.33) .. controls (230.27,46.52) and (235.8,50.48) .. (240.65,58.15) ;
\draw    (229.15,57.6) -- (236.52,45.64) ;
\draw   (233.2,47.46) -- (236.62,45.44) -- (235.67,50.68) ;

\draw [line width=1.5]    (220.65,334.15) .. controls (221.39,333.09) and (222.1,332.15) .. (222.8,331.33) .. controls (230.27,322.52) and (235.8,326.48) .. (240.65,334.15) ;
\draw    (229.15,333.6) -- (236.52,321.64) ;
\draw   (233.2,323.46) -- (236.62,321.44) -- (235.67,326.68) ;

\draw [line width=1.5]    (220.65,304.15) .. controls (221.39,303.09) and (222.1,302.15) .. (222.8,301.33) .. controls (230.27,292.52) and (235.8,296.48) .. (240.65,304.15) ;
\draw    (229.15,303.6) -- (236.52,291.64) ;
\draw   (233.2,293.46) -- (236.62,291.44) -- (235.67,296.68) ;

\draw [line width=1.5]    (220.65,252.15) .. controls (221.39,251.09) and (222.1,250.15) .. (222.8,249.33) .. controls (230.27,240.52) and (235.8,244.48) .. (240.65,252.15) ;
\draw    (229.15,251.6) -- (236.52,239.64) ;
\draw   (233.2,241.46) -- (236.62,239.44) -- (235.67,244.68) ;

\draw [line width=1.5]    (220.65,222.15) .. controls (221.39,221.09) and (222.1,220.15) .. (222.8,219.33) .. controls (230.27,210.52) and (235.8,214.48) .. (240.65,222.15) ;
\draw    (229.15,221.6) -- (236.52,209.64) ;
\draw   (233.2,211.46) -- (236.62,209.44) -- (235.67,214.68) ;

\draw    (198,298.33) .. controls (241.59,280.73) and (232.54,287.26) .. (264.75,297.02) ;
\draw [shift={(266.25,297.47)}, rotate = 196.39] [color={rgb, 255:red, 0; green, 0; blue, 0 }  ][line width=0.75]    (10.93,-3.29) .. controls (6.95,-1.4) and (3.31,-0.3) .. (0,0) .. controls (3.31,0.3) and (6.95,1.4) .. (10.93,3.29)   ;
\draw    (198,327.33) .. controls (241.59,309.73) and (232.54,316.26) .. (264.75,326.02) ;
\draw [shift={(266.25,326.47)}, rotate = 196.39] [color={rgb, 255:red, 0; green, 0; blue, 0 }  ][line width=0.75]    (10.93,-3.29) .. controls (6.95,-1.4) and (3.31,-0.3) .. (0,0) .. controls (3.31,0.3) and (6.95,1.4) .. (10.93,3.29)   ;
\draw    (198,216.33) .. controls (241.59,198.73) and (232.54,205.26) .. (264.75,215.02) ;
\draw [shift={(266.25,215.47)}, rotate = 196.39] [color={rgb, 255:red, 0; green, 0; blue, 0 }  ][line width=0.75]    (10.93,-3.29) .. controls (6.95,-1.4) and (3.31,-0.3) .. (0,0) .. controls (3.31,0.3) and (6.95,1.4) .. (10.93,3.29)   ;
\draw    (198,245.33) .. controls (241.59,227.73) and (232.54,234.26) .. (264.75,244.02) ;
\draw [shift={(266.25,244.47)}, rotate = 196.39] [color={rgb, 255:red, 0; green, 0; blue, 0 }  ][line width=0.75]    (10.93,-3.29) .. controls (6.95,-1.4) and (3.31,-0.3) .. (0,0) .. controls (3.31,0.3) and (6.95,1.4) .. (10.93,3.29)   ;
\draw    (197,164.33) .. controls (240.59,146.73) and (231.54,153.26) .. (263.75,163.02) ;
\draw [shift={(265.25,163.47)}, rotate = 196.39] [color={rgb, 255:red, 0; green, 0; blue, 0 }  ][line width=0.75]    (10.93,-3.29) .. controls (6.95,-1.4) and (3.31,-0.3) .. (0,0) .. controls (3.31,0.3) and (6.95,1.4) .. (10.93,3.29)   ;
\draw    (197,135.33) .. controls (240.59,117.73) and (231.54,124.26) .. (263.75,134.02) ;
\draw [shift={(265.25,134.47)}, rotate = 196.39] [color={rgb, 255:red, 0; green, 0; blue, 0 }  ][line width=0.75]    (10.93,-3.29) .. controls (6.95,-1.4) and (3.31,-0.3) .. (0,0) .. controls (3.31,0.3) and (6.95,1.4) .. (10.93,3.29)   ;
\draw    (197,53.33) .. controls (240.59,35.73) and (231.54,42.26) .. (263.75,52.02) ;
\draw [shift={(265.25,52.47)}, rotate = 196.39] [color={rgb, 255:red, 0; green, 0; blue, 0 }  ][line width=0.75]    (10.93,-3.29) .. controls (6.95,-1.4) and (3.31,-0.3) .. (0,0) .. controls (3.31,0.3) and (6.95,1.4) .. (10.93,3.29)   ;
\draw    (197,82.33) .. controls (240.59,64.73) and (231.54,71.26) .. (263.75,81.02) ;
\draw [shift={(265.25,81.47)}, rotate = 196.39] [color={rgb, 255:red, 0; green, 0; blue, 0 }  ][line width=0.75]    (10.93,-3.29) .. controls (6.95,-1.4) and (3.31,-0.3) .. (0,0) .. controls (3.31,0.3) and (6.95,1.4) .. (10.93,3.29)   ;

\draw (50,289.4) node [anchor=north west][inner sep=0.75pt]    {$\ket{0}$};
\draw (50,319.4) node [anchor=north west][inner sep=0.75pt]    {$\ket{0}$};
\draw (101-3,291.17) node [anchor=north west][inner sep=0.75pt]   [align=left] {Hd};
\draw (158-2,305-4) node [anchor=north west][inner sep=0.75pt]   [align=left] {\begin{minipage}[lt]{24.28pt}\setlength\topsep{0pt}
\begin{center}
CNOT
\end{center}

\end{minipage}};
\draw (101-3,320.17) node [anchor=north west][inner sep=0.75pt]   [align=left] {Hd};
\draw (275-2,305-4) node [anchor=north west][inner sep=0.75pt]   [align=left] {\begin{minipage}[lt]{24.28pt}\setlength\topsep{0pt}
\begin{center}
CNOT
\end{center}

\end{minipage}};
\draw (50,208.4) node [anchor=north west][inner sep=0.75pt]    {$\ket{0}$};
\draw (50,236.4) node [anchor=north west][inner sep=0.75pt]    {$\ket{0}$};
\draw (158-2,223-4) node [anchor=north west][inner sep=0.75pt]   [align=left] {\begin{minipage}[lt]{24.28pt}\setlength\topsep{0pt}
\begin{center}
CNOT
\end{center}

\end{minipage}};
\draw (100-3,238.17-28) node [anchor=north west][inner sep=0.75pt]   [align=left] {Hd};
\draw (275-2,223-4) node [anchor=north west][inner sep=0.75pt]   [align=left] {\begin{minipage}[lt]{24.28pt}\setlength\topsep{0pt}
\begin{center}
CNOT
\end{center}

\end{minipage}};
\draw (50,126.4) node [anchor=north west][inner sep=0.75pt]    {$\ket{0}$};
\draw (50,154.4) node [anchor=north west][inner sep=0.75pt]    {$\ket{0}$};
\draw (100-3,129+28) node [anchor=north west][inner sep=0.75pt]   [align=left] {Hd};
\draw (158-2,142-4) node [anchor=north west][inner sep=0.75pt]   [align=left] {\begin{minipage}[lt]{24.28pt}\setlength\topsep{0pt}
\begin{center}
CNOT
\end{center}

\end{minipage}};
\draw (275-2,142-4) node [anchor=north west][inner sep=0.75pt]   [align=left] {\begin{minipage}[lt]{24.28pt}\setlength\topsep{0pt}
\begin{center}
CNOT
\end{center}

\end{minipage}};
\draw (50,44.4) node [anchor=north west][inner sep=0.75pt]    {$\ket{0}$};
\draw (50,72.4) node [anchor=north west][inner sep=0.75pt]    {$\ket{0}$};
\draw (158-2,60-4) node [anchor=north west][inner sep=0.75pt]   [align=left] {\begin{minipage}[lt]{24.28pt}\setlength\topsep{0pt}
\begin{center}
CNOT
\end{center}

\end{minipage}};
\draw (275-2,60-4) node [anchor=north west][inner sep=0.75pt]   [align=left] {\begin{minipage}[lt]{24.28pt}\setlength\topsep{0pt}
\begin{center}
CNOT
\end{center}

\end{minipage}};
\draw (91,62) node [anchor=north west][inner sep=0.75pt]   [align=left] {\begin{minipage}[lt]{19.18pt}\setlength\topsep{0pt}
\begin{center}
$\vb_1$
\end{center}

\end{minipage}};
\draw (125,145.33) node [anchor=north west][inner sep=0.75pt]   [align=left] {\begin{minipage}[lt]{19.18pt}\setlength\topsep{0pt}
\begin{center}
$\vb_2$
\end{center}

\end{minipage}};
\draw (123,226.33) node [anchor=north west][inner sep=0.75pt]   [align=left] {\begin{minipage}[lt]{19.18pt}\setlength\topsep{0pt}
\begin{center}
$\vb_3$
\end{center}

\end{minipage}};
\draw (125,309.33) node [anchor=north west][inner sep=0.75pt]   [align=left] {\begin{minipage}[lt]{19.18pt}\setlength\topsep{0pt}
\begin{center}
$\vb_4$
\end{center}
\end{minipage}};
\end{tikzpicture}
\end{centering}
\caption{QPT setup for a two-qubit CNOT gate, with $n_i = 4$ initial states and $n_s = 2$ time delays. In Section \ref{section:QPTsetup} we stated that we apply the process to be identified before we measure any state, and that it is a choice that we had to make in order to implement the setup experimentally. Indeed \cite{Note3} does not let us measure the actual qubits that have supposedly been set to $\ket{0}$ and have never been modified by a quantum gate.}
\label{fig:setup_exp}
\end{figure}
The four circuits of Fig. \ref{fig:setup_exp} are implemented. Each of them is an explicit version of Fig. \ref{fig:deux_td} (with $k=1$ to $k=4$) for $n_{qb} = 2$. The $n_t = 2 n_{qb}+1 = 5$ measurement types are performed $n_c = 250$ times on each one of the $8$ measured states. In total $n_c n_t n_s n_i = 250\times 5\times 2 \times 4 = 10000$ quantum measurements are performed. Table 2 (in Appendix \ref{section:tables}) records the experimental measurement counts for each outcome and each state. 

We want to use the QST algorithm described in Appendix \ref{section:DataModel} in order to get an estimate of the measured states. There is a small problem here, the measurement types defined in Appendix \ref{section:DataModel} are $ZZ, ZX, ZY, XX$ and $YX$ for two qubits, but the measurements that we have performed on the real quantum computer are $ZZ, ZX, ZY, XX$ and $YY$ (the last measurement type is not the same). It turns out that the QST algorithm of Appendix \ref{section:DataModel} also works on those measurements. The initial estimate of the measured state will be wrong (because it only works with the former set of measurements), but the QST algorithm has two steps (see Section \ref{section:DataModel3}), and the second step (fine tuning with maximum likelihood) corrects the initial estimate in the two-qubit case (because it can work with the types of measurements that are available). This is not a big problem for two reasons: (i) as just stated, the QST algorithm still works (ii) we aim to test the QPT algorithm not the QST, the former relies on the QST output, and does not use the measurements directly.

From the measurements, we estimate the $8$ measured states with the QST algorithm. We arrange the estimated states in the matrices $\widehat{\Xv}$ and $\widehat{\Yv}$ defined in Section \ref{section:QPT_ls} (they are estimates of $\Xv = [\Mv \vb_{1}, \Mv \vb_{2}, \Mv \vb_{3}, \Mv \vb_{4}]$ and $\Yv =  [\Mv^2 \vb_{1}, \Mv^2 \vb_{2}, \Mv^2 \vb_{3}, \Mv^2 \vb_{4}]$ respectively). Here are their numerical values:

\noindent \resizebox{\linewidth}{!}{\arraycolsep=2pt $\widehat{\Xv} = \begin{pmatrix} 1.00- 0.00 i & 0.76- 0.00 i & 0.70- 0.00 i & 0.49- 0.00 i \\ 
0.01- 0.01 i & 0.65- 0.01 i & -0.06+ 0.05 i & 0.45- 0.00 i \\ 
-0.02- 0.03 i & 0.02- 0.03 i & 0.06- 0.03 i & 0.54- 0.06 i \\ 
0.03- 0.06 i & -0.01- 0.08 i & 0.68- 0.19 i & 0.50- 0.02 i \end{pmatrix}$}

\noindent \resizebox{\linewidth}{!}{\arraycolsep=2pt $ \widehat{\Yv} = \begin{pmatrix} 1.00- 0.00 i & 0.72- 0.00 i & 0.70- 0.00 i & 0.54- 0.00 i \\ 
0.01+ 0.01 i & 0.70+ 0.06 i & -0.02- 0.00 i & 0.46+ 0.04 i \\ 
0.03- 0.00 i & -0.02- 0.01 i & 0.72- 0.02 i & 0.51+ 0.08 i \\ 
0.00- 0.01 i & 0.01- 0.01 i & 0.01- 0.02 i & 0.47+ 0.08 i \end{pmatrix}$} 

We then use the phase recovery algorithm of Section \ref{section:phase} to compute $\widetilde{\Yv}$ which is the same as $\widehat{\Yv}$ but with each column multiplied by a phase factor. 
We then use the method of Section \ref{section:idea} to find the unitary matrix $\widehat{\Mv}_{LS}$ that links $\widehat{\Xv}$ and $\widetilde{\Yv}$ (the re-phased version of $\widehat{\Yv}$) best. Finally, we change its global phase by a factor $e^{i\phi}$ in order to compare it with $\Mv_{tg}$: 

$\widehat{\Mv}_{LS} \longleftarrow e^{i\theta}\widehat{\Mv}_{LS} $ with $\theta = arg\left(tr(\widehat{\Mv}_{LS}^*\Mv_{tg})\right)$, like in (\ref{eqn:phaseM}). This last step is only possible (and useful) if we know $\Mv_{tg}$, which is the case here. If we had no idea of what the gate was supposed to do, we would not perform this step.

\noindent Here is the resulting (rounded) estimate:

\noindent \resizebox{\linewidth}{!}{\arraycolsep=2pt%
$\widehat{\Mv}_{LS}=\begin{pmatrix} 
0.98- 0.17 i & -0.02- 0.02 i & 0.02+ 0.02 i & 0.01+ 0.07 i \\ 
0.02- 0.02 i & 0.99- 0.09 i & 0.01+ 0.03 i & 0.03+ 0.01 i \\ 
0.00+ 0.07 i & -0.02+ 0.01 i & 0.08- 0.02 i & 0.99+ 0.08 i \\ 
-0.01+ 0.02 i & -0.01+ 0.03 i & 0.98+ 0.18 i & -0.07- 0.04 i
\end{pmatrix}$ }

The moduli are close to their target values but there are fairly significant errors on the phases that cannot be corrected with a global phase shift. This is particularly noticeable between the first and third columns. The distance between $\widehat{\Mv}_{LS}$ and the target $\Mv_{tg}$ can be defined as $\epsilon(\widehat{\Mv}_{LS}, \Mv_{tg}) \simeq 0.11$ with $\epsilon$ defined in (\ref{eqn:ourerr}). According to Fig. \ref{fig:eps1}, $0.11$ is a very reasonable error with $n_c=250$ in this setup, so we cannot reject the hypothesis that the gate is perfectly realized by simply looking at the error. This is not our objective however, we do not want to perform quantum gate benchmarking but quantum gate tomography.

\section{Conclusion and future work}
In this paper we introduced a quantum process tomography (QPT) method that is very flexible on the values of the initial states used (as long as they remain pure). We proposed a semi-blind setup (Fig. \ref{fig:setup_hard}) that splits the copies of each initial state into $n_s$ groups, and measures the copies of the $k$-th group (only once) after they go through the process $k\in\{1, ..., n_s\}$ times. This trick allows us to estimate the states that go through the process rather than assuming that the input states are correctly prepared. The resulting QPT setup is resilient to systematic errors on the initial states. We proved that, in the absence of QST errors, our QPT algorithm always gives a perfect estimate of the process to be identified if we satisfy the condition of (\ref{eqn:CNS}). We also showed that, if (\ref{eqn:CNS}) is not satisfied, then, it is not possible to find a QPT algorithm that works (because there are several different processes that yield the same measurement outcomes). 

As explained in Section \ref{section:SQPT} our algorithm would also work on a more standard QPT (SQPT) setup (the values of the initial states are known beforehand, and all copies of the initial states are processed the same way: they go through the process only once and then get measured). Our algorithm is attractive (running on the semi-blind and the SQPT setups) when we compare it to \cite{nearunitary} which is the best known unitary quantum process tomography algorithm in the literature (see the simulations in Section \ref{section:simuBaldwin}).  We also tested the resilience of our algorithm to different types of errors and we tested it experimentally to identify a CNOT gate on a trapped-ions qubit quantum computer.

Our QPT algorithm does not require an initial estimate of the process and is quite fast after the QST of the measured states has been performed, we only have to compute a few dot products (see Section \ref{section:phase}) and a singular value decomposition (see Section \ref{section:idea}). The limitation of our estimate is that it minimizes the least square error on the QST results (after the phase recovery), and as stated at the end of Section \ref{section:idea}, this is not optimal from a maximum likelihood standpoint. The least-square estimate of the process can be fine tuned with a slower but more precise gradient descent algorithm that starts at our estimates and finds the unitary matrix that maximizes the likelihood of the measurements without performing QST. Defining a model of the likelihood of the quantum measurements that takes into account the multinomial error as well as the potential non-purity of the measured states is a challenge in and of itself and we intend to study it in the future.

A proponent of gate set tomography (GST) could also criticize our QPT setup by pointing out the fact that it only works if the measurements we perform on the unknown states are known with precision. This is a fair point, and the fact that the measurements that we use might be flawed is a drawback of our algorithm (it is also a drawback of all QPT algorithms in the literature that do not use GST) that we tolerated because we need a frame of reference. We trust neither the values of the input states nor the process to be identified, but we have to trust the measurement process, otherwise we would be left with the gauge error that plagues GST. This flaw in our algorithm is somewhat mitigated by the fact that, we only rely on very simple unentangled measurements. We intend to mitigate it further by introducing a (blind) quantum detector tomography algorithm that can be used to estimate some parameters of the measurements that we propose to use on each qubit without using predetermined states. This is possible because the measurements that we perform are somewhat redundant.

\appendix
\section{States, measurements and QST}
\label{section:DataModel}
The current appendix quickly goes over the choices that we made for the measurements performed and the QST algorithm. We need it because the central QPT algorithm that we aim to present in this paper relies on some kind of QST. However, the choices we make in this section do not affect the QPT algorithm of Section \ref{section:QPT_ls}, and other types of measurements and QST can be used.

\subsection{Definition of the measurements}
\label{section:Pauli}
We chose to perform measurements that have $d$ outcomes ($d$-outcome measurements), as $d$ is the maximum number of outcomes for a type of quantum measurement in a $d$-dimensional Hilbert space. We are not interested in the actual values of the outcomes, they can be denoted as $1, ..., d$ or $0...0, 0...01, ..., 1...1$, it makes no difference to us. We are interested in the probabilities of the measurement outcomes. They are estimated by performing the measurements several times on copies of the considered state, and computing the frequencies of occurrence of each outcome.

If a projective measurement $\mathcal{M}$ has $d$ outcomes, then there exists a unitary matrix $\Ev_{\mathcal{M}}$ such that the probabilities of all outcomes when measuring any state $\vb$ are contained in the vector $|\Ev_{\mathcal{M}}^* \vb|^2$ (with the convention that $|.|^2$ is the element-wise squared modulus), this is known as the Born rule. The columns of the matrix $\Ev_{\mathcal{M}}$ (each of them is defined up to a global phase) entirely characterize the type of measurement (up to the values of the outcomes). We call it the eigenvector matrix, because its columns are the eigenvectors of the Hermitian matrix that is often used to characterize the measurements in the literature (see (2.102) in \cite{bookNC}).

We also choose to use unentangled measurements. Unentangled measurement are quantum measurements that can be performed in parallel on each qubit. To our knowledge unentangled measurements are the only types of measurements that can be performed on the current version of quantum computers without using entangling gates (that have to be characterized with QPT).

If the measurement $\mathcal{M}$ is unentangled, then $\Ev_{\mathcal{M}}$ can be written as the tensor product of $n_{qb}$ matrices in $\mathds{U}_2(\mathds{C})$. Those $2\times 2$ matrices are the eigenvector matrices associated with each 1-qubit measurement.

For each qubit, the $2\times 2$ unitary matrices we use are among the following three: 
\begin{equation}
\label{eqn:Pauli2}
\Ev_X = \frac{1}{\sqrt{2}} \begin{pmatrix} 1 & 1\\
1 & -1
\end{pmatrix} 
\Ev_Y = \frac{1}{\sqrt{2}} \begin{pmatrix} 1 & 1\\
i & -i
\end{pmatrix} 
\Ev_Z = \begin{pmatrix} 1 & 0\\
0 & 1
\end{pmatrix}.
\end{equation}
If the qubit represents the spin of an electron, those eigenvector matrices represent the measurement of the spin component along directions $X, Y$ and $Z$.

For two or more qubits, measurements can be performed along directions  $X, Y$ or $Z$ for each qubit. It can be shown that the resulting eigenvector matrix is the tensor product of the 2-dimensional matrices of (\ref{eqn:Pauli2}). For example for two qubits, performing a measurement along $Z$ for the first qubit and along $X$ for the second one yields the following eigenvector matrix

$\Ev_{ZX} = \Ev_{Z} \otimes \Ev_{X} =  \frac{1}{\sqrt{2}} \begin{pmatrix} 1 & 1 & 0 & 0\\
1 & -1 & 0 & 0\\
0 & 0 & 1 & 1\\
0 & 0 & 1 & -1
\end{pmatrix}$.

In this example, if the qubits represent the spins of two electrons, then the measurement we perform is equivalent  to measuring the first spin component along $Z$ and the second along $X$. The spin pair measurement has 4 possible outcomes $(+\frac{1}{2}, +\frac{1}{2}), (+\frac{1}{2}, -\frac{1}{2}), (-\frac{1}{2}, +\frac{1}{2})$ and $(-\frac{1}{2}, -\frac{1}{2})$ in normalized units, and if $\vb$ represents the considered state, the probabilities of each outcome are in the vector $|\Ev_{ZX}^* \vb|^2$. 

There are $3^{n_{qb}}$ types of measurements with this definition. We will only use those characterized by the $2 n_{qb}+1$ eigenvector matrices contained in the following set:

\noindent \scalebox{1}[1]{$\mathcal{E} = \bigg\{\Ev_{\underbrace{Z...Z}_\textrm{$n_{qb}$ times}}, \Big\{ \Ev_{\underbrace{Z...Z}_\textrm{$n_{qb}-i$ times} \scalebox{1}[1]{S}  \underbrace{X...X}_\textrm{$i-1$ times}}, \begin{matrix}1\leq i \leq n_{qb} \\ S\in\{X, Y\} \end{matrix}\Big\} \bigg\}$.}

For example, with $n_{qb}=2$, we have: $\Ev_{ZZ}, \Ev_{ZX}, \Ev_{ZY},$ $\Ev_{XX}$ and $\Ev_{YX}$.
\subsection{Quantum state tomography}
\label{section:DataModel3}
In \cite{PRA} we showed how a quantum state can be estimated by performing measurements for $n_c$ copies of that state, successively with each measurement matrix of $\mathcal{E}$. The total number of measurements performed on copies of the state is $n_t n_c = (2n_{qb}+1)n_c$. This might seem unimpressive as, for example, \cite{Goyeneche} would only use $4 n_c$ or $5 n_c$ copies (depending on the used basis). But the 4 or 5 measurement types of \cite{Goyeneche} are not all unentangled, and we were unable to find fewer than $(2n_{qb}+1)$ types of unentangled measurements that make QST possible with a closed form explicit algorithm.

We here use the algorithm of Section 3 of \cite{PRA} fined tuned with the maximum likelihood method of Section 4 with the Gaussian version of the likelihood. We use the Gaussian version instead of the real likelihood or the mixed algorithm because we want an algorithm that will ``catch" types of errors that are not considered by the multinomial likelihood model (typically decoherence).
\section{mathematical demonstrations for the identifiability condition}
\label{section:math}
Our identifiability condition is (\ref{eqn:CNS}). It is necessary and sufficient for $\Mv$ to be identifiable (up to a global phase and with a given QPT setup) in the set of unitary matrices. In Section \ref{section:math1}, we show that it is a sufficient condition. Importantly we do this by showing that our algorithm is then able to retrieve $\Mv$ up to a global phase. This means that our QPT algorithm works in any situation where QPT is theoretically possible. In Section \ref{section:math2}, we show that it is a necessary condition. Finally, in Section \ref{section:math3}, we show that (\ref{eqn:CNS}) is equivalent to another condition of the literature: (\ref{eqn:CNSreich}).  In all the current appendix, we assume that there is no QST error (the columns of $\widehat{\Xv}$ and of $\widehat{\Yv}$ are the same as those of $\Xv$ and of $\Yv$ respectively up to global phases).


\subsection{Proof that (\ref{eqn:CNS}) is sufficient}
\label{section:math1}

In order to show that (\ref{eqn:CNS}) is sufficient for QPT to be possible (or for ``$\Mv$ to be identifiable") with the setup of Fig. \ref{fig:setup_hard}, we need to define what we mean by ``QPT is possible" (or ``$\Mv$ is identifiable"). Formally, this means that $\Xv$ (defined by $\Mv$ and the $\vb_i$, see (\ref{eqn:XY0}) and (\ref{eqn:QST})) is such that the ensemble $\mathcal{U} (\widehat{\Xv}, \widehat{\Yv})$ of unitary matrices that are compatible with the QST results $\widehat{\Xv}$ and $\widehat{\Yv}$:

\small
\begin{equation}
\label{eqn:defU}
\mathcal{U} (\widehat{\Xv}, \widehat{\Yv})= \left\{\Uv \in \mathds{U}_d , \exists  \xib \in \mathds{R}^{n_x}, \Uv \ \widehat{\Xv} \ \Dv(\xib)^* = \widehat{\Yv}\right \} 
\end{equation}
\normalfont

\noindent contains only $\Mv$ and matrices that are the same as $\Mv$ up to a global phase: $\mathcal{U} (\widehat{\Xv}, \widehat{\Yv}) = \{e^{i \phi}\Mv\}_\phi$.

(\ref{eqn:CNS}) is a sufficient condition for QPT to be possible because if (\ref{eqn:CNS}) is true, then, essentially, the algorithms of Sections \ref{section:idea} and \ref{section:phase} yield a unique $\Mv$ up to a global phase:
\begin{itemize} [leftmargin=*]
\item The algorithm of Section \ref{section:phase} always succeeds, i.e. we exit the algorithm at Step 4 or 6. 
Indeed, if $rank\left( \Fv_S^{n_x}(x_{\ell_0}) \right) = d$, then, even if the algorithm starts poorly and we have to set $b_{orth} = 0$ at Step 7, then the condition of Step 6 will eventually be satisfied after going through Step 5 at most $n_x$ times. This is because of the equality we pointed out in Section \ref{section:CNS} below (\ref{eqn:CNS}) between the rank of $\Fv_S^{k}(\xb_{\ell_0})$ and the dimension of the subspace spanned by the columns of $\mathcal{S}$ after going through Step 5  $k$ times with $b_{orth} = 0$. The reader could think that we could exit the algorithm prematurely (i.e. before the set $\mathcal{S}$ spans $\mathds{C}^d$) because the condition that makes us loop from Step 5 to Step 3 stops being satisfied. This is a non-issue, because if the condition is not satisfied then it is pointless to continue as the elements of $\mathcal{S}$ will not increase even if we were to go to Step 3.
\item The phase differences between the columns of $\widetilde{\Yv}$ that we exit the algorithm of Section \ref{section:phase} with are the unique  solution of (\ref{eqn:propriete1}), because (\ref{eqn:propriete1}) has a unique solution if and only if $\widehat{\yb}_{\ell_1}\not\perp \widehat{\yb}_{\ell_2}$ for the $\ell_1, \ell_2$ that we use to compute (\ref{eqn:argument}) at Step 3.(c) of Section \ref{section:phase}. And we made sure that this is the case in the phase recovery algorithm. 
\item The ranks of the matrices $\widetilde{\Yv}$ and $\widehat{\Xv}$ that we exit the algorithm of Section \ref{section:phase} with are full ($d$), because (i) $\Fv_S^{n_x}(\xb_{\ell})$ only contains columns of $\Xv$ by definition, thus (\ref{eqn:CNS}) $\Rightarrow rank(\Xv)=d$, (ii) $\Mv$ is unitary, thus $rank(\Yv)=rank(\Mv \Xv)=d$, (iii) There is no systematic error, thus $rank(\widehat{\Xv}) = rank(\Xv) = d$ and $rank(\widetilde{\Yv}) = rank(\Yv) = d$. Therefore, the total least square problem with a unitarity constraint has a single solution: $\Mv_{LS}$ (as discussed at the end of Section \ref{section:idea}). But any matrix that is the same as $\Mv_{LS}$ up to a global phase is also a solution of the QPT problem. This is because an arbitrary choice has been made for the global phase of $\widetilde{\Yv}$ in Section \ref{section:phase} ($\ell_0$ has been chosen with (\ref{eqn:maxl0}) and $\xi_{\ell_0}$ has been set to $0$).
\end{itemize}

Therefore $\mathcal{U} (\widehat{\Xv}, \widehat{\Yv}) = \{e^{i \phi}\Mv\}_\phi$.
\subsection{Proof that (\ref{eqn:CNS}) is necessary}
\label{section:math2}
Let us now show that (\ref{eqn:CNS}) is a necessary condition. We assume that $\Xv$ is such that (\ref{eqn:CNS}) is false, i.e.  

\begin{equation}
\label{eqn:nonCNS}
\exists \ell\in \{1, ..., n_x\}, \  rank\left(\Fv_S^{n_x}(\xb_{\ell})\right) < d
\end{equation}
 We want to show that there exists a matrix $\Mv_2$ that differs from $\Mv$ by more than a global phase such that 
$\Mv_2 \in \mathcal{U} (\Xv, \Mv \Xv)$ 
 with the same definition for $\mathcal{U}$ as in (\ref{eqn:defU}). Its is straightforward to show that, since there is no QST error, $\mathcal{U} (\Xv, \Mv \Xv) = \mathcal{U} (\widehat{\Xv}, \widehat{\Yv})$.
This means that the processes represented $\Mv_2$ and $\Mv$ are different quantum processes (they differ by more than a global phase) but our measurements are not sufficient to distinguish them. 

Let us first define the function  $\Gv_{S}$ and list its useful properties:

$\Gv_{S}: k \longrightarrow \Fv_S^{k}(\xb_{\ell})$:
\begin{itemize} [leftmargin=*]
\item The columns of $\Gv_{S}(k)$ are all also columns of $\Gv_{S}(k+1)$.
\item If $k_0$ is the smallest integer such that $\Gv_{S}(k_0) = \Gv_{S}(k_0+1)$, then $\forall k \geq k_0$, $\Gv_{S}(k_0) = \Gv_{S}(k)$.
\item The number of columns of $\Gv_{S}(k)$ is upper bounded by $n_x$ (the number of columns of $\Xv$) $\forall k$.
\end{itemize}
Therefore the number of columns of $\Gv_{S}(k)$ increases strictly with $k$ for the first $k_0$ iterations, and for $k \geq k_0$ $\Gv_{S}$ becomes constant. And $k_0$ has to be smaller than $n_x$ because the number of columns increases by at least 1 at each iteration before $k_0$ and is bounded by $n_x$.

For any $\ell$, and in particular for the $\ell$ of (\ref{eqn:nonCNS}), $\Fv_S^{n_x}(\xb_{\ell}) = \Fv_S^{n_x+1}(\xb_{\ell})$. Therefore, we can split the columns of $\Xv$ into two groups $\Xv_s$ (defined as $\Fv_S^{n_x}(\xb_{\ell})$) and $\Xv_f$ (defined as the matrix that contains the other columns of $\Xv$ in order). The matrix $\Xv_f$ can be empty, but if it is not, the columns it contains are all orthogonal to the columns of $\Xv_s$ (since $\Fv_S(\Xv_s) = \Xv_s$). According to (\ref{eqn:nonCNS}), $\Xv_s$ is not of full rank.

From there we consider the only two possible implications of (\ref{eqn:CNS}) being false:
\begin{enumerate}
\item $\Xv_f$ is empty, $\Xv_s = \Xv$ and $rank(\Xv)<d$.
\item $\Xv_f$ is not empty and $\Xv$ can be decomposed: $\Xv = [\Xv_s, \Xv_f] \Qv_{per}$, where $\Qv_{per}$ is an $n_x \times n_x$ permutation matrix.
\end{enumerate}
And either one of these conditions makes QPT impossible because:
\begin{itemize} [leftmargin=*]
\item If the condition of Case 1 is true, there exists a unit norm vector called $\vb_{ker}$ in the null-space of $\Xv^*$, i.e. $\Xv^* \vb_{ker} = \boldsymbol{0}$ and $\vb_{ker}^* \Xv = \boldsymbol{0}$. We call $\Vv_{hker}$ a $d\times (d-1)$ matrix such that $\Pv_v = [\Vv_{hker}, \vb_{ker}]$ is a unitary matrix ($\Vv_{hker}$ and $\Pv_v $ are not unique for a given $\vb_{ker}$, $\Vv_{hker}$ can be any orthogonal basis of the subspace orthogonal to $\vb_{ker}$). We define $\begin{bmatrix} \Cv_1 & \cb_2  \end{bmatrix} = \Mv \Pv_v$ ($\Cv_1 \in \mathds{C}^{d\times (d-1)}, \cb_2 \in \mathds{C}^{d}$) and a straightforward calculation shows that for any angle $0<\phi<2 \pi$, the unitary matrix $\Mv_2 = \begin{bmatrix}\Cv_1 & \cb_2 e^{i\phi} \end{bmatrix} \Pv_v^*$ is not the same as $\Mv$ (even up to a global phase since $d>1$) and $\Mv$ and $\Mv_2$ are both in $\mathcal{U} (\Xv, \Mv\Xv )$ since $\Mv\Xv = \Mv_2 \Xv$. 
\item If the condition of Case 2 is true, we only have to consider the case when $\Xv$ is of full rank (because otherwise, we use the reasoning above), and $vect(\Xv_f)$ is the orthogonal complement of $vect(\Xv_s)$ ($vect$ is the subspace spanned by the column vectors of a matrix). We define $\Pv_s$ as an orthonormal basis of the subspace $vect(\Xv_s)$ and $\Pv_f$ as an orthonormal basis of $vect(\Xv_f)$. 
We define the matrix $\Mv_2$ as $\Mv_2 = \Mv (\Pv_s \Pv_s^* e^{i\phi} + \Pv_f \Pv_f^*)$, and define $\Xv_{alt} = \begin{bmatrix} \Xv_s e^{-i\phi} & \Xv_f \end{bmatrix}\Qv_{per}$ ($\Qv_{per}$ defined above). With those definitions $\Mv_2$ is unitary (this is easy to check), and a straightforward calculation shows that $\Mv\Xv = \Mv_2 \Xv_{alt}$, $\Xv_{alt} $ has the same columns as $\Xv$ up to global phases. Let $\xib$ be the vector that contains those phases and $\Dv(\xib)$ be the diagonal matrix such that $\Xv_{alt} = \Xv \Dv(\xib)$. We have $\Mv\Xv = \Mv_2 \Xv_{alt} = \Mv_2\Xv \Dv(\xib)$, thus $\Mv$ and $\Mv_2$ are both in $\mathcal{U} (\Xv, \Mv\Xv )$  but $\Mv_2$ and $\Mv$ differ by more than a global phase.
\end{itemize}
\subsection{Equivalence between (\ref{eqn:CNS}) and (\ref{eqn:CNSreich})}
\label{section:math3}

The condition of Reich et al. has been rewritten in (\ref{eqn:CNSreich}), it guarantees the identifiability of the process represented by $\Mv$ among all processes. This is stronger than identifiability among unitary processes, that we guarantee with (\ref{eqn:CNS}). Therefore (\ref{eqn:CNSreich}) implies (\ref{eqn:CNS}). Let us show that (\ref{eqn:CNS}) implies (\ref{eqn:CNSreich}): 

Let $\Cv \in \mathds{U}_d (\mathds{C})$ be a unitary matrix in $Com \left( \left\{ \xb_{\ell} \xb_{\ell}^*\right\}_\ell \right)$. 
Let us show that, if (\ref{eqn:CNS}) is met, then: 
$\exists \theta, \Cv = e^{i\theta} \Iv_d.$

If two matrices commute, then there exists a basis that diagonalizes both of them (see Theorem 1.3.12 in \cite{Horn}). Therefore, the fact that $\Cv$ commutes with all $\left\{ \xb_{\ell} \xb_{\ell}^*\right\}_\ell$ implies that all $\left\{ \xb_{\ell}\right\}_\ell$ are eigenvectors of $\Cv$ (because any basis that diagonalizes $\xb_{\ell} \xb_{\ell}^*$ contains $\xb_{\ell}$ up to a global phase). Let us call $e^{i \lambda_\ell}$ the eigenvalue of $\Cv$ associated with $\xb_{\ell}$ (they are of unit modulus because $\Cv$ is unitary). 

Let us consider two indices $\ell_1$ and $\ell_2$, and let us show that $\xb_{\ell_1} \not\perp \xb_{\ell_2}\Rightarrow e^{i \lambda_{\ell_1}} = e^{i \lambda_{\ell_2}}$: 

\noindent $\xb_{\ell_1}^* \xb_{\ell_2}=(\Cv \xb_{\ell_1})^* \Cv\xb_{\ell_2} = e^{i (\lambda_{\ell_2}-\lambda_{\ell_1}} \xb_{\ell_1}^* \xb_{\ell_2}$. If $\xb_{\ell_1} \not\perp \xb_{\ell_2}$, then we can divide both sides of the equation by $\xb_{\ell_1}^* \xb_{\ell_2}$ and we have $1=e^{i (\lambda_{\ell_2}-\lambda_{\ell_1})}\Rightarrow e^{i \lambda_{\ell_1}} = e^{i \lambda_{\ell_2}}$.

Let $\ell_1$ be in $\{ 1, ..., n_x\}$. $\xb_{\ell_1}$ is not orthogonal to any column of $\Fv_S(\xb_{\ell_1})$ (by definition of $\Fv_S$), and the columns of $\Fv_S(\xb_{\ell_1})$ are also in $\left\{ \xb_{\ell}\right\}_\ell$. Thus, all columns of $\Fv_S(\xb_{\ell_1})$ have $e^{i \lambda_{\ell_1}}$ as the associated eigenvalue. The same is true for the columns of $\Fv_S^k(\xb_{\ell_1})$ for any $k\geq 1$ (straightforward by mathematical induction). In particular, all columns of $\Fv_S^{n_x}(\xb_{\ell_1})$ have the same associated eigenvalue: $e^{i\lambda_{\ell_1}}$. But (\ref{eqn:CNS}) guarantees that $\Fv_S^{n_x}(\xb_{\ell_1})$ has rank $d$. Thus there are $d$ linearly independent columns of $\Fv_S^{n_x}(\xb_{\ell_1})$ that form a basis, they are also eigenvectors of $\Cb$ (like all columns of $\Fv_S^{n_x}(\xb_{\ell_1})$). This basis is therefore an eigenbasis of $\Cb$, and there is only one associated eigenvalue: $e^{i\lambda_{\ell_1}}$. This means that $\Cv = e^{i\lambda_{\ell_1}} \Iv_d$.
\section{Table of the measurement counts}
\label{section:tables}
\begin{table}[h]
\centering
\caption{Measurement counts on the actual quantum computer. For example the value $243$ in the row called $ZZ$ $00$ and the column called $\Mv \vb_1$ means that when measuring $\Mv \vb_1$ with measurement type $ZZ$ we obtained the first outcome (called $00$ here) $243$ times}
\setlength{\tabcolsep}{1pt}
 \begin{tabular}{|l|l|l|l|l|l|l|l|l| }
\hline
- &$\Mv\vb_1$&$\Mv\vb_2$&$\Mv\vb_3$& $\Mv\vb_4$&$\Mv^2\vb_1$& $\Mv^2\vb_2$&$\Mv^2\vb_3$&$\Mv^2\vb_4$\\ \hline 
ZZ 00&243 & 139 & 123 & 58 & 249 & 128 & 123 & 75 \\ \hline 
ZZ 01&6 & 107 & 4 & 52 & 1 & 122 & 0 & 45 \\ \hline
ZZ 10&0 & 1 & 1 & 74 & 0 & 0 & 125 & 71 \\ \hline 
ZZ 11&1 & 3 & 122 & 66 & 0 & 0 & 2 & 59 \\ \hline  
ZX 00&126 & 244 & 54 & 107 & 129 & 249 & 52 & 129 \\ \hline 
ZX 01&122 & 4 & 71 & 1 & 121 & 1 & 64 & 0 \\ \hline 
ZX 10&2 & 2 & 82 & 142 & 0 & 0 & 72 & 121 \\ \hline 
ZX 11&0 & 0 & 43 & 0 & 0 & 0 & 62 & 0 \\ \hline 
ZY 00&120 & 123 & 70 & 55 & 138 & 132 & 64 & 78 \\ \hline
ZY 01&129 & 124 & 59 & 58 & 112 & 118 & 61 & 54 \\ \hline 
ZY 10&1 & 2 & 63 & 73 & 0 & 0 & 61 & 63 \\ \hline 
ZY 11&0 & 1 & 58 & 64 & 0 & 0 & 64 & 55 \\ \hline 
XX 00&63 & 127 & 118 & 248 & 69 & 123 & 125 & 248 \\ \hline  
XX 01&55 & 1 & 1 & 2 & 64 & 0 & 125 & 1 \\ \hline
XX 10&65 & 122 & 4 & 0 & 57 & 127 & 0 & 1 \\ \hline 
XX 11&67 & 0 & 127 & 0 & 60 & 0 & 0 & 0 \\ \hline 
YY 00&54 & 61 & 5 & 59 & 61 & 66 & 59 & 68 \\ \hline  
YY 01&63 & 54 & 112 & 50 & 63 & 56 & 62 & 71 \\ \hline 
YY 10&72 & 62 & 127 & 72 & 56 & 74 & 62 & 58 \\ \hline 
YY 11&61 & 73 & 6 & 69 & 70 & 54 & 67 & 53 \\ \hline 
\end{tabular}
\end{table}

\bibliography{mybibfile}

\end{document}